\documentclass[11pt, twoside, titlepage]{book}
 \usepackage{axodraw}

\usepackage{amsfonts}
\usepackage{amsmath}
\usepackage{amssymb}
\usepackage{url}
\usepackage{graphicx}   
\usepackage{epsfig}     

\usepackage{psfrag}
\usepackage{palatino} 
\usepackage{enumerate}
\usepackage{moreverb}
\usepackage{latexsym}

\def\DashPhoton(#1,#2)(#3,#4)#5#6#7{
\put(\axoxoff,\axoyoff){
}
}

\makeatletter
\def\ps@headings{\let\@mkboth\markboth
  \def\@oddfoot{}
  \def\@evenfoot{}
  \def\@evenhead{\protect\underline{
     \hbox to\hsize{\bf \thepage\hfil \leftmark}}}
  \def\@oddhead{\protect\underline{
     \hbox to\hsize{\bf \rightmark \hfil\thepage}}}
  \def\chaptermark##1{\markboth{##1}{}}
  \def\sectionmark##1{\markright{\ifnum \c@secnumdepth
     >\z@\thesection \ \fi ##1}}}
  \makeatother

\newcommand{\beq}{\begin{equation}}
\newcommand{\eeq}{\end{equation}}
\newcommand{\bdm}{\begin{displaymath}}
\newcommand{\edm}{\end{displaymath}}
\newcommand{\beqna}{\begin{eqnarray}}
\newcommand{\eeqna}{\end{eqnarray}}

\newcommand{\commentt}[1]{}

\begin{document}

\newtheorem{prop}{Proposition}[section]
\newtheorem{lemma}{Lemma}[section]

\frontmatter

\begin{titlepage}
\mbox{}
\vspace{3cm}
\begin{center} \bf \LARGE Quantized Equations of Motion and Currents \\ in Noncommutative Theories
\end{center}
\vspace{3cm}
\begin{center}
Diplomarbeit \\
\vspace{2cm}
{\bf Tobias Reichenbach}\\
\vspace{1cm}
Institut f\"ur Theoretische Physik \\Fakult\"at f\"ur Physik und Geowissenschaften\\Universit\"at Leipzig \\ \vspace{4cm}
Leipzig, im November 2004
\end{center}

\end{titlepage}

\thispagestyle{empty}
\mbox{}\\
\vspace{17cm}
\mbox{}\\
\begin{tabular}{ll}
Betreuer:&Prof. Dr. Klaus Sibold\\
\\
Gutachter:&Prof. Dr. Klaus Sibold\\
&Prof. Dr. Manfred Salmhofer
\end{tabular}
\newpage

\thispagestyle{empty}
\vspace{10cm}
\begin{center}
\begin{minipage}[b][7cm]{7cm}
Immer die eine, die Pappel \\
am Saum des Gedankens. \\
Immer der Finger, der aufragt \\
am Rain.\\
\end{minipage} \\
\vspace{0.5cm}
\hspace{0.8cm}\textsc{Paul Celan}
\end{center}

\newpage
\thispagestyle{empty}
\mbox{}       
\newpage

\thispagestyle{plain}
\begin{center}
\bf Abstract
\end{center}

\vspace{0.5cm}
We study quantized equations of motion and currents, that means equations on the level of Green's functions, in three different approaches to noncommutative quantum field theories. \\
At first, the case of only spatial noncommutativity is investigated in which the modified Feynman rules can be applied. The classical equations of motion and currents are found to be also valid on the quantized level, and the BRS current for NCQED is derived.\\
We then turn to the more complicated case of time-space noncommutativity and consider the approach of TOPT. Additional terms depending on $\theta^{0i}$, which are not present on the classical level, appear in the quantized equations of motion. We conclude that the same terms arise in quantized currents and cause the violation of Ward identities in NCQED. The question of remaining Lorentz symmetry is also discussed and found to be violated in a simple scattering process. \\
Another approach to time-space noncommutative theories uses retarded functions. We present this formalism and discuss the question of unitarity, as well as equations of motion, and currents. The problems that emerge for $\theta^{0i}\neq 0$ are seen to arise from a certain type of diagrams. We propose a modified theory which is unitary and preserves the classical equations of motion and currents on the quantized level.

\newpage
\thispagestyle{plain}
\mbox{}       
\newpage

\tableofcontents

\mainmatter

\addcontentsline{toc}{chapter}{Introduction}

\chapter*{Introduction}

\markboth{Introduction}{}

The nature of spacetime at short distances represents one of the fundamental problems in physics since the establishment of general relativity and quantum theory in the early decades of the last century. It is this period of time when the idea of a discrete spacetime was formulated by Heisenberg \cite{Heisenberg}, being the first step towards noncommutative theories. For a couple of reasons, these theories have received renewed attention and enjoy wide popularity these days\footnote{See \cite{DouglasNikita} and \cite{Szabo} for a review.}.

The original motivation that led Heisenberg to the suggestion of a lattice world with smallest length \cite{Heisenberg} was the belief that in this way he could overcome the problem of infinities met in quantum field theory, for example in the calculation of the electrons self energy. Although having been successful in the latter point, he soon considered this approach as too radical to progress it further. But he was not the only one to think about modifications of microscopic spacetime. Schr\"odinger \cite{Schroedinger} also gave arguments in favour building on the uncertainty principle which is characteristic for quantum theories and which might be in contradiction with a geometry allowing  precise localization.

It was Snyder \cite{Snyder} who proposed in 1947 an implementation of discrete spacetime by replacing the usual coordinates by operators satisfying certain commutation relations, a way that is followed by the modern approach. Introducing this model he, as Heisenberg, aimed to avoid the ultraviolet divergences in quantum field theories. To improve the UV behaviour was also the hope when noncommutative theories appeared in the context of string theory, where they have been shown to arise as low-energy limit of open string theories on $D$-brane configurations in background magnetic fields. Important insight is gained here by the Seiberg-Witten map \cite{SeibergWitten} giving to a classical gauge theory on the ordinary Minkowski space a corresponding noncommutative theory.

Perhaps the most fundamental motive for noncommutative theories comes from the interplay between general relativity and quantum theory. While the first assumes that the geometry of spacetime locally resembles the one of $\mathbb{R}^4$, the second casts doubt on this assumption and suggests drastic alterations on distances near the Planck length $\lambda_P\simeq 1.6\cdot 10^{-35}$m.  Considerations taking up the idea of spacetime uncertainty relations have been carried out in \cite{DFR} and can be comprehended from the following argument. If we want to localize a particle in a region $\Delta x_0, \Delta x_1,...,\Delta x_3$, Heisenberg's uncertainty principle tells us that we need an energy transfer of the order $E\sim \frac{\hbar c}{\Delta x}$ with $\Delta x$ being the smallest of the $\Delta x_\mu$'s. According to general relativity, this energy modifies the spacetime by curving it and may for extremely high values, which would emerge in very accurate position measurements, cause a black hole of radius $R\sim \frac{EG}{c^3}$. In the latter situation, which occurs when the $\Delta x_\mu$'s are of the order of  the Planck length $\lambda_P$, the interesting region would be shielded, implying that we cannot arbitrarily shrink it and face uncertainties for position measurements. These uncertainty relations were derived in \cite{DFR} and read
\beq
\Delta x_0(\Delta x_1+\Delta x_2+\Delta x_3) \ge \lambda_P^2~, \quad \Delta x_1\Delta x_2+\Delta x_1+\Delta x_3+ \Delta x_2\Delta x_3  \ge \lambda_P^2  \quad .
\nonumber
\eeq
It was found that they may be obtained in the same way as we get the usual uncertainty relations from commutation relations if we identify the coordinates with operators $\hat{x}^\mu$ which satisfy the commutation relations
\beq
[\hat{x}^\mu,\hat{x}^\nu]=i\theta^{\mu\nu}\quad
\nonumber
\eeq
where $\theta^{\mu\nu}$ commutes with the $\hat{x}^\mu$'s. From these noncommuting operators $\hat{x}^\mu$, noncommutative quantum field theories got their name.
The central object is the commutator $\theta^{\mu\nu}$, which we want to comment on briefly. Throughout this thesis, we will assume it to be a real constant tensor, which explicitly breaks Lorentz invariance as it singles out distinct directions in spacetime. It obviously serves as a parameter for noncommutativity,  for vanishing $\theta^{\mu\nu}$ we want to recover the commutative case.  \\

In physics, the role of time belongs to the not yet satisfactorily answered questions. Being incorporated in special and general relativity in a similar manner as space, this similarity might be inadequate if viewed from a point at the edge of physics and philosophy as von Weizs\"acker did. He pointed out the fundamental meaning of time for quantum theory, as opposed to space \cite{Weizsaecker}.

In the context of noncommutative theories, the case that noncommutativity is restricted to the spatial operators $\hat{x}^i$, meaning that the time operator $\hat{x}^0$ does commute with them, corresponding to $\theta^{0i}=0$, is remarkably different from the more general case that also $\hat{x}^0$ does not commute, i.e. $\theta^{0i}\neq 0$. For perturbation theory in the first case, modified Feynman rules were proposed in \cite{Filk}, which differ from the ordinary only by the appearance of momentum dependent oscillatory functions at the vertices, so called phase factors. However, it was recognized that this approach, if applied to the second, more general case, violates unitarity of the S-matrix \cite{GomisMehen}. Successful work has been carried out to solve this problem, and two ways of unitary perturbation theory have been proposed for the general case. These are TOPT \cite{LiaoSiboldTOPT} and the Yang-Feldman formalism which is discussed in \cite{BahnsDiss}. The diagrammatic rules turn out to be more complicated in these theories as one does not longer have the ordinary propagators. Both approaches simplify to the modified Feynman rules if $\theta^{0i}$ vanishes.

However, further problems arise. Already in \cite{BahnsDiss}, it has been argued that the classical equations of motion are violated in TOPT. Another problem occurs in \cite{ORZ} where it has been shown that the Ward identity in noncommutative QED (NCQED) is not longer valid if one applies TOPT in the general case of $\theta^{0i}\neq 0$. This surprising result is the starting point for the considerations in this thesis.

The Ward identity normally results from the internal symmetries in a gauge theory, in particular the existence of a current and the related Slavnov-Taylor identities. For noncommutative QED, which resembles a nonabelian gauge theory, this is the BRS current, which was investigated on the classical level e.g. in \cite{double}. Difficulties could not be seen from there. For this reason, we  investigate currents on the quantized level. The troubles met there are related to the ones which emerge in the consideration of quantized equations of motion. The classical equations are found to be not realised on the quantized level but are disturbed by the appearance of terms which depend on $\theta^{0i}$. These additional terms are  most likely responsible for the violation of the Ward identity.

Another approach based on retarded functions is then studied which is neither unitary for nonvanishing $\theta^{0i}$ nor preserves the classical equations of motion and currents on the quantized level, but it allows both violations to be assigned to a certain type of diagrams. This insight suggests a modified theory which is unitary and respects the classical equations of motion and currents  on the quantized level.

We also analyze the Ward identity for Compton scattering in double gauged NCQED \cite{double}, a theory which has a richer structure than the ordinary NCQED. The validity of the Ward-identity is derived in the case $\theta^{0i}=0$. However, in the general case its violation is clear from \cite{ORZ} as the ordinary NCQED discussed there can be obtained as a special case of double gauged NCQED.

Another result of this thesis concerns remaining Lorentz symmetry. We investigate a scattering process in time-ordered perturbation theory and find the amplitude to be not invariant under the remaining Lorentz symmetries. This failure is explained by the typical form of the phase factors in TOPT.

The Yang-Feldman approach is not studied here as it does not naturally allow for the definition of time-ordered Green's functions, which are central for our considerations of quantized equations of motion and currents. Whether or not this formalism preserves the classical equations on the quantized level remains unclear.

All calculations presented in this thesis are unrenormalized, we only consider the tree level or the imaginary part of the one loop level.\\

This thesis is structured as follows.

The first chapter gives an introduction to noncommutative field theories and introduces the star product.

In the second one, we investigate the easier case of only spatial noncommutativity, review the modified Feynman rules, and investigate easy examples for quantized currents and equations of motion. The classical equations are found to hold also on the quantized level. We proceed with the more difficult example of the  quantized BRS current in noncommutative QED which is relevant for the validity of Ward identities. Again the classical BRS current is recovered, which includes terms due to the noncommutativity of the star product.

The general case of time-space noncommutativity in the approach of TOPT is presented in the third chapter. We develop Feynman rules for double gauged NCQED, the difficulty lies in the fact that we do not longer have the usual propagators. Then, these are applied to Compton scattering, and we show the Ward identity for $\theta^{0i}=0$. Equations of motion are studied in an easy example and are seen to exhibit additional terms depending on $\theta^{0i}$, these terms will also appear for quantized currents. Finally we demonstrate the violation of remaining Lorentz invariance.

The fourth chapter contains the approach via retarded functions. This formalism is at first discussed in the commutative case in such a way that conclusions can easily be drawn to noncommutative theories, diagrammatic rules are derived. In the noncommutative case, the violation of unitarity as well as of the naive equations of motion and currents on the quantized level is analyzed and seen to result from the appearance of a certain type of diagrams, which leads to a modified theory.

We conclude the thesis by an outlook.

\newpage
\chapter{Implementing noncommutativity: The star product}

How can we implement the information coming from the noncommuting coordinates into a quantum field theory? The answer is the introduction of a new, nonlocal product: the star product. In the following we will show how this naturally arises as correspondence between the algebra of coordinate operators, describing noncommutativity, and the algebra of functions on the Minkowski space, in terms of which a quantum field theory is usually formulated. This procedure follows the idea of Weyl quantization and may also be read in the review on noncommutative quantum field theory \cite{Szabo}.\\
\\
We start with the algebra generated by the hermitian operators $\hat{x}^\mu$, for which we have the commutation relations
\beq
[\hat{x}^\mu, \hat{x}^\nu]=i\theta^{\mu\nu} \quad .
\eeq
In the simplest case, which we will restrict ourselves to in this thesis, $\theta^{\mu\nu}$ is constant and real-valued, obviously it is antisymmetric and commutes with $\hat{x}^\rho$.

Our aim is to transport this structure to the algebra of (possibly complex valued) functions on the Minkowski space, where we consider sufficiently smooth and at infinity rapidly enough decreasing ones.
They allow to be described by its Fourier transforms:
\beq
\tilde{f}(k)=\int d^4k e^{-ik_\mu x^\mu }f(x)  \quad,
\eeq
with whose help we introduce a map $\hat{\mathcal{W}}$ from the algebra of sufficiently well-behaved functions on the Minkowski space to the algebra generated by the operators $\hat{x}^\mu$: for a function $f$ out of the first algebra we define
\beq
\hat{\mathcal{W}}[f]=\int \frac{d^4k}{(2\pi)^4}\tilde{f}(k)e^{ik_\mu\hat{x}^\mu} \quad .
\eeq
Which product, to be denoted by $\star$, do we have to define for the functions on the Minkowski space in order to make this map $\hat{\mathcal{W}}$ a homomorphism with respect to multiplication, i.e. to have the relation
\beq
\hat{\mathcal{W}}[f]\hat{\mathcal{W}}[g]=\hat{\mathcal{W}}[f\star g]
\label{homo_cond}
\eeq
fulfilled?

Let us start by evaluating the left hand side:
\begin{align}
\hat{\mathcal{W}}[f]\hat{\mathcal{W}}[g]=\int \frac{d^4k}{(2\pi)^4}\int \frac{d^4l}{(2\pi)^4}\tilde{f}(k)\tilde{g}(l)e^{ik_\mu\hat{x}^\mu}e^{il_\mu\hat{x}^\mu}
\end{align}

Using the Baker-Campbell-Hausdorff formula
\beq
e^Ae^B=e^{A+B}e^{\frac{1}{2}[A,B]}
\eeq
which is valid for operators $A,B$ satisfying $\big[A,[A,B]\big]=\big[B,[B,A]\big]=0$
we can write this as
\begin{align}
\hat{\mathcal{W}}[f]\hat{\mathcal{W}}[g]=\int \frac{d^4k}{(2\pi)^4}\int \frac{d^4l}{(2\pi)^4}\tilde{f}(k)\tilde{g}(l)e^{i(k_\mu+l_\mu)\hat{x}^\mu}e^{-\frac{i}{2}\theta^{\mu\nu}k_\mu l_\nu} \quad.
\end{align}

Substituting $l$ by $q=k+l$ this becomes
\begin{align}
\hat{\mathcal{W}}[f]\hat{\mathcal{W}}[g]=\int \frac{d^4k}{(2\pi)^4}\int \frac{d^4q}{(2\pi)^4}\tilde{f}(k)\tilde{g}(q-k)e^{iq_\mu\hat{x}^\mu}e^{-\frac{i}{2}\theta^{\mu\nu}k_\mu q_\nu} \quad,
\end{align}

allowing us to turn the condition (\ref{homo_cond}) into

\beq
\widehat{f\star g}(q)=\int \frac{d^4k}{(2\pi)^4}\tilde{f}(k)\tilde{g}(q-k)e^{-\frac{i}{2}\theta^{\mu\nu}k_\mu q_\nu} \quad .
\eeq

We have thus found the solution:
\beq
(f\star g)(x)=\int \frac{d^4k}{(2\pi)^4} \frac{d^4q}{(2\pi)^4} \tilde{f}(k)\tilde{g}(q-k)e^{-\frac{i}{2}\theta^{\mu\nu}k_\mu q_\nu}e^{iq_\mu
 x^\mu}
\label{int_star_product}
 \eeq

which is called the \emph{Groenewold-Moyal star-product} (we will simply refer to it as \emph{star product}). The quantity $e^{-\frac{i}{2}\theta^{\mu\nu}k_\mu q_\nu}$ is called the noncommutative phase factor, and with the definition of the wedge product $k\wedge q=\frac{1}{2}\theta^{\mu\nu}k_\mu q_\nu$ conveniently written in the form  $e^{-ik\wedge q}$.

 One easily checks that the star product may also be written as
\beq
(f\star g)(x)=e^{\frac{i}{2}\theta^{\mu\nu}\partial_\mu^\xi\partial_\nu^\eta}f(x+\xi)g(x+\eta)\Big|_{\xi=\eta=0}
\quad ,
\label{diff_star}
\eeq
a form which we will most often prefer in this thesis. The expression $e^{\frac{i}{2}\theta^{\mu\nu}\partial_\mu^\xi\partial_\nu^\eta}$ is to be understood as its formal series expansion.\\

One may derive some properties for the star product. The first observation is its nonlocality, as derivatives to arbitrary high order are involved. Obviously it reduces to the ordinary pointwise product of functions for $\theta^{\mu\nu}=0$. Further it is associative:
\beq
(f\star(g\star h))(x)=((f\star g)\star h)(x) \equiv (f\star g\star h)(x)
\eeq

and the quantity
\beq
\int dx ~(f_1\star ... \star f_n)(x)
\eeq
is invariant under cyclic (but not arbitrary!) permutations of the functions $f_i$, this is often referred to as \emph{trace property}. Also, one may omit one star under the integral:
\beq
\int dx ~ (f\star g)(x)=\int dx~ f(x)g(x) \quad .
\label{nostar}
\eeq

The Groenewold-Moyal star product is a typical example of products which are defined in deformation quantization, see e.g. \cite{Kontsevich}, and may also be studied in this context.\\

To transform  an ordinary quantum field theory into a noncommutative one, we
 modify the action and replace the ordinary product by the star product, e.g. in $\phi^3$-theory
\beq
S=\int dx~\big(\frac{1}{2}\partial^\mu\phi\star\partial_\mu\phi - \frac{1}{2}m^2\phi\star\phi +\frac{g}{3!}\phi\star\phi\star\phi\big)(x)
\quad .
\eeq
According to rule (\ref{nostar}) we can leave out the star in the free part of the action:
\beq
S=\int dx~\big(\frac{1}{2}\partial^\mu\phi\partial_\mu\phi - \frac{1}{2}m^2\phi\phi +\frac{g}{3!}\phi\star\phi\star\phi\big)(x)
\quad ,
\eeq

having the effect that the free theory is not affected by noncommutativity in this way of introducing it. We are thus allowed to use the the usual expressions for free fields, their expansion in terms of creation and annihilation operators, and the Hilbert space of the free theory. They will be as usual the starting point for perturbation theory, the star product then enters via the interaction.
s

\newpage
\thispagestyle{plain}
\mbox{}       
\newpage

\newpage

\chapter{The case of only spatial noncommutativity\label{spatial}}

The restriction to the case that time \emph{does} commute with space, meaning that $\theta^{0i}$ vanishes, causes considerably fewer problems and can be treated in an easier way than the general one. The modified Feynman rules, to be presented in the following, can be applied. A justification of these will be given in chapter \ref{topt}, where we will recover them for the case $\theta^{0i}=0$.

\section{The modified Feynman rules}

A first suggestion for perturbation theory in noncommutative theories was made in \cite{Filk}. Starting point is the Fourier transform of the interaction part of the action, which in the case of $S_{\text{int}}=\int dx (\phi_1\star...\star\phi_n)(x)$ with interacting fields $\phi_1,...,\phi_n$ may be written as
\beq
S_{\text{int}}=\int \frac{d^4p_1}{(2\pi)^4}\cdots \frac{d^4p_n}{(2\pi)^4}\tilde{\phi_1}(p_1)\cdots\tilde{\phi_n}(p_n)V(p_1,...,p_n)\delta^{(4)}(p_1+...+p_n)
\eeq

which suggests to modify the usual Feynman rules by associating the noncommutative phase factor $V(p_1,...,p_n)$ in addition to the momentum conservation $\delta^{(4)}(p_1+...+p_n)$ to a vertex with incoming momenta $p_1,...,p_n$ in a diagram. The phase factor, as may be checked by looking at (\ref{int_star_product}), has the form
\beq
V(p_1,...,p_n)=e^{-i(p_1,p_2,...,p_n)}
\label{phase_factor}
\eeq

where we introduced the short-hand notation
\beq
(p_1,p_2,...,p_n)=\sum_{i<j\leq n}p_i\wedge p_j \quad .
\eeq

Notice that in the commutative case ($\theta^{\mu\nu}=0$) the phase factor reduces to $1$, such that we obtain the ordinary Feynman rules.

\begin{figure}[h]
\begin{center}
\begin{picture}(130,50)(0,0)
\Vertex(30,25){1}
\Vertex(70,25){1}
\ArrowLine(5,25)(30,25)
\ArrowLine(70,25)(95,25)
\ArrowArcn(50,25)(20,180,0)
\ArrowArc(50,25)(20,180,360)
\Text(-3,25)[]{$p$}
\Text(103,25)[]{$p$}
\Text(50,55)[]{$k$}
\Text(50,-5)[]{$p-k$}
\end{picture}
\parbox{8cm}{\caption{\label{bubble} A planar diagram: the overall phase factor is trivial}}
\end{center}
\end{figure}
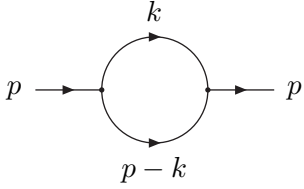

It is crucial to observe that the phase factor together with the momentum conservation is invariant under cyclic, but not arbitrary permutations of the momenta $p_1,...,p_n$. We have to fix the in- and outgoing lines at vertices in their cyclic ordering, and consider all possible diagrams. The effect is a certain averaging over the different possible orderings, e.g. in noncommutative $\phi^n$-theory we could instead of (\ref{phase_factor}) directly use
\beq
\bar{V}(p_1,...,p_n)=\frac{1}{n!}\sum_{\pi\in S_n}e^{-i(p_{\pi(1)},...,p_{\pi(n)})}
\label{av_phase_factor}
\eeq
and would no longer need to take different ordering of momenta in diagrams into account. We will follow this idea in the coming sections, but here stay with \cite{Filk}.
  The influence of the phase factors to graphs is there further elaborated, it leads to the distinction of two types of diagrams. The first one, called \emph{planar}, does not depend on internal momenta through phase factors. Such a planar diagram is the one in Fig. \ref{bubble}, we calculate the overall phase to be
\begin{align}
V(p, -k, -p+k)V(k,-p,p-k)&=e^{-i(p,-k,-p+k)}e^{-i(k,-p,p-k)} \cr
&=e^{-i(-p\wedge k+p\wedge k+k\wedge p-k\wedge p+k\wedge p+p\wedge k)}  \cr
&=1\quad . \nonumber
\end{align}

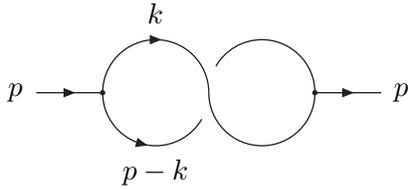
\begin{figure}[h]
\begin{center}
\begin{picture}(130,50)(0,0)
\Vertex(10,25){1}
\Vertex(90,25){1}
\ArrowLine(-15,25)(10,25)
\ArrowLine(90,25)(115,25)
\ArrowArcn(30,25)(20,180,0)
\ArrowArc(30,25)(20,180,330)
\CArc(70,25)(20,0,150)
\CArc(70,25)(20,180,360)
\Text(-23,25)[]{$p$}
\Text(123,25)[]{$p$}
\Text(30,55)[]{$k$} 
\Text(30,-5)[]{$p-k$}
\end{picture}
\parbox{8cm}{\caption{\label{twobubbles}A nonplanar diagram: the internal momentum appears in the overall phase factor}}
\end{center}
\end{figure}

If we twist the internal momenta to obtain a different ordering at the right vertex, we encounter a nonplanar diagram: the phase factor yields
\begin{align}
V(p,-k,-p+k)V(p-k,-p,k)&=e^{-i(p,-k,-p+k)}e^{-i(p-k,-p,k)} \cr
&=e^{-i(-p\wedge k+p\wedge k+k\wedge p+k\wedge p+p\wedge k-p\wedge k)}  \cr
&=e^{-2ik\wedge p}\quad  \nonumber
\end{align}
and does depend on the internal momenta. \\

However, the distinction into planar and nonplanar diagrams will not be of further interest in this thesis. Instead we are going to work directly with averaged phase factors of the type (\ref{av_phase_factor}).

\section{Quantized operator identities: Equations of motion and currents}

In this section we want to derive easy examples for equations of motion and current conservation laws on the quantized level, and in particular show that the classical ones hold in the case of spatial noncommutativity. Our aim is to illustrate the mechanism of deriving both types of equations and show the similarity. \\

To consider the level of Green's functions means that instead of an  operator identity at the classical level
\beq
\mathcal{O}_1=\mathcal{O}_2
\label{class_id}
\eeq
we want to show it on the quantized level, i.e. for Green's functions
\beq
\langle 0|T(\mathcal{O}_1 X)|0\rangle =\langle 0|T(\mathcal{O}_2 X)|0\rangle + \text{c.t.}
\label{quant_id}
\eeq
where $X$ is an arbitrary collection of fields, $X=\Phi_1(x_1)\cdots \Phi_n(x_n)$ and c.t stands for contact terms. Identity (\ref{quant_id}) gives then rise to identity (\ref{class_id}), as all matrix elements of the operators $\mathcal{O}_1$ and $\mathcal{O}_2$ may be obtained from the left and right hand side of (\ref{quant_id}) by the LZS reduction formula, the contact terms vanish when this procedure is applied.

For showing an identity of type (\ref{quant_id}), we have to be proleptic and use a result from chapter \ref{topt}, namely that Filks approach agrees in the case of only spatial noncommutativity with the formula
\beq
\langle 0|T\big(\mathcal{O}~\Phi(x_{1})...\Phi(x_n)\big)|0\rangle
=\langle 0|T\Big(\mathcal{O}~\Phi^{(0)}(x_{1})...\Phi^{(0)}(x_n)e^{i\int dx \mathcal{L}^{(0)}_{\text{int}}(x)}\Big)|0\rangle\quad ,   \label{pert_green1}
\eeq
the index $^{(0)}$ on the right hand side means that all fields are the free ones.
We will benefit from this well-known formula, allowing us to calculate Green's functions of type (\ref{quant_id})  in the following considerations.\\

As an illustrating example, we consider a field theory with a complex field $\phi$ and a real field $\sigma$, the Lagrangian reads
\beq
\mathcal{L}=\partial^\mu\phi^\dag\partial_\mu\phi -m^2\phi^\dag\phi+ \frac{1}{2}\partial^\mu\sigma\partial_\mu\sigma -\frac{1}{2}m^2\sigma^2+g\phi^\dag\star\sigma\star\phi
\eeq

and is left invariant under the infinitesimal $U(1)$-transformation
\begin{align}
\delta\phi&=i\alpha\phi \cr
\delta\phi^\dag&=-i\alpha\phi^\dag  \quad .
\end{align}

On the classical level we can use the principle of least action to obtain the equations of motion
\begin{align}
\partial^\mu\partial_\mu\phi+m^2\phi-g\sigma\star\phi=0 \cr
\partial^\mu\partial_\mu\phi^\dag+m^2\phi^\dag-g\phi^\dag\star\sigma=0\cr
\partial^\mu\partial_\mu\sigma+m^2\sigma-g\phi\star\phi^\dag=0
\label{eq_mo}
\end{align}
and with their help find that the current
\beq
j_\mu=i\phi^\dag\star\partial_\mu\phi - i\partial_\mu\phi^\dag\star\phi
\label{curr}
\eeq
is conserved: $\partial^\mu j_\mu=0$. \\

The equations (\ref{eq_mo}) and (\ref{curr}) are of the classical type (\ref{class_id}), we now want to derive them on the quantized level in the sense of (\ref{quant_id}).

\subsection{Equations of motion}

Let us start with the equation of motion for the real scalar field $\sigma$, our aim is to derive
\beq
\langle 0|T\big\{(\partial^\mu\partial_\mu\sigma+m^2\sigma-g\phi\star\phi^\dag)(x)X\big\}|0\rangle+\textrm{c.t.} =0 \quad ,
\label{eq_mo1}
\eeq
where  $X$ is a collection of field operators:
\beq
X=\mathcal{O}_1(x_1)...\mathcal{O}_n(x_n)~;\quad
\mathcal{O}_i\epsilon\big\{\phi,\phi^\dag,\sigma\}  \quad .
\label{X}
\eeq

We know that the difference between having a derivative inside or outside the time-ordering can at most be a contact term\footnote{In the case of a time derivative, it is a commutator times a one-dimensional $\delta$-distribution depending on the time argument. For $\theta^{0i}=0$ the commutator vanishes at spacelike distances, such that is only different from zero at coinciding points and thus a contact term. }, so let us study the expression
\beq
(\Box_x+m^2) \langle 0|T\big\{\sigma(x)~X\big\}|0\rangle   \quad .
\eeq
Recalling that according to formula (\ref{pert_green1}) the field $\sigma(x)$ can either be contracted with another field $\sigma(y)$ coming from $\mathcal{L}_{\text{int}}(y)$ or with a field $\sigma$ at one of the points $x_1,...,x_n$, we may write the above expression diagrammatically. Symbolize the contraction of $\phi$ and $\phi^\dag$ by a continuous and the contraction of two $\sigma$-fields by a dashed line. Let $L_{\sigma}\subset \{1,...,n\}$ be such that for $l\in L_\sigma$ there is a field $\sigma$ at place $x_l$, we then find
\begin{align}
&(\Box_x+m^2) \langle 0|T\big\{\sigma(x)~X\big\}|0\rangle
=\begin{picture}(100,40)(15,40)
\Vertex(50,85){1}
\Vertex(30,0){1}
\Vertex(70,0){1}
\GOval(50,42.5)(17,30)(0){0.75}
\DashLine(50,85)(50,60){2}
\Line(30,15)(30,0)
\DashLine(30,30)(30,15){2}
\Line(70,15)(70,0)
\DashLine(70,30)(70,15){2}
\Text(50,95)[]{$x$}
\Text(30,-10)[]{$x_1$}
\Text(50,-10)[]{...}
\Text(70,-10)[]{$x_n$}
\Text(55,73)[l]{$(\Box+m^2)$}
\end{picture}\cr
=&~
\int dy~
\begin{picture}(100,120)(15,40)
\Vertex(50,88){1}
\Vertex(30,0){1}
\Vertex(70,0){1}
\Vertex(50,73){1}
\GOval(50,42.5)(17,30)(0){0.75}
\Line(30,15)(30,0)
\DashLine(30,30)(30,15){2}
\Line(70,15)(70,0)
\DashLine(70,30)(70,15){2}
\DashLine(50,88)(50,73){2}
\Line(50,73)(35,57)
\Line(50,73)(65,57)
\Text(40,73)[]{$y$}
\Text(50,98)[]{$x$}
\Text(30,-10)[]{$x_1$}
\Text(50,-10)[]{...}
\Text(70,-10)[]{$x_n$}
\Text(55,83)[l]{$(\Box+m^2)$}
\end{picture}
+\qquad \sum_{l\in L_\sigma}~
\begin{picture}(100,120)(15,40)
\Vertex(90,85){1}
\Vertex(30,0){1}
\Vertex(70,0){1}
\Vertex(90,0){1}
\GOval(50,42.5)(17,30)(0){0.75}
\Line(30,15)(30,0)
\DashLine(30,30)(30,15){2}
\Line(70,15)(70,0)
\DashLine(70,30)(70,15){2}
\DashLine(90,85)(90,0){2}
\Text(90,95)[]{$x$}
\Text(30,-10)[]{$x_1$}
\Text(50,-10)[]{...}
\Text(50,0)[]{$\check{x}_l$}
\Text(70,-10)[]{$x_n$}
\Text(90,-10)[]{$x_l$}
\Text(95,50)[l]{$(\Box+m^2)$}
\end{picture}
\cr
\end{align}
where the half continuous and half dashed lines connecting the points $x_1,...,x_n$ symbolize that these lines can either be dashed or continuous, depending on the type of field at that point. The idea is now to use the fact that the lines correspond to propagators, such that $(\Box+m^2)$ acting on the dashed line gives a $\delta$-distribution and allows to simplify the terms. \\

To derive the analytic expressions for the above diagrams, we carefully start again from (\ref{pert_green1}) and use (\ref{diff_star}) to rewrite the star product. The contraction of two fields, which gives the propagator, is denoted by
$\begin{picture}(30,10)(-5,5)
\Line(0,10)(0,16)
\Line(0,16)(20,16)
\Line(20,16)(20,10)
\end{picture} $
and we compute
\begin{align}
&(\Box_x+m^2) \langle 0|T\big\{\sigma(x)~X\big\}|0\rangle  \cr
=&\int dy~(\Box_x+m^2)\langle 0|T\big\{
\begin{picture}(0,20)(,0)
\Line(4,18)(59,18)
\Line(4,18)(4,8)
\Line(59,18)(59,8)
\end{picture}
\sigma(x)ig(\phi^\dag\star\sigma\star\phi)(y)X\big\}|0\rangle \cr
&+
\begin{picture}(0,20)(-3,0)
\Line(97,18)(118,18)
\Line(97,18)(97,8)
\Line(118,18)(118,8)
\end{picture}\sum_{l\in L_\sigma}(\Box_x+m^2)\langle 0|T\big\{\sigma(x)\sigma(x_l)X_{\check{l}}\big\}|0\rangle \cr
=&~
ig\int dy~(\Box_x+m^2)\cr
&\qquad \langle 0|T\Big\{
\begin{picture}(0,20)(0,0)
\Line(4,18)(155,18)
\Line(4,18)(4,8)
\Line(155,18)(155,8)
\end{picture}
\sigma(x)e^{\frac{i}{2}\theta^{\mu\nu}\partial_\mu^\xi\partial_\nu^\eta}e^{\frac{i}{2}\theta^{\rho\sigma}\partial_\rho^\zeta\partial_\sigma^\chi}\phi^\dag(y+\xi)\sigma(y+\eta+\zeta)\phi(y+\eta+\chi)X\Big\}|0\rangle\Big|_{\xi,\eta,\zeta,\chi=0}   \cr
&+
\begin{picture}(0,20)(-3,0)
\Line(97,18)(118,18)
\Line(97,18)(97,8)
\Line(118,18)(118,8)
\end{picture}\sum_{l\in L_\sigma}(\Box_x+m^2)\langle 0|T\big\{\sigma(x)\sigma(x_l)X_{\check{l}}\big\}|0\rangle
\nonumber
  \end{align}
\begin{align}
 =&~ig\int dy~(\Box_x+m^2)e^{\frac{i}{2}\theta^{\mu\nu}\partial_\mu^\xi\partial_\nu^\eta}e^{\frac{i}{2}\theta^{\rho\sigma}\partial_\rho^\zeta\partial_\sigma^\chi}\Delta_C(x-y-\eta-\zeta) \cr
 &\qquad\qquad\langle 0|T\big\{\phi^\dag(y+\xi)\phi(y+\eta+\chi)X\big\}|0\rangle \Big|_{\xi,\eta,\zeta,\chi=0}   \cr
 &+\sum_{l\in L_\sigma}(\Box_x+m^2)\Delta_C(x-x_l)\langle 0|T\big\{X_{\check{l}}\big\}|0\rangle
 \quad .
\end{align}
As already mentioned, we now use the relation $(\Box+m^2)\Delta_C(x-y)=-i\delta^{(4)}(x-y)$ such that we can further simplify these terms:
\begin{align}
&(\Box_x+m^2) \langle 0|T\big\{\sigma(x)~X\big\}|0\rangle  \cr
=&~g\int dy~e^{\frac{i}{2}\theta^{\mu\nu}\partial_\mu^\xi\partial_\nu^\eta}e^{\frac{i}{2}\theta^{\rho\sigma}\partial_\rho^\zeta\partial_\sigma^\chi}\delta^{(4)}(x-y-\eta-\zeta)
 \langle 0|T\big\{\phi^\dag(y+\xi)\phi(y+\eta+\chi)X\big\}|0\rangle \Big|_{\xi,\eta,\zeta,\chi=0}   \cr
 &-i\sum_{l\in L_\sigma}\delta^{(4)}(x-x_l)\langle 0|T\big\{X_{\check{l}}\big\}|0\rangle \quad .
\end{align}
The second expression on the right hand side is from the appearance of $\delta^{(4)}(x-x_l)$ seen to be a contact term. In the first term we can apply the $\delta$-distribution to carry out the integration over $y$ and arrive at
\begin{align}
&(\Box_x+m^2) \langle 0|T\big\{\sigma(x)~X\big\}|0\rangle  \cr
=&~g e^{\frac{i}{2}\theta^{\mu\nu}\partial_\mu^\xi\partial_\nu^\eta}e^{\frac{i}{2}\theta^{\rho\sigma}\partial_\rho^\zeta\partial_\sigma^\chi}
 \langle 0|T\big\{\phi^\dag(x-\eta-\zeta+\xi)\phi(x-\eta-\zeta+\eta+\chi)X\big\}|0\rangle \Big|_{\xi,\eta,\zeta,\chi=0}    \quad
  +~\text{c.t.}
  \nonumber
\end{align}
We recognize that the two $\eta$ appearing in the argument of the field $\phi$ cancel each other, such that $e^{\frac{i}{2}\theta^{\mu\nu}\partial_\mu^\xi\partial_\nu^\eta}$ yields the identity. $e^{\frac{i}{2}\theta^{\rho\sigma}\partial_\rho^\zeta\partial_\sigma^\chi}$ gives now the star product between the two $\phi$ fields: $e^{\frac{i}{2}\theta^{\rho\sigma}\partial_\rho^\zeta\partial_\sigma^\chi}\phi^\dag(x-\zeta)\phi(x-\zeta+\chi)\Big|_{\zeta,\chi=0}=(\phi\star\phi^\dag)(x)$ and we find
\begin{align}
(\Box_x+m^2) \langle 0|T\big\{\sigma(x)~X\big\}|0\rangle
=~g
 \langle 0|T\big\{(\phi\star\phi^\dag)(x)X\big\}|0\rangle    \quad
  +~\text{c.t.} \quad .
  \label{quant_eq_mo}
\end{align}
But this is the equation of motion (\ref{eq_mo1}), such that our calculation has come to the desired result.

\subsection{Current conservation laws}

We now turn to the derivation of the current conservation law
\beq
\partial^\mu\langle 0|T\big\{j_\mu(x)X\big\}|0\rangle+\textrm{c.t.} =0 \quad ,
\eeq
with $X$ in the form (\ref{X}). The calculation will be similar to the one in the previous subsection, but slightly more challenging as we will have to deal with a composite operator at place $x$, namely the star product of two fields. \\

Consider the expression
\beq
\begin{picture}(0,20)(0,0)
\Line(8,10)(8,17)
\Line(8,17)(101,17)
\LongArrow(101,17)(101,10)
\end{picture}
(\Box+m^2)\langle 0|T\big\{(\phi^\dag\star\phi)(x)X\big\}|0\rangle
 -
\begin{picture}(0,20)(0,0)
\Line(8,10)(8,17)
\Line(8,17)(80,17)
\LongArrow(80,17)(80,10)
\end{picture}
 (\Box+m^2)\langle 0|T\big\{(\phi^\dag\star\phi)(x)X\big\}|0\rangle  \quad ,
\label{curr_start}
 \eeq
the arrow denotes the field where the derivative acts on. We focus on the first term which,
again symbolizing the contraction of $\phi^\dag$ and $\phi$ by a continuous line and the contraction of two $\sigma$-fields by a dashed line, can be cast into the diagrammatic form
\begin{align}
\begin{picture}(0,10)(0,0)
\Line(8,10)(8,17)
\Line(8,17)(101,17)
\LongArrow(101,17)(101,10)
\end{picture}
&(\Box+m^2)\langle 0|T\big\{(\phi^\dag\star\phi)(x)X\big\}|0\rangle
=\begin{picture}(100,60)(15,40)
\Vertex(50,85){1}
\Vertex(30,0){1}
\Vertex(70,0){1}
\GOval(50,42.5)(17,30)(0){0.75}
\Line(30,55)(50,85)
\Line(50,85)(70,55)
\Line(30,15)(30,0)
\DashLine(30,30)(30,15){2}
\Line(70,15)(70,0)
\DashLine(70,30)(70,15){2}
\Text(50,95)[]{$x$}
\Text(30,-10)[]{$x_1$}
\Text(50,-10)[]{...}
\Text(70,-10)[]{$x_n$}
\Text(65,77)[l]{$(\Box+m^2)$}
\end{picture}
\cr
=&
\quad  \int dy~
\begin{picture}(100,120)(15,40)
\Vertex(50,85){1}
\Vertex(30,0){1}
\Vertex(70,0){1}
\Vertex(60,70){1}
\GOval(50,42.5)(17,30)(0){0.75}
\Line(30,55)(50,85)
\Line(50,85)(60,70)
\Line(30,15)(30,0)
\DashLine(30,30)(30,15){2}
\Line(70,15)(70,0)
\DashLine(70,30)(70,15){2}
\DashLine(60,70)(50,60){2}
\Line(60,70)(70,55)
\Text(50,95)[]{$x$}
\Text(68,70)[]{$y$}
\Text(30,-10)[]{$x_1$}
\Text(50,-10)[]{...}
\Text(70,-10)[]{$x_n$}
\Text(60,85)[l]{$(\Box+m^2)$}
\end{picture} +
\quad \sum_{l\in L_{\phi^\dag}}
\begin{picture}(100,120)(10,40)
\Vertex(65,85){1}
\Vertex(30,0){1}
\Vertex(70,0){1}
\Vertex(110,0){1}
\GOval(50,42.5)(17,30)(0){0.75}
\Line(30,55)(65,85)
\Line(65,85)(110,0)
\Line(30,15)(30,0)
\DashLine(30,30)(30,15){2}
\Line(70,15)(70,0)
\DashLine(70,30)(70,15){2}
\Text(65,95)[]{$x$}
\Text(30,-10)[]{$x_1$}
\Text(50,-10)[]{...}
\Text(50,0)[]{$\check{x}_l$}
\Text(70,-10)[]{$x_n$}
\Text(115,-10)[]{$x_l$}
\Text(93,50)[l]{$(\Box+m^2)$}
\end{picture}
\cr
\cr
\nonumber
 \end{align}
 The two diagrams in the last line arise because the line from point $x$ carrying the differential operator can either go to an internal vertex or to one of the outer points. The index set $L_{\phi^\dag}$ refers to all numbers $l\in\{1,...,n\}$ for which a field $\phi^\dag$ is located at
 $x_l$.

 To calculate the analytic expressions we apply formula (\ref{pert_green1}), write the star-product in the form (\ref{diff_star}) and find
\begin{align}
\begin{picture}(0,20)(0,0)
\Line(8,10)(8,17)
\Line(8,17)(101,17)
\LongArrow(101,17)(101,10)
\end{picture}
&(\Box+m^2)\langle 0|T\big\{(\phi^\dag\star\phi)(x)X\big\}|0\rangle  \cr
=&\int dy~
\begin{picture}(0,10)(0,0)
\Line(8,10)(8,17)
\Line(8,17)(99,17)
\LongArrow(99,17)(99,10)
\end{picture}
\begin{picture}(0,20)(0,0)
\Line(102,10)(102,16)
\Line(102,16)(139,16)
\Line(139,16)(139,10)
\end{picture}
(\Box+m^2)\langle 0|T\big\{(\phi^\dag\star\phi)(x)ig(\phi^\dag\star\sigma\star\phi)(y) X\big\}|0\rangle  \cr
&+\sum_{l\in L_{\phi^\dag}}
\begin{picture}(0,10)(0,0)
\Line(8,10)(8,17)
\Line(8,17)(99,17)
\LongArrow(99,17)(99,10)
\end{picture}
\begin{picture}(0,20)(0,0)
\Line(102,10)(102,16)
\Line(102,16)(126,16)
\Line(126,16)(126,10)
\end{picture}
(\Box+m^2)\langle 0|T\big\{(\phi^\dag\star\phi)(x)\phi^\dag(x_l) X_{\check{l}}\big\}|0\rangle  \cr
=&~
\begin{picture}(0,10)(0,0)
\Line(133,10)(133,17)
\Line(133,17)(244,17)
\LongArrow(244,17)(244,10)
\end{picture}
ig\int dy~  e^{\frac{i}{2}\theta^{\rho\sigma}\partial_\rho^\xi\partial_\sigma^\eta}e^{\frac{i}{2}\theta^{\rho\sigma}\partial_\rho^\zeta\partial_\sigma^\chi}(\Box+m^2)\langle0|T\big\{\phi^\dag(x+\xi)\Delta_C(x+\eta-y-\zeta)
 (\sigma\star\phi)(y+\chi)X\big\}|0\rangle  \cr
\begin{picture}(-45,20)(0,0)
\Line(45,10)(45,17)
\Line(45,17)(163,17)
\LongArrow(163,17)(163,10)
\end{picture}
&+g\sum_{l\in L_{\phi^\dag}} e^{\frac{i}{2}\theta^{\rho\sigma}\partial_\rho^\xi\partial_\sigma^\eta}(\Box+m^2)\langle 0|  T\big\{\phi^\dag(x+\xi)\Delta_C(x+\eta-x_l)X_{\check{l}}\big\}|0\rangle \Big|_{\xi,\eta=0}  \quad ,
\end{align}
the symbol$\begin{picture}(30,10)(-5,5)
\Line(0,10)(0,16)
\Line(0,16)(20,16)
\Line(20,16)(20,10)
\end{picture} $
denotes the contraction between two field operators. Remembering
\beq
(\Box+m^2)\Delta_C(x-y)=-i\delta^{(4)}(x-y)
\eeq
we obtain
\begin{align}
\begin{picture}(0,20)(0,0)
\Line(8,10)(8,17)
\Line(8,17)(101,17)
\LongArrow(101,17)(101,10)
\end{picture}
&(\Box+m^2)\langle 0|T\big\{(\phi^\dag\star\phi)(x)X\big\}|0\rangle  \cr
=&~
g\int dy~  e^{\frac{i}{2}\theta^{\rho\sigma}\partial_\rho^\xi\partial_\sigma^\eta}e^{\frac{i}{2}\theta^{\rho\sigma}\partial_\rho^\zeta\partial_\sigma^\chi}\delta^{(4)}(x+\eta-y-\zeta)\langle0|T\big\{\phi^\dag(x+\xi)
 (\sigma\star\phi)(y+\chi)X\big\}|0\rangle  \Big|_{\xi,\eta,\zeta,\chi=0} \cr
&-ig\sum_{l\in L_{\phi^\dag}} e^{\frac{i}{2}\theta^{\rho\sigma}\partial_\rho^\xi\partial_\sigma^\eta}\delta^{(4)}(x+\eta-x_l)\langle 0|  T\big\{\phi^\dag(x+\xi)X_{\check{l}}\big\}|0\rangle \Big|_{\xi,\eta=0}   \quad .
\end{align}
The four-dimensional $\delta$-distribution in the second term on the right hand side may, as we are in the case $\theta^{0i}=0$, be written as $\delta^{(4)}(x+\eta-x_l)=\delta^{(1)}(x^0-x^0_l)\delta^{(3)}(x^i+\eta^i-x^i_l)$ such that it is seen to be local in time and we recognize it as a contact term. The  $\delta$-distribution in the other term allows us to carry out the integration over y:
\begin{align}
\begin{picture}(0,20)(0,0)
\Line(8,10)(8,17)
\Line(8,17)(101,17)
\LongArrow(101,17)(101,10)
\end{picture}
&(\Box+m^2)\langle 0|T\big\{(\phi^\dag\star\phi)(x)X\big\}|0\rangle  \cr
=&~
g e^{\frac{i}{2}\theta^{\rho\sigma}\partial_\rho^\xi\partial_\sigma^\eta}e^{\frac{i}{2}\theta^{\rho\sigma}\partial_\rho^\zeta\partial_\sigma^\chi}\langle0|T\big\{\phi^\dag(x+\xi)
 (\sigma\star\phi)(x+\eta-\zeta+\chi)X\big\}|0\rangle\Big|_{\xi,\eta,\zeta,\chi=0} \quad +~ \text{c.t.}
 \nonumber
\end{align}
The derivatives of $e^{\frac{i}{2}\theta^{\rho\sigma}\partial_\rho^\zeta\partial_\sigma^\chi}$ both act on the same function ($\sigma\star\phi$) and so $e^{\frac{i}{2}\theta^{\rho\sigma}\partial_\rho^\zeta\partial_\sigma^\chi}$ yields only the identity. $e^{\frac{i}{2}\theta^{\rho\sigma}\partial_\rho^\xi\partial_\sigma^\eta}$ is then seen to give the star product between $\phi^\dag$ and $\sigma\star\phi$, such that the final result is
\begin{align}
\begin{picture}(0,20)(0,0)
\Line(8,10)(8,17)
\Line(8,17)(101,17)
\LongArrow(101,17)(101,10)
\end{picture}
(\Box+m^2)\langle 0|T\big\{(\phi^\dag\star\phi)(x)X\big\}|0\rangle
=~
g\langle0|T\big\{(\phi^\dag\star\sigma\star\phi)(x)X\big\}|0\rangle \quad +~ \text{c.t} \quad .
\end{align}
The second term of (\ref{curr_start}) is computed analogously:
\begin{align}
\begin{picture}(0,20)(0,0)
\Line(8,10)(8,17)
\Line(8,17)(80,17)
\LongArrow(80,17)(80,10)
\end{picture}
 (\Box+m^2)\langle 0|T\big\{(\phi^\dag\star\phi)(x)X\big\}|0\rangle
 =
g\langle0|T\big\{(\phi^\dag\star\sigma\star\phi)(x)X\big\}|0\rangle \quad +~ \text{c.t} \quad ,
 \end{align}
such that together
\beq
\begin{picture}(0,20)(0,0)
\Line(8,10)(8,17)
\Line(8,17)(101,17)
\LongArrow(101,17)(101,10)
\end{picture}
(\Box+m^2)\langle 0|T\big\{(\phi^\dag\star\phi)(x)X\big\}|0\rangle
 -
\begin{picture}(0,20)(0,0)
\Line(8,10)(8,17)
\Line(8,17)(80,17)
\LongArrow(80,17)(80,10)
\end{picture}
 (\Box+m^2)\langle 0|T\big\{(\phi^\dag\star\phi)(x)X\big\}|0\rangle
+ \text{c.t.} =0 \quad .
\label{nearly_curr}
 \eeq

We now try to rewrite the left hand side. Obviously we can leave the $m^2$
in both terms away, they cancel each other. Furthermore, we encounter
\begin{align}
&\begin{picture}(0,20)(0,0)
\Line(8,10)(8,17)
\Line(8,17)(101,17)
\LongArrow(101,17)(101,10)
\end{picture}
(\Box+m^2)\langle 0|T\big\{(\phi^\dag\star\phi)(x)X\big\}|0\rangle
 -
\begin{picture}(0,20)(0,0)
\Line(8,10)(8,17)
\Line(8,17)(80,17)
\LongArrow(80,17)(80,10)
\end{picture}
 (\Box+m^2)\langle 0|T\big\{(\phi^\dag\star\phi)(x)X\big\}|0\rangle  + \text{c.t.}\cr
 =~&
\begin{picture}(0,30)(0,0)
\Line(11,10)(11,17)
\Line(11,17)(81,17)
\LongArrow(81,17)(81,10)
\end{picture}
 \begin{picture}(0,30)(0,0)
\Line(11,10)(11,20)
\Line(11,20)(101,20)
\LongArrow(101,20)(101,10)
\end{picture}
\partial^\mu\partial_\mu\langle 0|T\big\{\big(\phi^\dag\star\phi-\phi^\dag\star\phi\big)(x)X\big\}|0\rangle+ \text{c.t.} \cr
=~&\partial^\mu\langle 0|T\big\{\big(\phi^\dag\star\partial_\mu\phi-\partial^\mu\phi^\dag\star\phi\big)(x)X\big\}|0\rangle+ \text{c.t.} \quad ,
\end{align}
and this vanishes according to (\ref{nearly_curr}). For the operator (\ref{curr})
\beq
j_\mu=i\phi^\dag\star\partial_\mu\phi - i\partial_\mu\phi^\dag\star\phi
\eeq
which we derived as the classical current,
we have thus now proven the conservation law on the level of Green's functions
\beq
\partial^\mu\langle 0|T\big\{j_\mu(x)X\big\}|0\rangle+\textrm{c.t.} =0  \quad .
\eeq

\section{BRS current and Slavnov-Taylor identities in NCQED \label{BRS}}

In this section we want to consider a physically relevant example of a quantized current and derive the  BRS current conservation on the level of Green's functions. Integration  yields the Slavnov-Taylor identities, from which the Ward identities for NCQED may be proven. They allow to show that the physical Hilbert space is stable under time evolution \cite{SiboldSkript}. \\

 The calculation follows the lines of the derivation of  a current in the previous easy example, but we will have more terms to consider. Also, the current will not look as nice as the previous one nor like the BRS current in commutative theories. We face additional terms that come from star-commutators of fields and are due to the fact that only the action but not the Lagrangian is invariant under the BRS-transformations. This last issue was already discussed on the classical level in \cite{MicuJabbari} and \cite{Okumura}, where the same type of additional terms emerged.   \\

In \cite{double} a noncommutative theory of electrodynamics has been constructed which is invariant under combined left- and right BRS transformations. For simplicity, we only take the part of this theory which remains invariant under BRS transformations from the left, this theory is usually referred to as NCQED.
Its full Lagrangian, including gauge-fixing and Faddeev-Popov terms, reads
\beq
\mathcal{L}=i\bar{\psi}\gamma^\mu\partial_\mu\psi-m\bar{\psi}\psi+g\bar{\psi}\star\gamma^\mu A_\mu\star\psi-\frac{1}{4}F_{\mu\nu}\star F^{\mu\nu} +\frac{\xi}{2}B^2+
B\partial^\mu A_\mu-\bar{c}\star \partial^\mu D_\mu c
\eeq
where
\begin{align}
F_{\mu\nu}&=\partial_\mu A_\nu-\partial_\nu A_\mu -ig[A_\mu,A_\nu]_\star \\
D_\mu c&=\partial_\mu c+ig[c,A_\mu]_\star           \qquad.
\end{align}
As stated above, the corresponding action is invariant under the left BRS transformations which read
\begin{align}
s\psi&=igc\star \psi \\
s\bar{\psi}&=-ig\bar{\psi}\star c \\
sc&=igc\star c \\
s\bar{c}&= B\\
sB&=0 \\
sA_\mu&=D_\mu c=\partial_\mu c+ig[c,A_\mu]_\star  \quad .
\end{align}
We have to use the trace property of the star product to show the invariance of the theory under these transformations which we can only apply for the action but not for the Lagrangian, the latter turns in fact out to be modified. The calculations are lengthy and have been performed in other contexts, see e.g. \cite{double}, such that we do not repeat them here. \\

To find the conserved current associated to these transformations, i.e. an operator $j^{BRS}_\mu(x)$ satisfying the conservation law
\beq
\partial^\mu\langle 0|T\{j^{BRS}_\mu(x)X\}|0\rangle + \text{c.t.} =0
\label{current}
\eeq
where $X$ is a collection of field operators:
\beq
X=\mathcal{O}_1(x_1)...\mathcal{O}_n(x_n)~;\quad
\mathcal{O}_i\epsilon\big\{\bar{\psi},\psi,A_\mu,\bar{c},c\big\}
\eeq
we consider the expression
\begin{align}
i
\begin{picture}(0,10)(0,0)
\Line(12,10)(12,17)
\Line(12,17)(116,17)
\LongArrow(116,17)(116,10)
\end{picture}
(i\partial_\mu-m)\langle0&|T\{(s\bar{\psi}\star\gamma^\mu \psi)(x)X\}|0\rangle \cr
-i
\begin{picture}(0,15)(0,0)
\Line(12,10)(12,17)
\Line(12,17)(82,17)
\LongArrow(82,17)(82,11)
\end{picture}
(i\partial_\mu+m)\langle0&|T\{(\bar{\psi}\star\gamma^\mu s{\psi})(x)X\}|0\rangle\cr
-i
\begin{picture}(0,15)(0,0)
\Line(5,10)(5,17)
\Line(5,17)(73,17)
\LongArrow(73,17)(73,10)
\end{picture}
\Box\langle0&|T\{(sA_\mu\star A^\mu)(x)X\}|0\rangle \cr
+i
\begin{picture}(0,15)(0,0)
\Line(5,10)(5,17)
\Line(5,17)(62,17)
\LongArrow(62,17)(62,10)
\end{picture}
\Box\langle 0&|T\{(s\bar{c}\star c)(x)X\}|0\rangle \cr
+i
\begin{picture}(0,15)(0,0)
\Line(5,10)(5,17)
\Line(5,17)(41,17)
\LongArrow(41,17)(41,10)
\end{picture}
\Box\langle 0&|T\{(\bar{c}\star sc)(x)X\}|0\rangle \quad .   \label{start}
\end{align}

The derivatives are meant to only act on the field where the arrow points to.

Let us begin with the detailed calculation of the first term, which resembles our considerations in the previous section. Diagrammatically it is
\begin{align}
i
\begin{picture}(0,10)(0,0)
\Line(12,10)(12,17)
\Line(12,17)(116,17)
\LongArrow(116,17)(116,10)
\end{picture}
&(i\partial_\mu-m)\langle0|T\{(s\bar{\psi}\star\gamma^\mu \psi)(x)X\}|0\rangle
=
\begin{picture}(0,10)(0,0)
\Line(17,10)(17,17)
\Line(17,17)(132,17)
\LongArrow(132,17)(132,10)
\end{picture}
g(i\partial_\mu-m)\langle0|T\{(\bar{\psi}\star c \star\gamma^\mu \psi)(x)X\}|0\rangle \cr
=&\quad g \int dy~
\begin{picture}(100,70)(15,40)
\Vertex(50,85){1}
\Vertex(30,0){1}
\Vertex(70,0){1}
\Vertex(60,70){1}
\GOval(50,42.5)(17,30)(0){0.75}
\ArrowLine(30,55)(50,85)
\ArrowLine(50,85)(60,70)
\Line(30,30)(30,0)
\Line(70,30)(70,0)
\Photon(60,70)(50,60){2}{3}
\ArrowLine(60,70)(70,55)
\Text(50,95)[]{$x$}
\Text(70,70)[]{$y$}
\Text(30,-10)[]{$x_1$}
\Text(50,-10)[]{...}
\Text(70,-10)[]{$x_n$}
\Text(60,85)[l]{$(i\!\!\not\!\partial-m)$}
\end{picture} +
\quad g~\sum_{l\in L_{\bar{\psi}}}
\begin{picture}(100,70)(15,40)
\Vertex(65,85){1}
\Vertex(30,0){1}
\Vertex(70,0){1}
\Vertex(110,0){1}
\GOval(50,42.5)(17,30)(0){0.75}
\ArrowLine(30,55)(65,85)
\ArrowLine(65,85)(110,0)
\ArrowLine(30,30)(30,0)
\ArrowLine(70,30)(70,0)
\Text(65,95)[]{$x$}
\Text(30,-10)[]{$x_1$}
\Text(50,-10)[]{...}
\Text(50,0)[]{$\check{x}_l$}
\Text(70,-10)[]{$x_n$}
\Text(110,-10)[]{$x_l$}
\Text(93,50)[l]{$(i\!\!\not\!\partial-m)$}
\end{picture}
\cr
\end{align}
where $L_{\bar\psi}$ refers to the set of all indices $l\in \{1,...,n\}$ for which we have a field $\bar\psi$ at place $x_l$, analogously we will use the notations $L_\psi,L_{A_\mu},L_c,L_{\bar c}$.

The analytic expressions are
\begin{align}
\begin{picture}(0,10)(0,0)
\Line(17,10)(17,17)
\Line(17,17)(131,17)
\LongArrow(131,17)(131,10)
\end{picture}
&g(i\partial_\mu-m)\langle0|T\{(\bar{\psi}\star c \star\gamma^\mu \psi)(x)X\}|0\rangle \cr
=&
\begin{picture}(0,20)(0,0)
\Line(43,10)(43,17)
\Line(43,17)(156,17)
\LongArrow(156,17)(156,10)
\end{picture}
\begin{picture}(0,20)(0,0)
\Line(160,11)(160,16)
\Line(160,16)(198,16)
\Line(198,16)(198,11)
\end{picture}
g\int dy ~(i\partial_\mu-m)\langle 0|  T\{(\bar{\psi}\star c \star\gamma^\mu \psi)(x) ig(\bar\psi\star\gamma^\nu A_\nu\star\psi)(y)X\}|0\rangle                 \cr\
&
\begin{picture}(0,10)(0,0)
\Line(48,10)(48,17)
\Line(48,17)(166,17)
\LongArrow(166,17)(166,10)
\end{picture}
\begin{picture}(0,10)(0,0)
\Line(170,11)(170,16)
\Line(170,16)(194,16)
\Line(194,16)(194,11)
\end{picture}
+g\sum_{l\in L_{\bar{\psi}}} (i\partial_\mu-m)\langle 0|  T\big\{(\bar{\psi}\star c \star\gamma^\mu \psi)(x) \bar\psi(x_l)X_{\check{l}}\big\}|0\rangle  \cr
=&
\begin{picture}(0,10)(0,0)
\Line(138,10)(138,17)
\Line(138,17)(282,17)
\LongArrow(282,17)(282,10)
\end{picture}
ig^2\int dy ~ e^{\frac{i}{2}\theta^{\rho\sigma}\partial_\rho^\xi\partial_\sigma^\eta}e^{\frac{i}{2}\theta^{\rho\sigma}\partial_\rho^\zeta\partial_\sigma^\chi}(i\partial_\mu-m)\langle0|T\big\{(\bar\psi\star c\gamma^\mu)(x+\xi)S_C(x+\eta-y-\zeta)\cr
&\qquad (\gamma^\nu A_\nu\star \psi)(y+\chi)X\big\}|0\rangle\Big|_{\xi,\eta,\zeta,\chi=0} \cr
\begin{picture}(-45,20)(0,0)
\Line(48,10)(48,17)
\Line(48,17)(195,17)
\LongArrow(195,17)(195,10)
\end{picture}
&+g\sum_{l\in L_{\bar{\psi}}} e^{\frac{i}{2}\theta^{\rho\sigma}\partial_\rho^\xi\partial_\sigma^\eta}(i\partial_\mu-m)\langle 0|  T\big\{(\bar{\psi}\star c \gamma^\mu )(x+\xi)S_C(x+\eta-x_l)X_{\check{l}}\big\}|0\rangle\Big|_{\xi=\eta=0}
\end{align}
which by using $(i\!\!\not\!\partial-m)S_C(x-y)=i\delta^{(4)}(x-y)$ allow us to carry out the integration over y and turn into
\begin{align}
\begin{picture}(0,10)(0,0)
\Line(17,10)(17,17)
\Line(17,17)(130,17)
\LongArrow(130,17)(130,10)
\end{picture}
&g(i\partial_\mu-m)\langle0|T\{(\bar{\psi}\star c \star\gamma^\mu \psi)(x)X\}|0\rangle \cr
=&
-g^2 e^{\frac{i}{2}\theta^{\rho\sigma}\partial_\rho^\xi\partial_\sigma^\eta}e^{\frac{i}{2}\theta^{\rho\sigma}\partial_\rho^\zeta\partial_\sigma^\chi}\langle0|T\big\{(\bar\psi\star c)(x+\xi)(\gamma^\nu A_\nu\star \psi)(x+\eta-\zeta+\chi)X\big\}|0\rangle\Big|_{\xi,\eta,\zeta,\chi=0} \cr
&+ig\sum_{l\in L_{\bar{\psi}}}e^{\frac{i}{2}\theta^{\rho\sigma}\partial_\rho^\xi\partial_\sigma^\eta}\delta^{(4)}(x+\eta-x_l)\langle 0|  T\big\{(\bar{\psi}\star c )(x+\xi)X_{\check{l}}\big\}|0\rangle\Big|_{\xi=\eta=0}\quad .
\end{align}
Exploiting the condition $\theta^{0i}=0$ we find that we can write the four-dimensional $\delta$-distribution in the sum as $\delta^{(4)}(x+\eta-x_l)=\delta^{(0)}(x^0-x_l^0)\delta^{(3)}(x^i+\eta^i-x_l^i)$, from the first one-dimensional $\delta$-distribution we conclude that the terms in the sum are local in time and recognize them as contact terms.\\
The first term can be recast in the form of a Moyal-product, if we take \\
$e^{\frac{i}{2}\theta^{\rho\sigma}\partial_\rho^\zeta\partial_\sigma^\chi}f(x-\zeta+\chi)\Big|_{\zeta=\chi=0}=f(x)$ into account:
\begin{align}
\begin{picture}(0,10)(0,0)
\Line(17,10)(17,17)
\Line(17,17)(130,17)
\LongArrow(130,17)(130,10)
\end{picture}
&g(i\partial_\mu-m)\langle0|T\{(\bar{\psi}\star c \star\gamma^\mu \psi)(x)X\}|0\rangle \cr
=&
-g^2 e^{\frac{i}{2}\theta^{\rho\sigma}\partial_\rho^\xi\partial_\sigma^\eta}\langle0|T\big\{(\bar\psi\star c)(x+\xi)(\gamma^\nu A_\nu\star \psi)(x+\eta)X\big\}|0\rangle\Big|_{\xi=\eta=0} \quad + \text{c.t.}\cr
=&
-g^2\langle0|T\big\{(\bar\psi\star c\gamma^\mu\star A_\mu\star \psi)(x)X\big\}|0\rangle \quad + \text{c.t.} \quad .
\end{align}

The remaining terms are calculated in a similar manner, they yield
\begin{align}
&-i
\begin{picture}(0,20)(0,0)
\Line(12,10)(12,17)
\Line(12,17)(82,17)
\LongArrow(82,17)(82,11)
\end{picture}
(i\partial_\mu+m)\langle0|T\{(\bar{\psi}\star\gamma^\mu s{\psi})(x)X\}|0\rangle
=
\begin{picture}(0,20)(0,0)
\Line(17,10)(17,17)
\Line(17,17)(87,17)
\LongArrow(87,17)(87,11)
\end{picture}
g(i\partial_\mu+m)\langle0|T\{(\bar{\psi}\star\gamma^\mu \star c \star \psi)(x)X\}|0\rangle \cr
&\quad =g^2\langle0|T\big\{(\bar\psi\star \gamma^\mu A_\mu\star c\star \psi)(x)X\big\}|0\rangle \quad + \text{c.t.}
\end{align}
\begin{align}
&-i
\begin{picture}(0,20)(0,0)
\Line(5,10)(5,17)
\Line(5,17)(73,17)
\LongArrow(73,17)(73,10)
\end{picture}
\Box\langle0|T\{(sA_\mu\star A^\mu)(x)X\}|0\rangle
=-i
\begin{picture}(0,20)(0,0)
\Line(5,10)(5,17)
\Line(5,17)(137,17)
\LongArrow(137,17)(137,10)
\end{picture}
\Box\langle0|T\{\big((\partial_\mu c+ig[c,A_\mu]_\star)\star A^\mu\big)(x)X\}|0\rangle \cr
&\quad=-ig\langle0|T\{((\partial_\mu c+ig[c,A_\mu]_\star)\star \psi \gamma^{\mu T}\star\bar\psi)(x)X\}|0\rangle  \cr
&\qquad +g\langle0|T\{\big((\partial_\mu c+ig[c,A_\mu]_\star)\star(\partial^\mu B+ig\partial^\mu\bar c\star c)\big)(x)X\}|0\rangle  \cr
&\qquad +ig\langle0|T\{\big((\partial_\mu c+ig[c,A_\mu]_\star)\star c\star\partial^\mu\bar c\big)(x)X\}|0\rangle  \cr
&\qquad +g\langle0|T\{(-i\partial^\mu(D_\nu c\star F^{\mu\nu})+\frac{1}{2}g[\partial_\mu A_\nu-\partial_\nu A_\mu,c\star F^{\mu\nu}]_\star+g[\partial^\nu c\star F^{\mu\nu},A_\mu]_\star) \cr
&\qquad\qquad  +ig^2[c\star F^{\mu\nu}\star A^\mu,A_\nu]_\star+ig^2[A^\nu,c\star A^\mu\star F^{\mu\nu}]_\star)(x)X\}|0\rangle  \quad + \text{c.t.}\cr
&i
\begin{picture}(0,20)(0,0)
\Line(5,10)(5,17)
\Line(5,17)(62,17)
\LongArrow(62,17)(62,10)
\end{picture}
\Box\langle 0|T\{(s\bar{c}\star c)(x)X\}|0\rangle
=i
\begin{picture}(0,20)(0,0)
\Line(5,10)(5,17)
\Line(5,17)(60,17)
\LongArrow(60,17)(60,10)
\end{picture}
\Box\langle 0|T\{(B\star c)(x)X\}|0\rangle  \cr
&\quad=ig\langle 0|T\{\big(B\star \partial^\mu[c,A_\mu]_\star\big)(x)X\}|0\rangle  \quad + \text{c.t.} \cr
&i
\begin{picture}(0,20)(0,0)
\Line(5,10)(5,17)
\Line(5,17)(41,17)
\LongArrow(41,17)(41,10)
\end{picture}
\Box\langle 0|T\{(\bar{c}\star sc)(x)X\}|0\rangle
=
-g\begin{picture}(0,20)(0,0)
\Line(5,10)(5,17)
\Line(5,17)(41,17)
\LongArrow(41,17)(41,10)
\end{picture}
\Box\langle 0|T\{(\bar{c}\star c\star c)(x)X\}|0\rangle \cr
&\quad =g^2\langle 0|T\{([A_\mu,\partial^\mu \bar c]_\star \star c\star c)(x)X\}|0\rangle\quad + \text{c.t.}
\end{align}
Adding them up we find the identity
\begin{align}
\begin{picture}(0,10)(0,0)
\Line(17,10)(17,17)
\Line(17,17)(130,17)
\LongArrow(130,17)(130,10)
\end{picture}
&g(i\partial_\mu-m)\langle0|T\{(\bar{\psi}\star c \star\gamma^\mu \psi)(x)X\}|0\rangle
+
\begin{picture}(0,20)(0,0)
\Line(17,10)(17,17)
\Line(17,17)(87,17)
\LongArrow(87,17)(87,11)
\end{picture}
g(i\partial_\mu+m)\langle0|T\{(\bar{\psi}\star\gamma^\mu \star c \star \psi)(x)X\}|0\rangle \cr
&-i
\begin{picture}(0,20)(0,0)
\Line(5,10)(5,17)
\Line(5,17)(137,17)
\LongArrow(137,17)(137,10)
\end{picture}
\Box\langle0|T\{\big((\partial_\mu c+ig[c,A_\mu]_\star)\star A^\mu\big)(x)X\}|0\rangle  \cr
&+i
\begin{picture}(0,20)(0,0)
\Line(5,10)(5,17)
\Line(5,17)(60,17)
\LongArrow(60,17)(60,10)
\end{picture}
\Box\langle 0|T\{(B\star c)(x)X\}|0\rangle
-g\begin{picture}(0,20)(0,0)
\Line(5,10)(5,17)
\Line(5,17)(41,17)
\LongArrow(41,17)(41,10)
\end{picture}
\Box\langle 0|T\{(\bar{c}\star c\star c)(x)X\}|0\rangle \cr
=&-g^2\langle0|T\big\{(\bar\psi\star c\gamma^\mu\star A_\mu\star \psi)(x)X\big\}|0\rangle
+
g^2\langle0|T\big\{(\bar\psi\star \gamma^\mu A_\mu\star c\star \psi)(x)X\big\}|0\rangle\cr
&-ig\langle0|T\{((\partial_\mu c+ig[c,A_\mu]_\star)\star \psi \gamma^{\mu T}\star\bar\psi)(x)X\}|0\rangle  \cr
& +g\langle0|T\{\big((\partial_\mu c+ig[c,A_\mu]_\star)\star(\partial^\mu B+ig\partial^\mu\bar c\star c)\big)(x)X\}|0\rangle  \cr
&+ig\langle0|T\{\big((\partial_\mu c+ig[c,A_\mu]_\star)\star c\star\partial^\mu\bar c\big)(x)X\}|0\rangle  \cr
&+g\langle0|T\{(-i\partial^\mu(D_\nu c\star F^{\mu\nu})+\frac{1}{2}g[\partial_\mu A_\nu-\partial_\nu A_\mu,c\star F^{\mu\nu}]_\star+g[\partial^\nu c\star F^{\mu\nu},A_\mu]_\star) \cr
&\qquad  +ig^2[c\star F^{\mu\nu}\star A^\mu,A_\nu]_\star+ig^2[A^\nu,c\star A^\mu\star F^{\mu\nu}]_\star)(x)X\}|0\rangle  \cr
&+
ig\langle 0|T\{\big(B\star \partial^\mu[c,A_\mu]_\star\big)(x)X\}|0\rangle
+g^2\langle 0|T\{([A_\mu,\partial^\mu \bar c]_\star \star c\star c)(x)X\}|0\rangle\quad + \text{c.t.}
\end{align}
which we may put into the form
\begin{align}
&\partial_\mu\langle 0| T\big\{\big(ig\bar\psi\gamma^\mu\star c\star \psi+iD^\mu c\star B+iD_\nu c\star F^{\mu\nu}-g\partial^\mu\bar c\star c\star c\cr
&\qquad\quad  +gc\star\partial^\nu\partial_\nu A_\mu -gc\star\partial_\mu\partial^\nu A_\nu \big)(x)X\big\}|0\rangle \quad + \text{c.t.} \cr
=&g^2\langle0|T\big\{\big([A_\mu\star c\star c,\partial^\mu\bar c]_\star+[A_\mu\star\partial^\mu \bar c\star c,c]_\star +[A_\mu\star c\star \partial^\mu\bar c,c]_\star\big)(x)X\big\}|0\rangle\cr
&+ig\langle0|T\big\{\big[\bar\psi_\alpha\gamma^\mu_{\alpha\beta},(\partial_\mu c+ig[c,A_\mu]_\star)\star \psi_\beta\big]_\star(x)X\big\}|0\rangle \cr
&+g\langle0|T\big\{\big([\partial^\mu c,\partial_\mu\bar c\star c]_\star + [\partial^\mu c\star c, \partial_\mu\bar c]_\star\big)(x)X\big\}|0\rangle
\nonumber
\end{align}
\begin{align}
&+g\langle0|T\big\{\big(\frac{1}{2}[\partial_\nu A_\mu-\partial_\mu A_\nu,c\star F^{\mu\nu}]_\star +[\partial^\nu c\star F^{\mu\nu}, A_\mu]_\star + ig[c\star F^{\mu\nu}\star A_\mu, A_\nu]_\star  \cr
&\quad +ig[A_\nu, c\star A_\mu\star F^{\mu\nu}]_\star \big)(x)X\big\}|0\rangle   \quad .
\label{preBRS}
\end{align}
How can we read off a current fulfilling the conservation law (\ref{current}) from this identity? The left hand side is already in the form of a total derivative. We notice the right hand side to be a sum of star-commutators, these are the additional terms mentioned in the beginning of this section. Each of them can be written as a total derivative, which we will show in general for the star-commutator of two functions $f$ and $g$:
\begin{align}
[f,g]_\star(x)&=(f\star g)(x)-(g\star f)(x) \cr
&=\Big(e^{\frac{i}{2}\theta^{\mu\nu}\partial_\mu^\xi\partial_\nu^\eta}-e^{\frac{i}{2}\theta^{\mu\nu}\partial_\mu^\eta\partial_\nu^\xi}\Big)f(x+\xi)g(x+\eta)\Big|_{\xi=\eta=0}  \cr
&=\Big(\sum_{n=1}^\infty\frac{1}{n!}\big(\frac{i}{2}\theta^{\mu\nu}\partial_\mu^\xi\partial_\nu^\eta\big)^n-\sum_{n=1}^\infty\frac{1}{n!}\big(\frac{i}{2}\theta^{\mu\nu}\partial_\mu^\eta\partial_\nu^\xi\big)^n\Big)f(x+\xi)g(x+\eta)\Big|_{\xi=\eta=0}  \cr
&=\frac{i}{2}\theta^{\mu\nu}\partial_\mu^\xi\partial_\nu^\eta\Big(\sum_{n=1}^\infty\frac{1}{n!}\big(\frac{i}{2}\theta^{\mu\nu}\partial_\mu^\xi\partial_\nu^\eta\big)^{n-1}+\sum_{n=1}^\infty\frac{1}{n!}\big(-\frac{i}{2}\theta^{\mu\nu}\partial_\mu^\xi\partial_\nu^\eta\big)^{n-1}\Big)  \cr
&\hspace{5cm} f(x+\xi)g(x+\eta)\Big|_{\xi=\eta=0}  \cr
&=\frac{i}{2}\theta^{\mu\nu}(\partial_\mu^\xi+\partial_\mu^\eta)\partial_\nu^\eta\Big(\sum_{n=1}^\infty\frac{1}{n!}\big(\frac{i}{2}\theta^{\mu\nu}\partial_\mu^\xi\partial_\nu^\eta\big)^{n-1}+\sum_{n=1}^\infty\frac{1}{n!}\big(-\frac{i}{2}\theta^{\mu\nu}\partial_\mu^\xi\partial_\nu^\eta\big)^{n-1}\Big)  \cr
&\hspace{5cm} f(x+\xi)g(x+\eta)\Big|_{\xi=\eta=0}  \cr
&=\partial_\mu^x\frac{i}{2}\theta^{\mu\nu}\partial_\nu^\eta2\sum_{n=0}^\infty\frac{1}{(2n+1)!}\big(\frac{i}{2}\theta^{\mu\nu}\partial_\mu^\xi\partial_\nu^\eta\big)^{2n}
 f(x+\xi)g(x+\eta)\Big|_{\xi=\eta=0}  \cr
&=\partial_\mu H^\mu_{[f,g]_\star} (x)
\end{align}
where we defined the object
\beq
H^\mu_{[f,g]_\star} (x) \equiv
i\theta^{\mu\nu}\partial_\nu^\eta\sum_{n=0}^\infty\frac{1}{(2n+1)!}\big(\frac{i}{2}\theta^{\mu\nu}\partial_\mu^\xi\partial_\nu^\eta\big)^{2n}
 f(x+\xi)g(x+\eta)\Big|_{\xi=\eta=0}   \quad .
\eeq
We use this notation to find the conserved current
\begin{align}
j^\mu_{BRS}=&ig\bar\psi\gamma^\mu\star c\star \psi+iD^\mu c\star B+iD_\nu c\star F^{\mu\nu}-g\partial^\mu\bar c\star c\star c +gc\star\partial^\nu\partial_\nu A_\mu \cr &-gc\star\partial_\mu\partial^\nu A_\nu - g^2H^\mu_{[A_\mu\star c\star c,\partial^\mu\bar c]_\star} -g^2 H^\mu_{[A_\mu\star\partial^\mu\bar c\star c,c]_\star}-g^2H^\mu_{[A_\mu\star c\star \partial^\mu\bar c,c]_\star} \cr
&-igH^\mu_{[\bar\psi\gamma^\mu,(\partial_\mu c+ig[c,A_\mu]_\star)\star \psi]_\star}
-gH^\mu_{[\partial_\mu c,\partial^\mu\bar c\star c]_\star}-gH^\mu_{[\partial_\mu c\star c, \partial^\mu\bar c]_\star}  \cr
&-\frac{1}{2}gH^\mu_{[\partial_\nu A_\mu-\partial_\mu A_\nu,c\star F^{\mu\nu}]_\star}-gH^\mu_{[\partial^\nu c\star F^{\mu\nu}, A_\mu]_\star}-ig^2H^\mu_{[c\star F^{\mu\nu}\star A_\mu, A_\nu]_\star} \cr
&-ig^2H^\mu_{[A_\nu, c\star A_\mu\star F^{\mu\nu}]_\star}
\label{brs_curr}
\end{align}

which according to identity (\ref{preBRS}) indeed fulfills the conservation law
\beq
\partial_\mu\langle 0|T\{j^\mu_{BRS}(x)X\}|0\rangle + \text{c.t.} =0   \quad.
\eeq

Let us briefly comment on the terms appearing in the BRS current (\ref{brs_curr}). The first six resemble the ones which are also present in the nonabelian commutative case, while the remaining ones are due to the non-invariance of the Lagrangian under BRS transformations and only appear in noncommutative theories. \\

We want to go on and derive the Slavnov-Taylor identities
\beq
\langle0|T(sX)|0\rangle=0 \quad , \label{STI}
\eeq

where $s$ stands for the BRS-transformations on the collection of field operators $X$. We do this by integrating identity (\ref{preBRS}) over the point $x$, the terms on the right hand side then all dissappear, as they are total derivatives, and on the left hand side only the contact terms remain. How do these look like? They all come from evaluating the expression (\ref{start}), at the beginning we derived the one met in calculating
\beq
i
\begin{picture}(0,20)(0,0)
\Line(12,10)(12,17)
\Line(12,17)(116,17)
\LongArrow(116,17)(116,10)
\end{picture}
(i\partial_\mu-m)\langle0|T\{(s\bar{\psi}\star\gamma^\mu \psi)(x)X\}|0\rangle\Big|_{\xi=\eta=0} \quad ,
\eeq
we found
\beq
ig\sum_{l\in L_{\bar{\psi}}}e^{\frac{i}{2}\theta^{\rho\sigma}\partial_\rho^\xi\partial_\sigma^\eta}\delta^{(4)}(x+\eta-x_l)\langle 0|  T\big\{(\bar{\psi}\star c )(x+\xi)X_{\check{l}}\big\}|0\rangle \quad .
\eeq 
If integrated over $x$, we replace $x$ by $x_l-\eta$ and skip the infinitesimal parameters $\xi,\eta$, such that we obtain
\beq
ig\sum_{l\in L_{\bar{\psi}}}\langle 0|  T\big\{(\bar{\psi}\star c )(x_l)(X_{\check{l}}\big\}|0\rangle =-\sum_{l\in L_{\bar{\psi}}}\langle 0|  T\big\{s\bar{\psi} (x_l)X_{\check{l}}\big\}|0\rangle \quad .
\eeq
For the remaining contact terms coming from the other terms in expression (\ref{start}) we obtain similar results, the integrated identity (\ref{preBRS}) finally yields
\begin{align}
\sum_{l\in L_{\bar{\psi}}}\langle 0|  T\big\{s\bar\psi (x_l)X_{\check{l}}\big\}|0\rangle
+\sum_{l\in L_{\psi}}\langle 0|  T\big\{s\psi (x_l)X_{\check{l}}\big\}|0\rangle
+\sum_{l\in L_{A_\mu}}\langle 0|  T\big\{sA_\mu (x_l)X_{\check{l}}\big\}|0\rangle  &\cr
+\sum_{l\in L_{\bar{c}}}\langle 0|  T\big\{s\bar{c}(x_l)X_{\check{l}}\big\}|0\rangle
+\sum_{l\in L_{c}}\langle 0|  T\big\{s c (x_l)X_{\check{l}}\big\}|0\rangle \quad =&~0
\end{align}
which is the Slavnov-Taylor identity (\ref{STI}) at full length. It has just the same form as in the commutative case, neither a star product nor additional terms appear. The proof of stability of the physical Hilbert space under time evolution, which substantially builds on the Slavnov-Taylor identity, can thus be done just as in the commutative case. In fact, unitarity in NCQED is valid for the case of only spatial noncommutativity.

\newpage
\chapter{The approach of TOPT\label{topt}}

We now go over to the case of time-space noncommutativity, i.e. $\theta^{0i}\neq 0$.
The question how to introduce perturbative methods in this case, even on the unrenormalized level, is not yet satisfactorily answered. The approach of modified Feynman rules has been shown to violate unitarity \cite{GomisMehen}, other formalisms have been proposed to cure this problem. In this section we want to discuss the concept of TOPT, which was elaborated in noncommutative scalar theories by Liao and Sibold \cite{LiaoSiboldTOPT}. It is unitary by construction, and  we will show how it simplifies to the approach of modified Feynman rules in the case of $\theta^{0i}=0$. Equations of motion are discussed in  a detailed example, and we investigate the question of remaining Lorentz invariance.\\

\section{The concept}

In TOPT a Green's function for fields $\phi_1,...,\phi_n$ is calculated by the formula
\begin{align}
\langle 0|T\big(\phi_1(x_1)...\phi_n(x_n)\big)|0\rangle
=\langle 0|T\Big(\phi_1^{(0)}(x_1)...\phi_n^{(0)}(x_n)e^{i\int dx \mathcal{L}^{(0)}_{\text{int}}(x)}\Big)|0\rangle\quad,  \label{Green1}
\end{align}
the index $^{(0)}$ referring to free fields. The subtlety to observe is that the time-ordering is carried out \emph{after} the star product, which appears in $\mathcal{L}_{\text{int}}$, has been taken. Just as in the commutative case, an S-matrix based on the so defined Green's functions is unitary.

The calculation of examples for (\ref{Green1}) has been performed in \cite{LiaoSiboldTOPT} in detail, here, we only want to outline the characteristic steps. Let us evaluate the three-point function
\beq
\langle 0|T\big(\phi(x_1)\phi(x_2)\phi(x_3)\big)|0\rangle
\eeq
in the case of a cubic interaction for one scalar field $\phi$,
\begin{align}
\mathcal{L}_{\text{int}}(x)&=\frac{g}{3!}(\phi\star \phi\star \phi)(x) \quad .
\end{align}
In the first order of $g$ we have according to (\ref{Green1})
\begin{align}
\langle 0|T\big(\phi(x_1)\phi(x_2)\phi(x_3)\big)|0\rangle=i\frac{g}{3!}\int dx~\langle 0|T\Big(\phi^{(0)}(x_1)\phi^{(0)}(x_2)\phi^{(0)}(x_3)\big(\phi^{(0)}\star \phi^{(0)}\star\phi^{(0)}\big)(x)\Big)|0\rangle
\end{align}
We will omit the superscript $^{(0)}$ in the following. \\
The free fields may be contracted in different ways, we consider one possibility and obtain the other (without disconnected diagrams) by permutations of $x_1,x_2,x_3$.
\begin{align}
\begin{picture}(0,20)(0,0)
\Line(75,18)(158,18)
\Line(75,18)(75,10)
\Line(158,18)(158,10)
\end{picture}
\begin{picture}(0,20)(0,0)
\Line(101,22)(175,22)
\Line(101,22)(101,10)
\Line(175,22)(175,10)
\end{picture}
\begin{picture}(0,20)(0,0)
\Line(127,26)(192,26)
\Line(127,26)(127,10)
\Line(192,26)(192,10)
\end{picture}
&i\frac{g}{3!}\int dx~\langle 0|T\Big(\phi(x_1)\phi(x_2)\phi(x_3)\big(\phi\star \phi\star\phi\big)(x)\Big)|0\rangle  \cr
=~&i\frac{g}{3!}\int dx~\langle 0|T\Big(e^{\frac{i}{2}\theta^{\mu\nu}\partial_\mu^\xi\partial_\nu^\eta}e^{\frac{i}{2}\theta^{\rho\sigma}\partial_\rho^\zeta\partial_\sigma^\chi}\cr
&\qquad
\begin{picture}(0,20)(0,0)
\Line(4,18)(83,18)
\Line(4,18)(4,10)
\Line(83,18)(83,10)
\end{picture}
\begin{picture}(0,20)(0,0)
\Line(30,22)(122,22)
\Line(30,22)(30,10)
\Line(122,22)(122,10)
\end{picture}
\begin{picture}(0,20)(0,0)
\Line(56,26)(181,26)
\Line(56,26)(56,10)
\Line(181,26)(181,10)
\end{picture} \phi(x_1)\phi(x_2)\phi(x_3)\phi(x+\xi)\phi(x+\eta+\zeta)\phi(x+\eta+\chi)\Big|_{\xi,\eta,\zeta,\chi=0} \Big)|0\rangle   \cr
=~&i\frac{g}{3!}\int dx~e^{\frac{i}{2}\theta^{\mu\nu}\partial_\mu^\xi\partial_\nu^\eta}e^{\frac{i}{2}\theta^{\rho\sigma}\partial_\rho^\zeta\partial_\sigma^\chi}\cr
&\qquad
\big\{\vartheta(x_1^0-x^0)i\Delta_+(x_1^0-x^0-\xi^0)+\vartheta(x^0-x_1^0)i\Delta_+(x^0+\xi^0-x_1^0)\big\}\cr
&\qquad \big\{\vartheta(x_2^0-x^0)i\Delta_+(x_2^0-x^0-\eta^0-\zeta^0)+\vartheta(x^0-x_2^0)i\Delta_+(x^0+\eta^0+\zeta^0-x_2^0)\big\}\cr
&\qquad \big\{\vartheta(x_3^0-x^0)i\Delta_+(x_3^0-x^0-\eta^0-\chi^0)+\vartheta(x^0-x_3^0)i\Delta_+(x^0+\eta^0+\chi^0-x_3^0)\big\}\Big|_{\xi,\eta,\zeta,\chi=0}  \cr \label{noprop}
\end{align}
where we defined $i\Delta_+(x-y)=\langle 0|\phi(x)\phi(y)|0\rangle$, $\vartheta$ is the step function. Notice that the infinitesimal variables $\xi, \eta, \zeta, \chi$ which define the star product appear in the functions $\Delta_+$ but not in the step functions, such that $\Delta_+$ has another argument than its assigned step function. This is due to the prescription of taking the star product before time-ordering. The usual way to obtain propagators via the relation
\beq
\Delta_C(x-y)=\vartheta(x^o-y^0)i\Delta_+(x-y)+\vartheta(y^0-x^0)i\Delta_+(y-x)
\eeq

can thus not be applied unless $\theta^{0i}=0$ (only then it makes no difference whether or not the infinitesimal variables appear in the step functions). This is  why in the case $\theta^{0i}\neq 0$ \emph{we do not obtain the ordinary propagators}.

To simplify expression (\ref{noprop}) we use the Fourier transformations
\begin{align}
\vartheta(x_i^0-x^0)&=\frac{i}{2\pi}\int ds_i~\frac{e^{-is_i(x_i^0-x^0)}}{s_i+i\epsilon}\cr
\vartheta(x^0-x_i^0)&=-\frac{i}{2\pi}\int ds_i~\frac{e^{-is_i(x_i^0-x^0)}}{s_i-i\epsilon}\cr
i\Delta_+(x_i-x)&=\int \frac{d^3p_i}{2E_{p_i}(2\pi)^3}e^{-ip_i^+(x_i-x)}\cr
i\Delta_+(x-x_i)&=\int \frac{d^3p_i}{2E_{p_i}(2\pi)^3}e^{-ip_i^-(x_i-x)}
\end{align}
with $p^\pm=(\pm E_p,{\bf p})$. As the star product is performed with respect to the $\Delta_+$-functions, the noncommutative phase factor will contain the on-shell momenta $p_i^\pm$. We further substitute
$p_i^0\equiv s_i\pm E_{p_i}$ and arrive after some short calculation at
\begin{align}
\begin{picture}(0,20)(0,0)
\Line(75,18)(158,18)
\Line(75,18)(75,10)
\Line(158,18)(158,10)
\end{picture}
\begin{picture}(0,20)(0,0)
\Line(101,22)(175,22)
\Line(101,22)(101,10)
\Line(175,22)(175,10)
\end{picture}
\begin{picture}(0,20)(0,0)
\Line(127,26)(192,26)
\Line(127,26)(127,10)
\Line(192,26)(192,10)
\end{picture}
&i\frac{g}{3!}\int dx~\langle 0|T\Big(\phi(x_1)\phi(x_2)\phi(x_3)\big(\phi\star \phi\star\phi\big)(x)\Big)|0\rangle  \cr
=~&i\frac{g}{3!}\sum_{\lambda_1,\lambda_2,\lambda_3 =\pm 1}\int \frac{d^4p_1}{(2\pi)^4} \frac{d^4p_2}{(2\pi)^4} \frac{d^4p_3}{(2\pi)^4} \int dx ~P_{\lambda_1}(p_1) P_{\lambda_2}(p_2) P_{\lambda_3}(p_3)\cr
&\qquad e^{-ip_1(x_1-x)}e^{-ip_2(x_2-x)}e^{-ip_3(x_3-x)}e^{-i(p_1^{\lambda_1},p_2^{\lambda_2},p_3^{\lambda_3})}
\end{align}
where we defined the quantities
\beq
P_\lambda(p)=\frac{i\lambda}{2E_p(p^0-\lambda(E_p-i\epsilon))}
\eeq
for $\lambda=\pm 1$
and the abbreviation for the noncommutative phase
\beq
(p_1,p_2,...,p_n)=\sum_{i<j\leq n}p_i\wedge p_j \quad .
\eeq
Performing the integration over $x$ yields the momentum conservation, $(2\pi)^4\delta^{(4)}(p_1+p_2+p_3)$. To compute the three-point function at first order, with which we started, we have to sum over all permutations of $x_1,x_2,x_3$, which can be equivalently done by permuting $p_1,p_2,p_3$ in the noncommutative phase factor:
\begin{align}
&\langle 0|T\big(\phi(x_1)\phi(x_2)\phi(x_3)\big)|0\rangle \cr
=~&i\frac{g}{3!}\int dx~\langle 0|T\Big(\phi(x_1)\phi(x_2)\phi(x_3)\big(\phi\star \phi\star\phi\big)(x)\Big)|0\rangle  \cr
=~&i\frac{g}{3!}\sum_{\lambda_1,\lambda_2,\lambda_3 =\pm 1}\int \frac{d^4p_1}{(2\pi)^4} \frac{d^4p_2}{(2\pi)^4} \frac{d^4p_3}{(2\pi)^4}~(2\pi)^4 \delta^{(4)}(p_1+p_2+p_3) ~P_{\lambda_1}(p_1) P_{\lambda_2}(p_2) P_{\lambda_3}(p_3)\cr
&\qquad e^{-ip_1x_1}e^{-ip_2x_2}e^{-ip_3x_3}\sum_{\pi\in S_3}e^{-i\big(p_{\pi(1)}^{\lambda_{\pi(1)}},~p_{\pi(2)}^{\lambda_{\pi(2)}},~p_{\pi(3)}^{\lambda_{\pi(3)}}\big)}   \quad .
\label{threepoint}
\end{align}

The structure that becomes visible in this example in fact generalizes to other Green's functions. They may be expressed diagrammatically if one assigns to each line a momentum $p$ and $\lambda=\pm1$, corresponding to the expression $P_\lambda(p)$. A vertex contributes momentum conservation and a noncommutative phase factor. In the case of  cubic interaction this is $i\frac{g}{3!}\sum_{\pi\in S_3}e^{-i\big(p_{\pi(1)}^{\lambda_{\pi(1)}},~p_{\pi(2)}^{\lambda_{\pi(2)}},~p_{\pi(3)}^{\lambda_{\pi(3)}}\big)}$, as calculated above. \\

For only spatial noncommutativity we need not specify the zero-component of the momenta in the phase factor as they do not contribute, such that we can carry out the summation over the $\lambda_i$'s by using
\beq
\sum_{\lambda=\pm 1} P_\lambda(p)=\tilde{\Delta}_C(p) \quad ,
\eeq

so for $\theta^{0i}=0$
\begin{align}
&\langle 0|T\big(\phi(x_1)\phi(x_2)\phi(x_3)\big)|0\rangle \cr
=&i\frac{g}{3!}\int dx~\langle 0|T\Big(\phi(x_1)\phi(x_2)\phi(x_3)\big(\phi\star \phi\star\phi\big)(x)\Big)|0\rangle  \cr
=&i\frac{g}{3!}\int \frac{d^4p_1}{(2\pi)^4} \frac{d^4p_2}{(2\pi)^4} \frac{d^4p_3}{(2\pi)^4}~(2\pi)^4 \delta^{(4)}(p_1+p_2+p_3) ~\tilde{\Delta}_C(p_1) \tilde{\Delta}_C(p_2)\tilde{\Delta}_C(p_3)\cr
&\qquad e^{-ip_1x_1}e^{-ip_2x_2}e^{-ip_3x_3}\sum_{\pi\in S_3}e^{-i(p_{\pi(1)},~p_{\pi(2)},~p_{\pi(3)})}
\end{align}

which is the result of the modified Feynman rules \cite{Filk}. \\

However, this is only obtained for vanishing $\theta^{0i}$. If this is not the case, we cannot carry out the summation over the $\lambda_i$'s and do not obtain propagators, but are left with expression (\ref{threepoint}) and the general prescriptions given subsequently.

\newpage

\section{A scattering process in double gauged noncommutative electrodynamics}

The idea of double gauging the global $U(1)$ symmetry on noncommutative space leading to a richer structure of interaction terms was introduced in \cite{double}. We want to apply this idea to Coulomb scattering and investigate the Ward identity in this case. The framework of TOPT will be applied, we therefore at first derive Feynman rules for this theory.

\subsection{Feynman rules}

In the previous section we derived TOPT for scalar particles. Dealing now with spin-$\frac{1}{2}$ and spin-$1$ particles, we have to make some modifications. \\
The propagators are in TOPT, in the scalar case, replaced by the quantities
\beq
P_\lambda(k)=\frac{i\lambda}{2E_k(k^0-\lambda(E_k-i\epsilon))}
\eeq
where $\lambda=\pm1$. They are the Fourier transforms of the positive and negative energy contribution of the propagator:
\begin{align}
\Delta_C(x-y)&=\vartheta(x^0-y^0)\langle0|\phi(x)\phi(y)|0\rangle+\vartheta(y^0-x^0)\langle0|\phi(y)\phi(x)|0\rangle\cr
&=\sum_{\lambda=\pm1}\frac{i\lambda}{2\pi}\int ds\frac{e^{-is(x^0-y^0)}}{s+i\epsilon\lambda} \int \frac{d^3k}{2E_k(2\pi)^3}e^{-ik^\lambda (x-y)} \cr
&=\sum_{\lambda=\pm1}\int \frac{d^4k}{(2\pi)^4}\frac{i\lambda}{2E_k}\frac{e^{-is(x^0-y^0)}}{s+i\epsilon\lambda}e^{-ik (x-y)}\Big|_{s=k^0-\lambda E_k} \cr
&=\sum_{\lambda=\pm1}\int \frac{d^4k}{(2\pi)^4}P_\lambda(k)e^{-ik(x-y)} \quad ,
\end{align}
we will denote $k^\lambda=(\lambda E_k,k_1,k_2,k_3)$.\\
For spin-$\frac{1}{2}$ particles we calculate in the same way
\begin{align}
S_C(x-y)&=\vartheta(x^0-y^0)\langle0|\psi(x)\bar{\psi}(y)|0\rangle-\vartheta(y^0-x^0)\langle0|\bar{\psi}(y)\psi(x)|0\rangle\cr
&=\sum_{\lambda=\pm1}\frac{i\lambda}{2\pi}\int ds\frac{e^{-is(x^0-y^0)}}{s+i\epsilon\lambda} \int \frac{d^3k}{2E_k(2\pi)^3}(\not\!k+m)e^{-ik^\lambda (x-y)} \cr
&=\sum_{\lambda=\pm1}\int \frac{d^4k}{(2\pi)^4}\frac{i\lambda}{2E_k}\frac{e^{-is(x^0-y^0)}}{s+i\epsilon\lambda}(\not\!k^\lambda+m)e^{-ik^\lambda (x-y)}\Big|_{s=k^0-\lambda E_k} \cr
&=\sum_{\lambda=\pm1}\int \frac{d^4k}{(2\pi)^4}S_\lambda(k)e^{-ik(x-y)}
\end{align}
where we defined
\beq
S_\lambda(k)=\frac{i\lambda(\not\!k^\lambda+m)}{2E_k(k^0-\lambda(E_k-i\epsilon))} \quad .
\eeq
We find analogously
\begin{align}
\Delta_C^{\mu\nu}(x-y)=-g^{\mu\nu}\Delta_C(x-y)=-g^{\mu\nu}\sum_{\lambda=\pm1}\int \frac{d^4k}{(2\pi)^4}P_\lambda(k)e^{-ik(x-y)} \quad ,
\end{align}
and associate these expressions to lines:
\begin{align}
\begin{picture}(65,10)(0,2)
\ArrowLine(0,5)(50,5)
\Text(25,-4)[]{$\scriptstyle k,\lambda$}
\end{picture}
&=S_\lambda(k)=\frac{i\lambda(\not\!k^\lambda+m)}{2E_k(k^0-\lambda(E_k-i\epsilon))} \cr
\begin{picture}(65,10)(0,2)
\Photon(0,5)(50,5){2}{4}
\LongArrow(15,-2)(35,-2)
\Text(-7,5)[]{$\scriptstyle\mu$}
\Text(57,5)[]{$\scriptstyle\nu$}
\Text(25,-10)[]{$\scriptstyle k,\lambda$}
\end{picture}
&=-g^{\mu\nu}P_\lambda(k)=-g^{\mu\nu}\frac{i\lambda}{2E_k(k^0-\lambda(E_k-i\epsilon))} \quad .
\end{align}
Which form do the vertices have? They are phase factors arising in the Fourier transformation of the interacting part of the Lagrangian, with the momenta being on-shell. \\
The Lagrangian of the system is (see \cite{double})
\beq
\mathcal{L}=i\bar{\psi}\star \gamma^\mu D_\mu\psi-m^2\bar{\psi}\star\psi-\frac{1}{4}L_{\mu\nu}\star L^{\mu\nu}-\frac{1}{4}R_{\mu\nu}\star R^{\mu\nu}
\eeq
with
\begin{align}
D_\mu\psi=~&\partial_\mu\psi-ig_L L_\mu\star\psi+ig_R\psi\star R_\mu \cr
L_{\mu\nu}=~&\partial_\mu L_\nu-\partial_\nu L_\mu -ig_L[L_\mu.L_\nu]_\star  \cr
R_{\mu\nu}=~&\partial_\mu R_\nu-\partial_\nu R_\mu -ig_R[R_\mu.R_\nu]_\star
\end{align}
where $L_\mu$ is a left and $R_\mu$ a right transforming vector field. Physical meaningful fields are
\beq
\begin{pmatrix}
A_\mu \\ B_\mu
\end{pmatrix}
=
\begin{pmatrix}
c & -s \\ s & c
\end{pmatrix}
\begin{pmatrix}
L_\mu \\ R_\mu
\end{pmatrix} \quad ;\quad c=\frac{g_L}{e}, s=\frac{g_R}{e}~~ \text{and}~~ e=\sqrt{g_L^2+g_R^2} \quad ,
\eeq
the field $A_\mu$ is considered as the physical one for $\theta^{\mu\nu}=0$ it couples to the spinor fields as usual, the field $B_\mu$ decouples in the commutative case.
The covariant derivatives expressed in terms of them become
\begin{align}
D_\mu\psi=&~\partial_\mu\psi-ie(c^2A_\mu\star\psi+s^2\psi\star A_\mu)+iecs[\psi,B_\mu]_\star \cr
L_{\mu\nu}=&~c(\partial_\mu A_\nu-\partial_\nu A_\mu-iec^2[A_\mu,A_\nu]_\star)  \cr
&+s(\partial_\mu B_\nu-\partial_\nu B_\mu-iecs[B_\mu,B_\nu]_\star)  \cr
&-iec^2s([A_\mu,B_\nu]_\star+[B_\mu,A_\nu]_\star)\cr
R_{\mu\nu}=~&c(\partial_\mu B_\nu-\partial_\nu B_\mu-iecs[B_\mu,B_\nu]_\star)  \cr
&-s(\partial_\mu A_\nu-\partial_\nu A_\mu+ies^2[A_\mu,A_\nu]_\star)  \cr
&+iecs^2([A_\mu,B_\nu]_\star+[B_\mu,A_\nu]_\star) \quad .
\end{align}
To express $\mathcal{L}$ in terms of $A_\mu$ and $B_\mu$ we have to perform some longer but straightforward calculations, and arrive at
\begin{align}
\mathcal{L}=~&i\bar{\psi}\star\gamma^\mu \partial_\mu\psi-m^2\bar{\psi}\star\psi  \\
&+ec^2\bar{\psi}\star\gamma^\mu A_\mu\star\psi +es^2\bar{\psi}\star\gamma^\mu \psi\star A_\mu-ecs\bar{\psi}\star\gamma^\mu[\psi,B_\mu]_\star  \label{psi} \\
&-\frac{1}{4}(\partial_\mu A_\nu-\partial_\nu A_\mu)\star(\partial^\mu A^\nu-\partial^\nu A^\mu) \label{AAA1} \\
&+\frac{ie}{4}(c^4-s^4)\big\{\partial_\mu A_\nu-\partial_\nu A_\mu,[A^\mu,A^\nu]_\star\big\}_\star \label{AAA2}\\
&+\frac{iecs}{4}\big\{\partial_\mu A_\nu-\partial_\nu A_\mu,[A^\mu,B^\nu]_\star+[B^\mu,A^\nu]_\star\big\}_\star \label{AAB}\\
&+\frac{iecs}{4}\big\{\partial_\mu B_\nu-\partial_\nu B_\mu,[A^\mu,A^\nu]_\star\big\}_\star  \label{ABB}\\
&+\frac{iecs}{4}\big\{\partial_\mu B_\nu-\partial_\nu B_\mu,[B^\mu,B^\nu]_\star\big\}_\star  \label{BBB}\\
&+\frac{e^2}{4}(c^4+s^4)[A_\mu,A_\nu]_\star\star[A^\mu,A^\nu]_\star \label{AAAA}\\
&+\frac{e^2cs}{4}(c^4-s^4)\big\{[A_\mu,A_\nu]_\star,[A^\mu,B^\nu]_\star+[B^\mu,A^\nu]_\star \big\}_\star \label{AAAB}\\
&+\frac{e^2c^2s^2}{4}\big([A_\mu,B_\nu]_\star+[B_\mu,A_\nu]_\star\big)\star\big([A^\mu,B^\nu]_\star+[B^\mu,A^\nu]\big) \label{AABB}\\
&+\frac{e^2c^2s^2}{4}\big\{[A_\mu,A_\nu]_\star,[B^\mu,B^\nu]_\star \big\}_\star   \label{BBBB}
\end{align}
We encounter different three- and four-point vertices, let us look at the one in line (\ref{psi}) between two spinor particles and one $A_\mu$ first:
\begin{align}
&ec^2\bar{\psi}\star\gamma^\mu A_\mu\star\psi(x) +es^2\bar{\psi}\star\gamma^\mu \psi\star A_\mu(x)\cr =&e\int\frac{dk_1}{(2\pi)^4}\frac{dk_2}{(2\pi)^4}\frac{dk_3}{(2\pi)^4}\tilde{\bar{\psi}}(k_1)\gamma^\mu \tilde{A}_\mu(k_2)\tilde{\psi}(k_3) \cr
&e^{-i(k_1+k_2+k_3)x}\underline{\big(c^2e^{-i(k_1,k_2,k_3)}+s^2e^{-i(k_1,k_3,k_2)}\big)}
\end{align}
where we introduced $(k_1,...,k_n)=\sum_{i<j}k_i\wedge k_j$. The underlined term is, if the momenta are put on-shell, the noncommutative phase factor. Denoting the $A_\mu$-line by
\begin{picture}(50,10)(0,2)
\Photon(5,5)(45,5){2}{4}
\end{picture}
and the $B_\mu$-line by
\begin{picture}(50,10)(0,2)
\DashPhoton(5,5)(45,5){2.2}{4}{3}
\end{picture}
we have found for the above vertex
\begin{align}
\begin{picture}(90,40)(0,37)
\Vertex(30,40){1}
\ArrowLine(30,40)(0,75)
\ArrowLine(0,5)(30,40)
\Photon(30,40)(75,40){2}{4}
\Text(82,40)[]{$\scriptstyle \mu$}
\Text(2,28)[]{$\scriptstyle k_3,\lambda_3$}
\Text(28,63)[]{$\scriptstyle k_1,\lambda_1$}
\LongArrow(60,33)(45,33)
\Text(52,25)[]{$\scriptstyle k_2,\lambda_2$}
\end{picture}
=~&ie\gamma^\mu \delta^{(4)}(k_1-k_2-k_3)\cr
&e^{-i(k_2^{\lambda_2}\wedge k_1^{\lambda_1}+k_3^{\lambda_3}\wedge k_1^{\lambda_1})}\big(c^2e^{-ik_2^{\lambda_2}\wedge k_3^{\lambda_3}}+s^2e^{-ik_3^{\lambda_3}\wedge k_2^{\lambda_2}}\big)\quad .\nonumber
\end{align}
The other vertices may be found similarly:
\begin{align}
\begin{picture}(90,40)(0,37)
\Vertex(30,40){1}
\ArrowLine(30,40)(0,75)
\ArrowLine(0,5)(30,40)
\DashPhoton(30,40)(75,40){2.2}{4}{3}
\Text(82,40)[]{$\scriptstyle \mu$}
\Text(2,28)[]{$\scriptstyle k_3,\lambda_3$}
\Text(28,63)[]{$\scriptstyle k_1,\lambda_1$}
\LongArrow(60,33)(45,33)
\Text(52,25)[]{$\scriptstyle k_2,\lambda_2$}
\end{picture}
=~&2ecs\gamma^\mu \delta^{(4)}(k_1-k_2-k_3)\cr
&e^{-i(k_2^{\lambda_2}\wedge k_1^{\lambda_1}+k_3^{\lambda_3}\wedge k_1^{\lambda_1})}\sin (k_2^{\lambda_2}\wedge k_3^{\lambda_3})
\cr
\cr
\begin{picture}(90,80)(0,37)
\Vertex(30,40){1}
\Photon(30,40)(0,75){2}{4}
\Photon(0,5)(30,40){2}{4}
\Photon(30,40)(75,40){2}{4}
\Text(-5,80)[]{$\scriptstyle\mu_1$}
\Text(82,40)[]{$\scriptstyle\mu_2$}
\Text(-5,0)[]{$\scriptstyle\mu_3$}
\LongArrow(4,21)(13,31)
\Text(-4,30)[]{$\scriptstyle k_3,\lambda_3$}
\LongArrow(15,67)(24,57)
\Text(32,66)[]{$\scriptstyle k_1,\lambda_1$}
\LongArrow(62,33)(47,33)
\Text(52,25)[]{$\scriptstyle k_2,\lambda_2$}
\end{picture}
=~&2e(c^4-s^4)\delta^{(4)}(k_1+k_2+k_3) \cr
&\Big\{(k_1^{\mu_2}g^{\mu_1\mu_3}-k_1^{\mu_3}g^{\mu_1\mu_2})\cos(k_1^{\lambda_1}\wedge k_2^{\lambda_2}+k_1^{\lambda_1}\wedge k_3^{\lambda_3}) \cr
& \quad \sin(k_2^{\lambda_2}\wedge k_3^{\lambda_3}) \cr
&+(k_2^{\mu_3}g^{\mu_2\mu_1}-k_2^{\mu_1}g^{\mu_2\mu_3})\cos(k_2^{\lambda_2}\wedge k_3^{\lambda_3}+k_2^{\lambda_2}\wedge k_1^{\lambda_1}) \cr
& \quad \sin(k_3^{\lambda_3}\wedge k_1^{\lambda_1}) \cr
&+(k_3^{\mu_1}g^{\mu_3\mu_2}-k_3^{\mu_2}g^{\mu_3\mu_1})\cos(k_3^{\lambda_3}\wedge k_1^{\lambda_1}+k_3^{\lambda_3}\wedge k_2^{\lambda_2}) \cr
& \quad \sin(k_1^{\lambda_1}\wedge k_2^{\lambda_2})\Big\}
\cr
\begin{picture}(90,80)(0,37)
\Vertex(30,40){1}
\Photon(30,40)(0,75){2}{4}
\Photon(0,5)(30,40){2}{4}
\DashPhoton(30,40)(75,40){2.2}{4}{3}
\Text(-5,80)[]{$\scriptstyle\mu_1$}
\Text(82,40)[]{$\scriptstyle\mu_2$}
\Text(-5,0)[]{$\scriptstyle\mu_3$}
\LongArrow(4,21)(13,31)
\Text(-4,30)[]{$\scriptstyle k_3,\lambda_3$}
\LongArrow(15,67)(24,57)
\Text(32,66)[]{$\scriptstyle k_1,\lambda_1$}
\LongArrow(62,33)(47,33)
\Text(52,25)[]{$\scriptstyle k_2,\lambda_2$}
\end{picture}
=~&2ecs\delta^{(4)}(k_1+k_2+k_3)\cr
&\Big\{(k_1^{\mu_2}g^{\mu_1\mu_3}-k_1^{\mu_3}g^{\mu_1\mu_2})\cos(k_1^{\lambda_1}\wedge k_2^{\lambda_2}+k_1^{\lambda_1}\wedge k_3^{\lambda_3}) \cr
& \quad \sin(k_2^{\lambda_2}\wedge k_3^{\lambda_3}) \cr
&+(k_2^{\mu_3}g^{\mu_2\mu_1}-k_2^{\mu_1}g^{\mu_2\mu_3})\cos(k_2^{\lambda_2}\wedge k_3^{\lambda_3}+k_2^{\lambda_2}\wedge k_1^{\lambda_1}) \cr
& \quad \sin(k_3^{\lambda_3}\wedge k_1^{\lambda_1}) \cr
&+(k_3^{\mu_1}g^{\mu_3\mu_2}-k_3^{\mu_2}g^{\mu_3\mu_1})\cos(k_3^{\lambda_3}\wedge k_1^{\lambda_1}+k_3^{\lambda_3}\wedge k_2^{\lambda_2}) \cr
& \quad \sin(k_1^{\lambda_1}\wedge k_2^{\lambda_2})\Big\}
\nonumber
\end{align}
\vspace{4cm}
\begin{align}
\begin{picture}(90,80)(0,37)
\Vertex(30,40){1}
\DashPhoton(30,40)(0,75){2.2}{4}{3}
\DashPhoton(0,5)(30,40){2.2}{4}{3}
\DashPhoton(30,40)(75,40){2.2}{4}{3}
\Text(-5,80)[]{$\scriptstyle\mu_1$}
\Text(82,40)[]{$\scriptstyle\mu_2$}
\Text(-5,0)[]{$\scriptstyle\mu_3$}
\LongArrow(4,21)(13,31)
\Text(-4,30)[]{$\scriptstyle k_3,\lambda_3$}
\LongArrow(15,67)(24,57)
\Text(32,66)[]{$\scriptstyle k_1,\lambda_1$}
\LongArrow(62,33)(47,33)
\Text(52,25)[]{$\scriptstyle k_2,\lambda_2$}
\end{picture}
=~&2ecs\delta^{(4)}(k_1+k_2+k_3)\cr
&\Big\{(k_1^{\mu_2}g^{\mu_1\mu_3}-k_1^{\mu_3}g^{\mu_1\mu_2})\cos(k_1^{\lambda_1}\wedge k_2^{\lambda_2}+k_1^{\lambda_1}\wedge k_3^{\lambda_3}) \cr
& \quad \sin(k_2^{\lambda_2}\wedge k_3^{\lambda_3}) \cr
&+(k_2^{\mu_3}g^{\mu_2\mu_1}-k_2^{\mu_1}g^{\mu_2\mu_3})\cos(k_2^{\lambda_2}\wedge k_3^{\lambda_3}+k_2^{\lambda_2}\wedge k_1^{\lambda_1}) \cr
& \quad \sin(k_3^{\lambda_3}\wedge k_1^{\lambda_1}) \cr
&+(k_3^{\mu_1}g^{\mu_3\mu_2}-k_3^{\mu_2}g^{\mu_3\mu_1})\cos(k_3^{\lambda_3}\wedge k_1^{\lambda_1}+k_3^{\lambda_3}\wedge k_2^{\lambda_2}) \cr
& \quad \sin(k_1^{\lambda_1}\wedge k_2^{\lambda_2})\Big\}  \cr
\begin{picture}(90,90)(0,37)
\Vertex(40,40){1}
\Photon(0,80)(80,0){2}{10}
\Photon(0,0)(80,80){2}{10}
\Text(-5,85)[]{$\scriptstyle\mu_1$}
\Text(85,85)[]{$\scriptstyle\mu_2$}
\Text(85,-5)[]{$\scriptstyle\mu_3$}
\Text(-5,-5)[]{$\scriptstyle\mu_4$}
\LongArrow(10,20)(20,30)
\Text(3,30)[]{$\scriptstyle k_4,\lambda_4$}
\LongArrow(20,70)(30,60)
\Text(38,70)[]{$\scriptstyle k_1,\lambda_1$}
\LongArrow(70,60)(60,50)
\Text(80,52)[]{$\scriptstyle k_2,\lambda_2$}
\LongArrow(60,10)(50,20)
\Text(44,10)[]{$\scriptstyle k_3,\lambda_3$}
\end{picture}
=~&e^2(c^4+s^4)\delta^{(4)}(k_1+k_2+k_3+k_4)\sum_{\pi\epsilon\mathcal{S}_4}g^{\mu_{\pi(1)}\mu_{\pi(3)}}g^{\mu_{\pi(2)}\mu_{\pi(4)}}\cr
&e^{-i\big(k_{\pi(1)}^{\lambda_{\pi(1)}}\wedge k_{\pi(2)}^{\lambda_{\pi(2)}}+ k_{\pi(1)}^{\lambda_{\pi(1)}}\wedge k_{\pi(3)}^{\lambda_{\pi(3)}}+ k_{\pi(1)}^{\lambda_{\pi(1)}}\wedge k_{\pi(4)}^{\lambda_{\pi(4)}}+k_{\pi(2)}^{\lambda_{\pi(2)}}\wedge k_{\pi(3)}^{\lambda_{\pi(3)}}+ k_{\pi(2)}^{\lambda_{\pi(2)}}\wedge k_{\pi(4)}^{\lambda_{\pi(4)}}\big)}\cr
&\sin\big(k_{\pi(3)}^{\lambda_{\pi(3)}}\wedge k_{\pi(4)}^{\lambda_{\pi(4)}}\big)
\cr
\begin{picture}(90,90)(0,37)
\Vertex(40,40){1}
\Photon(0,80)(80,0){2}{10}
\Photon(40,40)(80,80){2}{5}
\DashPhoton(0,0)(40,40){2.2}{5}{3}
\Text(-5,85)[]{$\scriptstyle\mu_1$}
\Text(85,85)[]{$\scriptstyle\mu_2$}
\Text(85,-5)[]{$\scriptstyle\mu_3$}
\Text(-5,-5)[]{$\scriptstyle\mu_4$}
\LongArrow(10,20)(20,30)
\Text(3,30)[]{$\scriptstyle k_4,\lambda_4$}
\LongArrow(20,70)(30,60)
\Text(38,70)[]{$\scriptstyle k_1,\lambda_1$}
\LongArrow(70,60)(60,50)
\Text(80,52)[]{$\scriptstyle k_2,\lambda_2$}
\LongArrow(60,10)(50,20)
\Text(44,10)[]{$\scriptstyle k_3,\lambda_3$}
\end{picture}
=~&2ie^2cs(c^4-s^4)\delta^{(4)}(k_1+k_2+k_3+k_4) \cr
&\Big\{g^{\mu_1\mu_3}g^{\mu_2\mu_4}\sin(k_3^{\lambda_3}\wedge k_4^{\lambda_4}) \cr
&\quad \sin(k_1^{\lambda_1}\wedge k_2^{\lambda_2}+k_1^{\lambda_1}\wedge k_3^{\lambda_3}+ k_1^{\lambda_1}\wedge k_4^{\lambda_4}+k_2^{\lambda_2}\wedge k_3^{\lambda_3}+ k_2^{\lambda_2}\wedge k_4^{\lambda_4})\cr
&+g^{\mu_1\mu_2}g^{\mu_3\mu_4}\sin(k_1^{\lambda_1}\wedge k_4^{\lambda_4}) \cr
&\quad \sin(k_2^{\lambda_2}\wedge k_3^{\lambda_3}+k_2^{\lambda_2}\wedge k_1^{\lambda_1}+ k_2^{\lambda_2}\wedge k_4^{\lambda_4}+k_3^{\lambda_3}\wedge k_1^{\lambda_1}+ k_3^{\lambda_3}\wedge k_4^{\lambda_4})\cr
&+g^{\mu_2\mu_3}g^{\mu_1\mu_4}\sin(k_2^{\lambda_2}\wedge k_4^{\lambda_4}) \cr
&\quad \sin(k_3^{\lambda_3}\wedge k_1^{\lambda_1}+k_3^{\lambda_3}\wedge k_2^{\lambda_2}+ k_3^{\lambda_3}\wedge k_4^{\lambda_4}+k_1^{\lambda_1}\wedge k_2^{\lambda_2}+ k_1^{\lambda_1}\wedge k_4^{\lambda_4})\Big\}
\nonumber
\end{align}
\begin{align}
\begin{picture}(90,90)(0,37)
\Vertex(40,40){1}
\Photon(0,80)(40,40){2}{5}
\DashPhoton(40,40)(80,0){2.2}{5}{3}
\DashPhoton(40,40)(80,80){2.2}{5}{3}
\Photon(0,0)(40,40){2}{5}
\Text(-5,85)[]{$\scriptstyle\mu_1$}
\Text(85,85)[]{$\scriptstyle\mu_2$}
\Text(85,-5)[]{$\scriptstyle\mu_3$}
\Text(-5,-5)[]{$\scriptstyle\mu_4$}
\LongArrow(10,20)(20,30)
\Text(3,30)[]{$\scriptstyle k_4,\lambda_4$}
\LongArrow(20,70)(30,60)
\Text(38,70)[]{$\scriptstyle k_1,\lambda_1$}
\LongArrow(70,60)(60,50)
\Text(80,52)[]{$\scriptstyle k_2,\lambda_2$}
\LongArrow(60,10)(50,20)
\Text(44,10)[]{$\scriptstyle k_3,\lambda_3$}
\end{picture}
=&-4ie^2c^2s^2\delta^{(4)}(k_1+k_2+k_3+k_4) \cr
&\Big\{(g^{\mu_1\mu_2}g^{\mu_3\mu_4}-g^{\mu_1\mu_3}g^{\mu_2\mu_4})\sin(k_1^{\lambda_1}\wedge k_3^{\lambda_3}) \sin(k_2^{\lambda_2}\wedge k_4^{\lambda_4}) \cr
&\qquad \cos(k_1^{\lambda_1}\wedge k_2^{\lambda_2}+k_1^{\lambda_1}\wedge k_4^{\lambda_4}+k_3^{\lambda_3}\wedge k_2^{\lambda_2}+k_3^{\lambda_3}\wedge k_4^{\lambda_4})\cr
&+(g^{\mu_1\mu_4}g^{\mu_2\mu_3}-g^{\mu_1\mu_2}g^{\mu_3\mu_4})\sin(k_2^{\lambda_2}\wedge k_3^{\lambda_3}) \sin(k_4^{\lambda_4}\wedge k_1^{\lambda_1}) \cr
&\qquad \cos(k_1^{\lambda_1}\wedge k_2^{\lambda_2}+k_1^{\lambda_1}\wedge k_3^{\lambda_3}+k_4^{\lambda_4}\wedge k_2^{\lambda_2}+k_4^{\lambda_4}\wedge k_3^{\lambda_3})\cr
&+(g^{\mu_1\mu_3}g^{\mu_2\mu_4}-g^{\mu_1\mu_4}g^{\mu_2\mu_3})\sin(k_1^{\lambda_1}\wedge k_2^{\lambda_2}) \sin(k_3^{\lambda_3}\wedge k_4^{\lambda_4}) \cr
&\qquad \cos(k_1^{\lambda_1}\wedge k_3^{\lambda_3}+k_1^{\lambda_1}\wedge k_4^{\lambda_4}+k_2^{\lambda_2}\wedge k_3^{\lambda_3}+k_2^{\lambda_2}\wedge k_4^{\lambda_4}) \Big\}\cr
\nonumber
\end{align}
The lines are, as already mentioned above
\begin{align}
\begin{picture}(65,10)(0,2)
\ArrowLine(0,5)(50,5)
\Text(25,-4)[]{$\scriptstyle k,\lambda$}
\end{picture}
&=S_\lambda(k)=\frac{i\lambda(\not\!k^\lambda+m)}{2E_k(k^0-\lambda(E_k-i\epsilon))} \cr
\begin{picture}(65,10)(0,2)
\Photon(0,5)(50,5){2}{4}
\LongArrow(15,-2)(35,-2)
\Text(-7,5)[]{$\scriptstyle\mu$}
\Text(57,5)[]{$\scriptstyle\nu$}
\Text(25,-10)[]{$\scriptstyle k,\lambda$}
\end{picture}
&=-g^{\mu\nu}P_\lambda(k)=-g^{\mu\nu}\frac{i\lambda}{2E_k(k^0-\lambda(E_k-i\epsilon))}
\cr
\begin{picture}(65,10)(0,2)
\DashPhoton(0,5)(50,5){2.2}{4}{3}
\LongArrow(15,-2)(35,-2)
\Text(-7,5)[]{$\scriptstyle\mu$}
\Text(57,5)[]{$\scriptstyle\nu$}
\Text(25,-10)[]{$\scriptstyle k,\lambda$}
\end{picture}
&=-g^{\mu\nu}P_\lambda(k)=-g^{\mu\nu}\frac{i\lambda}{2E_k(k^0-\lambda(E_k-i\epsilon))}
\quad .
\end{align}
To calculate a diagram contributing to a Green's function of type (\ref{Green1}), we associate these expressions to the lines and vertices, integrate over momenta and sum over the indices $\lambda_i=\pm 1$. S-matrix elements may be obtained via the LSZ-reduction formula, the amputation gives the standard outcome for on-shell external momenta:
\begin{align}
\sum_{\lambda=\pm1}(-i)(k^2-m^2)P_\lambda(k)\Big|_{k^{(0)}=E_k}
&=\sum_{\lambda=\pm1}\frac{(k^2-m^2)\lambda}{2E_k(k^{(0)}-\lambda(E_k-i\epsilon)}\Big|_{k^{(0)}=E_k}\cr
&=\sum_{\lambda=\pm1}\frac{(k^2-m^2)\lambda(k^{(0)}+\lambda E_k)}{2E_k(k^{(0)2}-E_k^2)}\Big|_{k^{(0)}=E_k}\cr
&=\sum_{\lambda=\pm1}\frac{\lambda k^{(0)}+E_k}{2E_k}\Big|_{k^{(0)}=E_k}\cr
&=1
\end{align}
\begin{align}
\sum_{\lambda=\pm1}(-i)(\not\!k-m)S_\lambda(k)\Big|_{k^{(0)}=E_k}
&=\sum_{\lambda=\pm1}\frac{(\not\!k-m)\lambda(\not\!k^{\lambda}+m)}{2E_k(k^{(0)}-\lambda(E_k-i\epsilon)}\Big|_{k^{(0)}=E_k} \cr
&=\sum_{\lambda=\pm1}\frac{(k^2-m^2)\lambda(k^{(0)}+\lambda E_k)}{2E_k(k^{(0)2}-E_k^2)}\Big|_{k^{(0)}=E_k} \cr
& \quad +\sum_{\lambda=\pm1}\underbrace{\frac{(\not\!k-m)\lambda\gamma^0(\lambda E_k-k^{(0)})}{2E_k(k^{(0)}-\lambda(E_k-i\epsilon)}\Big|_{k^{(0)}=E_k}}_{=0\footnotemark[1]} \cr
&=1 \quad .
\end{align}
\footnotetext[1]{Because of $\bar{u}(k,s)(\not\!k-m)=(\not\!k-m)u(k,s)=0$ this is true if sandwiched between spinor polarisation vectors $\bar{u}(k,s), u(k,s)$ as it is the case in the reduction formula.}
This means that for S-matrix elements we associate the ordinary polarisation vectors and on-shell momenta to external lines, and the above derived expressions to inner lines and vertices, internal momenta are to be integrated. We will apply this procedure in the following example.

\subsection{Compton scattering}

For the scattering of a photon by an electron, $e^- \gamma \rightarrow e^-\gamma$, we calculate the S-matrix at lowest nonvanishing order. Due to the richer structure of interaction terms in the double gauged noncommutative theory we have a larger number of contributing diagrams, the s-, u- and t-channel are

\begin{align}
\begin{picture}(110,40)(0,38)
\Vertex(30,40){1}
\Vertex(70,40){1}
\Photon(10,75)(30,40){2}{4}
\Photon(70,40)(90,75){2}{4}
\ArrowLine(10,5)(30,40)
\ArrowLine(70,40)(90,5)
\ArrowLine(30,40)(70,40)
\LongArrow(22,67)(29,55)
\LongArrow(82,49)(89,61)
\Text(10,82)[]{$\scriptstyle k_1$}
\Text(90,82)[]{$\scriptstyle k_2$}
\Text(10,-2)[]{$\scriptstyle p_1$}
\Text(90,-2)[]{$\scriptstyle p_2$}
\Text(50,48)[]{$\scriptstyle q,\lambda$}
\end{picture}
&=i\mathcal{M}^s  \qquad\qquad
\begin{picture}(110,40)(0,38)
\Vertex(30,40){1}
\Vertex(70,40){1}
\Photon(10,75)(30,40){2}{4}
\Photon(70,40)(90,75){2}{4}
\ArrowLine(10,5)(70,40)
\ArrowLine(30,40)(90,5)
\ArrowLine(70,40)(30,40)
\LongArrow(22,67)(29,55)
\LongArrow(82,49)(89,61)
\Text(10,82)[]{$\scriptstyle k_1$}
\Text(90,82)[]{$\scriptstyle k_2$}
\Text(10,-2)[]{$\scriptstyle p_1$}
\Text(90,-2)[]{$\scriptstyle p_2$}
\Text(50,48)[]{$\scriptstyle q,\lambda$}
\end{picture}
=i\mathcal{M}^u  \cr\cr
\begin{picture}(100,100)(0,38)
\Vertex(40,30){1}
\Vertex(40,70){1}
\ArrowLine(40,30)(75,10)
\Photon(40,70)(75,90){2}{4}
\ArrowLine(5,10)(40,30)
\Photon(40,70)(5,90){2}{4}
\Photon(40,30)(40,70){2}{4}
\LongArrow(47,58)(47,42)
\LongArrow(19,89)(31,82)
\LongArrow(49,82)(61,89)
\Text(82,7)[]{$\scriptstyle p_2$}
\Text(82,93)[]{$\scriptstyle k_2$}
\Text(-2,7)[]{$\scriptstyle p_1$}
\Text(-2,93)[]{$\scriptstyle k_1$}
\Text(57,50)[]{$\scriptstyle q,\lambda$}
\end{picture}
&=i\mathcal{M}^{t(A)}\qquad \qquad
\begin{picture}(100,100)(0,38)
\Vertex(40,30){1}
\Vertex(40,70){1}
\ArrowLine(40,30)(75,10)
\Photon(40,70)(75,90){2}{4}
\ArrowLine(5,10)(40,30)
\Photon(40,70)(5,90){2}{4}
\DashPhoton(40,30)(40,70){2.2}{4}{3}
\LongArrow(47,58)(47,42)
\LongArrow(19,89)(31,82)
\LongArrow(49,82)(61,89)
\Text(82,7)[]{$\scriptstyle p_2$}
\Text(82,93)[]{$\scriptstyle k_2$}
\Text(-2,7)[]{$\scriptstyle p_1$}
\Text(-2,93)[]{$\scriptstyle k_1$}
\Text(57,50)[]{$\scriptstyle q,\lambda$}
\end{picture}
=i\mathcal{M}^{t(B)} \quad .
\end{align}
\\
\\
The corresponding analytic expressions are obtained by straightforward application of our Feynman rules from the previous subsection:
\begin{align}
i\mathcal{M}^s=-e^2\sum_{\lambda=\pm1}&\epsilon^\mu(k_1)\epsilon^{\nu\star}(k_2)\bar{u}(p_2,s_2)\gamma^\nu S_\lambda(q)\gamma^\mu u(p_1,s_1)
e^{-i(q^\lambda\wedge p_2+p_2\wedge k_2+p_1\wedge q^\lambda+k_1\wedge q^\lambda)} \cr
&\big(c^2e^{-iq^\lambda\wedge k_2}+s^2e^{-ik_2\wedge q^\lambda}\big)\big(c^2e^{-ik_1\wedge p_1}+s^2e^{-ip_1\wedge k_1}\big)\Big|_{q=p_1+k_1=p_2+k_2}
\end{align}
where the first phase factor can be further simplified if we remember total momentum conservation $p_1+k_1=p_2+k_2$:
\begin{align}
&q^\lambda\wedge p_2+p_2\wedge k_2+p_1\wedge q^\lambda+k_1\wedge q^\lambda\cr
=&(p_1+k_1-p_2)\wedge q^\lambda+p_2\wedge k_2 \cr
=& k_2\wedge q^\lambda+p_2\wedge k_2
\nonumber
\end{align}
such that we have for the first diagram
\begin{align}
i\mathcal{M}^s=-e^2\sum_{\lambda=\pm1}&\epsilon^\mu(k_1)\epsilon^{\nu\star}(k_2)\bar{u}(p_2,s_2)\gamma^\nu S_\lambda(q)\gamma^\mu u(p_1,s_1)
e^{-i(k_2\wedge q^\lambda+p_2\wedge k_2)} \cr
&\big(c^2e^{-iq^\lambda\wedge k_2}+s^2e^{-ik_2\wedge q^\lambda}\big)\big(c^2e^{-ik_1\wedge p_1}+s^2e^{-ip_1\wedge k_1}\big)\Big|_{q=p_1+k_1=p_2+k_2}
\end{align}
The second looks similar
\begin{align}
i\mathcal{M}^u=-e^2\sum_{\lambda=\pm1}&\epsilon^\mu(k_1)\epsilon^{\nu\star}(k_2)\bar{u}(p_2,s_2)\gamma^\nu S_\lambda(q)\gamma^\mu u(p_1,s_1)
e^{-i(q^\lambda \wedge p_2+k_1\wedge p_2+p_1\wedge q^\lambda +q^\lambda\wedge k_2)} \cr
&\big(c^2e^{-ip_1\wedge k_2}+s^2e^{-ik_2\wedge p_1}\big)\big(c^2e^{-ik_1\wedge q^\lambda}+s^2e^{-iq^\lambda\wedge k_1}\big)\Big|_{q=p_1-k_2=p_2-k_1}\cr
=-e^2\sum_{\lambda=\pm1}&\epsilon^\mu(k_1)\epsilon^{\nu\star}(k_2)\bar{u}(p_2,s_2)\gamma^\nu S_\lambda(q)\gamma^\mu u(p_1,s_1)
e^{-i(q^\lambda \wedge k_1+k_1\wedge p_2)} \cr
&\big(c^2e^{-ip_1\wedge k_2}+s^2e^{-ik_2\wedge p_1}\big)\big(c^2e^{-ik_1\wedge q^\lambda}+s^2e^{-iq^\lambda\wedge k_1}\big)\Big|_{q=p_1-k_2=p_2-k_1}
\end{align}
and the remaining ones are
\begin{align}
i\mathcal{M}^{t(A)}=2ie^2(c^4-s^4)\sum_{\lambda=\pm1}&\epsilon_\mu(k_1)\epsilon_\nu^\star(k_2)\bar{u}(p_2,s_2)\gamma_\rho u(p_1,s_1)P_\lambda(q) \cr
&\Big\{(g^{\mu\rho}k_1^\nu-g^{\mu\nu}k_1^\rho)\sin(k_2\wedge q^\lambda)\cos(k_2\wedge k_1+q^\lambda\wedge k_1) \cr
&+(g^{\mu\nu}k_2^\rho-g^{\nu\rho}k_2^\mu)\sin(k_1\wedge q^\lambda)\cos(k_1\wedge k_2+k_2\wedge q^\lambda) \cr
&+(g^{\nu\rho}q^\mu-g^{\mu\rho}q^\nu)\sin(k_2\wedge k_1)\cos(q^\lambda \wedge k_2+k_1\wedge q^\lambda)\Big\} \cr
&\big(c^2e^{-iq^\lambda\wedge p_1}+s^2e^{-ip_1\wedge q^\lambda}\big)e^{-i(p_1\wedge p_2+q^\lambda \wedge p_2)}\Big|_{q=p_2-p_1=k_1-k_2}\cr
\end{align}
\begin{align}
i\mathcal{M}^{t(B)}=4e^2c^2s^2\sum_{\lambda=\pm1}&\epsilon_\mu(k_1)\epsilon_\nu^\star(k_2)\bar{u}(p_2,s_2) \gamma_\rho u(p_1,s_1)P_\lambda(q) \cr
&\Big\{(g^{\mu\rho}k_1^\nu-g^{\mu\nu}k_1^\rho)\sin(k_2\wedge q^\lambda)\cos(k_2\wedge k_1+q^\lambda\wedge k_1) \cr
&+(g^{\mu\nu}k_2^\rho-g^{\nu\rho}k_2^\mu)\sin(k_1\wedge q^\lambda)\cos(k_1\wedge k_2+k_2\wedge q^\lambda) \cr
&+(g^{\nu\rho}q^\mu-g^{\mu\rho}q^\nu)\sin(k_2\wedge k_1)\cos(q^\lambda \wedge k_2+k_1\wedge q^\lambda)
\Big\} \cr
&\sin(q^\lambda\wedge p_1)e^{-i(p_1\wedge p_2+q^\lambda \wedge p_2)}\Big|_{q=p_2-p_1=k_1-k_2}\quad .
\end{align}
The last two contributions can be combined to give the complete t-channel, the result is simple if we note that we only need to add
\begin{align}
&\underbrace{(c^4-s^4)\big(c^2e^{-iq^\lambda\wedge p_1}+s^2e^{-ip_1\wedge q^\lambda}\big)}_{\text{from} ~i\mathcal{M}^{t(A)}}+\underbrace{\frac{1}{i}2c^2s^2\sin(q^\lambda\wedge p)}_{\text{from}~ i\mathcal{M}^{t(B)}}\cr
=~&c^2(c^4-s^4+s^2)e^{-iq^\lambda\wedge p_1}+s^2(c^4-s^4-c^2)e^{-ip_1\wedge q^\lambda}\cr
=~&c^2(c^4-(1-c^2)^2+1-c^2)e^{-iq^\lambda\wedge p_1}+s^2((1-s^2)^2-s^4-1+s^2)e^{-ip_1\wedge q^\lambda}\cr
=~&c^4e^{-iq^\lambda\wedge p_1}-s^4e^{-ip_1\wedge q^\lambda}
\end{align}
and so
\begin{align}
i\mathcal{M}^{t}=2ie^2\sum_{\lambda=\pm1}&\epsilon_\mu(k_1)\epsilon_\nu^\star(k_2)\bar{u}(p_2,s_2)\gamma_\rho u(p_1,s_1)P_\lambda(q) \cr
&\Big\{(g^{\mu\rho}k_1^\nu-g^{\mu\nu}k_1^\rho)\sin(k_2\wedge q^\lambda)\cos(k_2\wedge k_1+q^\lambda\wedge k_1) \cr
&+(g^{\mu\nu}k_2^\rho-g^{\nu\rho}k_2^\mu)\sin(k_1\wedge q^\lambda)\cos(k_1\wedge k_2+k_2\wedge q^\lambda) \cr
&+(g^{\nu\rho}q^\mu-g^{\mu\rho}q^\nu)\sin(k_2\wedge k_1)\cos(q^\lambda \wedge k_2+k_1\wedge q^\lambda)
\Big\} \cr
&\big(c^4e^{-iq^\lambda\wedge p_1}-s^4e^{-ip_1\wedge q^\lambda}\big)e^{-i(p_1\wedge p_2+q^\lambda \wedge p_2)}\Big|_{q=p_2-p_1=k_1-k_2}\quad .
\end{align}

\subsection{The Ward identity in the case of only spatial noncommutativity}

The Ward identity for the above scattering process is the vanishing of the amplitude if one of the photon polarization vectors, say $\epsilon_\mu(k_1)$, is longitudinal: $\epsilon_\mu(k_1)=k_{1\mu}$, and the other one physical: $k_2^\mu\epsilon_\mu(k_2)=0$. This is a fundamental requirement, as an unphysical photon, like a longitudinal, may not be scattered into a physical one.\\

Our derivation will be inspired by the proof of the Ward-identity for  $e^+e^-\rightarrow \gamma\gamma$ in NCQED as done in \cite{Mariz}.

In the case of $\theta^{0i}=0$ the above amplitude simplifies significantly, as we may now omit the requirement that we have on-shell momenta in the phase factor, i.e. omit the index $^\lambda$ in the phases. This allows us to exploit momentum conservation for simplification of the noncommutative phase factors and to carry out the summation over the index $\lambda$, yielding propagators. \\

We finally find
\begin{align}
i\mathcal{M}^s
=&-e^2\epsilon^\mu(k_1)\epsilon^{\nu\star}(k_2)\bar{u}(p_2,s_2)\gamma^\nu \tilde{S}_C(q)\gamma^\mu u(p_1,s_1)\cr
&\quad \big(c^2e^{-ip_2\wedge k_2}+s^2e^{-ik_2\wedge p_2}\big)\big(c^2e^{-ik_1\wedge p_1}+s^2e^{-ip_1\wedge k_1}\big)\Big|_{q=p_1+k_1=p_2+k_2}\cr
i\mathcal{M}^u
=&-e^2\epsilon^\mu(k_1)\epsilon^{\nu\star}(k_2)\bar{u}(p_2,s_2)\gamma^\nu \tilde{S}_C(q)\gamma^\mu u(p_1,s_1)\cr
&\quad \big(c^2e^{-ip_1\wedge k_2}+s^2e^{-ik_2\wedge p_1}\big)\big(c^2e^{-ik_1\wedge p_2}+s^2e^{-ip_2\wedge k_1}\big)\Big|_{q=p_1-k_2=p_2-k_1}\cr
i\mathcal{M}^{t}
=&2ie^2\epsilon_\mu(k_1)\epsilon_\nu^\star(k_2)\bar{u}(p_2,s_2)\gamma_\rho u(p_1,s_1)\Delta_C(q) \cr
&\quad \Big\{g^{\mu\nu}(k_1^\rho-k_2^\rho)+g^{\nu\rho}(k_2^\mu-q^\mu)+g^{\mu\rho}(q^\nu-k_1^\nu)\Big\}\cr
&\quad \sin(k_1\wedge k_2)
\big(c^4e^{-ip_2\wedge p_1}-s^4e^{-ip_1\wedge p_2}\big)\Big|_{q=p_2-p_1=k_1-k_2} \quad .
\end{align}
To check the Ward identity, we define the amplitude $i\mathcal{M}_\mu$ as the total amplitude with the polarisation vector $\epsilon^\mu(k_1)$ taken out, i.e. $i\mathcal{M}=i\mathcal{M}^s+ i\mathcal{M}^u+ i\mathcal{M}^t=\epsilon^\mu(k_1) i\mathcal{M}_\mu$. The Ward identity reads in these terms
\beq
ik_1^\mu \mathcal{M}_\mu=0      \quad .
\eeq
Replacing the polarisation vector $\epsilon^\mu(k_1)$ by $k_1^\mu$ in $i\mathcal{M}^s$ and $i\mathcal{M}^u$, we obtain the expressions $\bar{u}(p_2,s_2)\gamma^\nu \tilde{S}_C(p_1+k_1)\gamma^\mu k_1^\mu u(p_1,s_1)$ and $\bar{u}(p_2,s_2)\gamma^\nu \tilde{S}_C(p_1-k_2)\gamma^\mu k_1^\mu u(p_1,s_1)$, which after some algebraic calculation and with the help of the identity $(\not\!p_1-m)u(p_1,s_1)=0$ as well as
$ \bar{u}(p_2,s_2)(\not\!p_2-m)=0$ simplify to
\begin{align}
\bar{u}(p_2,s_2)\gamma^\nu \tilde{S}_C(p_1+k_1)\gamma^\mu k_1^\mu u(p_1,s_1)&=i\bar{u}(p_2,s_2)\gamma^\nu u(p_1,s_1) \cr
\bar{u}(p_2,s_2)\gamma^\nu \tilde{S}_C(p_1-k_2)\gamma^\mu k_1^\mu u(p_1,s_1)&=-i\bar{u}(p_2,s_2)\gamma^\nu u(p_1,s_1) \quad .\nonumber
\end{align}
The amplitude $i\mathcal{M}^t$ contains the expression
\begin{align}
k_{1\mu}\epsilon_\nu^\star(k_2)\gamma_\rho\Big\{g^{\mu\nu}(k_1^\rho-k_2^\rho)+g^{\nu\rho}(2k_2^\mu-k_1^\mu)-g^{\mu\rho}k_2^\nu\Big\}
\end{align}
where the first term vanishes due to $\not\!k_1-\not\!k_2=\not\!p_2-\not\!p_1=(\not\!p_2-m)-(\not\!p_1
-m)=0$ when put between $\bar{u}(p_2,s_2)$ and $u(p_1,s_1)$. Further using $k_1^\mu k_{1\mu}=0$ and
$\epsilon(k_2)_\nu^\star k_2^\nu=0$ we arrive at
\begin{align}
\Delta_C(k_1-k_2)k_{1\mu}\epsilon_\nu^\star(k_2)\bar{u}(p_2,s_2)\gamma^\rho &u(p_1,s_1)\Big\{g^{\mu\nu}(k_1^\rho-k_2^\rho)+g^{\nu\rho}(2k_2^\mu-k_1^\mu)-g^{\mu\rho}k_2^\nu\Big\} \cr
=&-i\epsilon(k_2)_\nu^\star\bar{u}(p_2,s_2) \gamma^\nu u(p_1,s_1)
\end{align}
For the total amplitude we thus obtain
\begin{align}
ik_1^\mu\mathcal{M}_\mu=~&ie^2\epsilon(k_2)_\nu^\star\bar{u}(p_2,s_2) \gamma^\nu u(p_1,s_1) \cr
&\Big\{-\big(c^2e^{-ip_2\wedge k_2}+s^2e^{-ik_2\wedge p_2}\big)\big(c^2e^{-ik_1\wedge p_1}+s^2e^{-ip_1\wedge k_1}\big) \cr
&+ \big(c^2e^{-ip_1\wedge k_2}+s^2e^{-ik_2\wedge p_1}\big)\big(c^2e^{-ik_1\wedge p_2}+s^2e^{-ip_2\wedge k_1}\big) \cr
&-2i\sin(k_1\wedge k_2)
\big(c^4e^{-ip_2\wedge p_1}-s^4e^{-ip_1\wedge p_2}\big)\Big\}
\end{align}
where we use total momentum conservation $p_1+k_1=p_2+k_2$ to write the phase factors more appropriate:
\begin{align}
ik_1^\mu\mathcal{M}_\mu=~&ie^2\epsilon(k_2)_\nu^\star\bar{u}(p_2,s_2) \gamma^\nu u(p_1,s_1) \cr
&\Big\{-c^4e^{-i(p_2\wedge k_1+p_2\wedge p_1 +k_1\wedge p_1)}-c^2s^2e^{-i(p_2\wedge k_1+p_2\wedge p_1+p_1\wedge k_1)} \cr
&-c^2s^2e^{-i(k_1\wedge p_2+p_1\wedge p_2+k_1\wedge p_1)}-s^4e^{-i(k_1\wedge p_2+p_1\wedge p_2+p_1\wedge k_1 )}\cr
&+c^4e^{-i(k_1\wedge p_2 + p_1\wedge k_1 -p_1\wedge p_2)}+c^2s^2e^{-i(k_1\wedge p_2+k_1\wedge p_1-p_2\wedge p_1)}\cr
&+c^2s^2e^{-i(p_2\wedge k_1+p_1\wedge k_1-p_1\wedge p_2)}+s^4e^{-i(p_2\wedge k_1+k_1\wedge p_1-p_2\wedge p_1)}\cr
&-2i\sin(k_1\wedge k_2)
\big(c^4e^{-ip_2\wedge p_1}-s^4e^{-ip_1\wedge p_2}\big)\Big\}
 \cr
=~&ie^2\epsilon(k_2)_\nu^\star\bar{u}(p_2,s_2) \gamma^\nu u(p_1,s_1) \cr
&\Big\{
2ic^4\sin(p_2\wedge k_1+k_1\wedge p_1)e^{-ip_2\wedge p_1}
+2is^4\sin(k_1\wedge p_2 +p_1\wedge k_1)e^{-ip_1\wedge p_2} \cr
&-2i\sin(k_1\wedge k_2)
\big(c^4e^{-ip_2\wedge p_1}-s^4e^{-ip_1\wedge p_2}\big)\Big\}
 \cr
=~&ie^2\epsilon(k_2)_\nu^\star\bar{u}(p_2,s_2) \gamma^\nu u(p_1,s_1) \cr
&\Big\{2ic^4\sin(k_1\wedge k_2)e^{-ip_2\wedge p_1} - 2is^4\sin(k_1\wedge k_2)e^{-ip_1\wedge p_2}\cr
&-2i\sin(k_1\wedge k_2)
\big(c^4e^{-ip_2\wedge p_1}-s^4e^{-ip_1\wedge p_2}\big)\Big\}\cr
=~&0
\end{align}
which is our desired result.

\newpage

\section{Equations of motion}

How do quantized equations of motion look like in the approach of TOPT? In this section we will investigate this question in an easy example and encounter very different results depending on whether $\theta^{0i}=0$ or not. The first case was already treated in chapter \ref{spatial}. However, for safety we review it in the present somewhat different formalism and will find the expected result. For the second, more interesting case we obtain additional terms to the classical equation of motion which may either be written as a series expansion in $\theta^{0i}$ starting at first order or be formulated with the help of a modified star product.
\\

To have best comparability with the earlier derived results for $\theta^{0i}=0$ in chapter \ref{spatial} we consider the same example, namely a complex scalar field $\phi$ and a real scalar field $\sigma$ interacting through $\mathcal{L}_{\text{int}}=g\phi^\dag\star\sigma\star\phi$. We intend to study the equation of motion for $\sigma$, which on the classical level reads $(\Box+m^2)\sigma=g\phi\star\phi^\dag$.
\newline

On the quantized level we consider a Green's function
\beq
\langle 0|T\big(\sigma(x)~X\big)|0\rangle
\label{eq_mo_start}
\eeq
with $X$ being again a collection of field operators:
\beq
X=\mathcal{O}_1(x_1)...\mathcal{O}_n(x_n)~;\quad
\mathcal{O}_i\epsilon\big\{\phi,\phi^\dag,\sigma\}  \quad .
\eeq
In TOPT the Green's function (\ref{eq_mo_start}) is calculated by the formula
\begin{align}
\langle 0|T\big(\sigma(x)~X\big)|0\rangle
=\langle 0|T\Big(\sigma^{(0)}(x)~X~e^{i\int dx \mathcal{L}^{(0)}_{\text{int}}(x)}\Big)|0\rangle\quad ,   \label{pert_green}
\end{align}
the index $^{(0)}$ referring to free fields.
To obtain the equation of motion for $\sigma(x)$ we act on the above expression with $(\Box_x+m^2)$. Notice that $\sigma^{(0)}(x)$ can be contracted with $\sigma^{(0)}(y)$ appearing in $\mathcal{L}_{\text{int}}(y)$ or with one of the fields $\sigma^{(0)}(x_l)$ $\big($let $L_{\sigma}\subset\{1,...,n\}$ be such that for $l\in L_{\sigma}$ there is a field $\sigma$ at $x_l \big)$  , so diagrammatically
\begin{align}
&(\Box_x+m^2)\langle 0|T\big(\sigma(x)~X\big)|0\rangle\cr
&= \int dy
\begin{picture}(100,70)(10,40)
\Vertex(50,88){1}
\Vertex(30,0){1}
\Vertex(70,0){1}
\Vertex(50,73){1}
\GOval(50,42.5)(17,30)(0){0.75}
\DashLine(50,88)(50,73){2}
\Line(30,15)(30,0)
\DashLine(30,30)(30,15){2}
\Line(70,15)(70,0)
\DashLine(70,30)(70,15){2}
\Line(50,73)(35,57)
\Line(50,73)(65,57)
\Text(40,73)[]{$y$}
\Text(50,98)[]{$x$}
\Text(30,-10)[]{$x_1$}
\Text(50,-10)[]{...}
\Text(70,-10)[]{$x_n$}
\Text(55,83)[l]{$(\Box+m^2)$}
\end{picture}
+\qquad \sum_{l\in L_{\sigma}}
\begin{picture}(100,70)(15,40)
\Vertex(90,85){1}
\Vertex(30,0){1}
\Vertex(70,0){1}
\Vertex(90,0){1}
\GOval(50,42.5)(17,30)(0){0.75}
\Line(30,15)(30,0)
\DashLine(30,30)(30,15){2}
\Line(70,15)(70,0)
\DashLine(70,30)(70,15){2}
\DashLine(90,85)(90,0){2}
\Text(90,95)[]{$x$}
\Text(30,-10)[]{$x_1$}
\Text(50,-10)[]{...}
\Text(50,0)[]{$\check{x}_l$}
\Text(70,-10)[]{$x_n$}
\Text(90,-10)[]{$x_l$}
\Text(95,50)[l]{$(\Box+m^2)$}
\end{picture}
\cr\cr
\end{align}
where
\begin{picture}(50,10)(0,2)
\Line(5,5)(45,5)
\end{picture}
is the $\phi$-line and
\begin{picture}(50,10)(0,2)
\DashLine(5,5)(45,5){2}
\end{picture}
the $\sigma$-line.
In the second term we do not have a star product involved in the line where the derivative acts on, so this gives a propagator and we use $(\Box+m^2)\Delta_C(x)=-i\delta^{(4)}(x)$ to calculate
\begin{align}
&(\Box_x+m^2)\langle 0|T\big(\sigma(x)X\big)|0\rangle\cr
=&
\int dy~(\Box_x+m^2)\langle 0|T\big(
\begin{picture}(0,20)(0,0)
\Line(4,18)(59,18)
\Line(4,18)(4,8)
\Line(59,18)(59,8)
\end{picture}
\sigma(x)ig(\phi^\dag\star\sigma\star\phi)(y)X\big)|0\rangle \cr
&+\sum_{l\in L_\sigma}(\Box_x+m^2)\Delta_C(x-x_l)\langle 0|T\big(X_{\check l}\big)|0\rangle \cr
=~&
ig\int dy~(\Box_x+m^2)\cr
&\qquad \langle 0|T\Big(
\begin{picture}(0,20)(0,0)
\Line(4,18)(157,18)
\Line(4,18)(4,8)
\Line(157,18)(157,8)
\end{picture}
\sigma(x)e^{\frac{i}{2}\theta^{\mu\nu}\partial_\mu^\xi\partial_\nu^\eta}e^{\frac{i}{2}\theta^{\rho\sigma}\partial_\rho^\zeta\partial_\sigma^\chi}\phi^\dag(y+\xi)\sigma(y+\eta+\zeta)\phi(y+\eta+\chi)X\Big)|0\rangle\Big|_{\xi,\eta,\zeta,\chi=0}  \cr
&+\sum_{l\in L_\sigma}(-i)\delta^{(4)}(x-x_l)\langle 0|T\big(X_{\check l}\big)|0\rangle \cr
\nonumber
\end{align}
The second term are contact terms (c.t.), we now focus on the first one:
\begin{align}
&(\Box_x+m^2)\langle 0|T\big(\sigma(x)X\big)|0\rangle\cr
&=-g\int dy~(\Box_x+m^2)\langle 0|T\Big(
e^{\frac{i}{2}\theta^{\mu\nu}\partial_\mu^\xi\partial_\nu^\eta}e^{\frac{i}{2}\theta^{\rho\sigma}\partial_\rho^\zeta\partial_\sigma^\chi} \cr
&\qquad \big\{\vartheta(x^0-y^0)\Delta_+(x-y-\eta-\zeta)+\vartheta(y^0-x^0)\Delta_+(y+\eta+\zeta-x)\big\}\cr
&\qquad \phi^\dag(y+\xi)\phi(y+\eta+\chi)X\Big)|0\rangle\Big|_{\xi,\eta,\zeta,\chi=0} \qquad + \text{c.t.}
\label{last_line}
\end{align}
where the time-ordering followed by contraction of $\sigma(x),\sigma(y+\eta+\zeta)$ was performed, \\ $\langle 0 |\sigma(x)\sigma(y)|0\rangle= i\Delta_+(x-y)$.
\newline

It is now crucial to notice that the star product is carried out before time-ordering, having the effect that it does not act on the step functions $\vartheta(x^0-y^0),\vartheta(y^0-x^0)$ in the last line but on the $\Delta_+$-functions. Of course, whether or not the star product acts on the step functions only makes a difference if $\theta^{0i}\neq0$. We will therefore at first turn to the case that $\theta^{0i}$ vanishes.
\newline

\subsection{The case $\theta^{0i}=0$}


As in this case no time derivatives are involved in the star product, we are allowed to replace $\vartheta(x^0-y^0)$ resp. $\vartheta(y^0-x^0)$ by
$\vartheta(x^0-y^0-\eta^0-\zeta^0)$ resp. $\vartheta(y^0+\eta^0+\zeta^0-x^0)$ in (\ref{last_line}).
Use \beq
(\Box_x+m^2)\big\{\vartheta(x^0)\Delta_+(x)+\vartheta(-x^0)\Delta_+(-x)\big\}=(\Box_x+m^2)\{-i\Delta_C(x)\}=-\delta(x)\eeq
to find
\begin{align}
&(\Box_x+m^2)\big\{\vartheta(x^0-y^0-\eta^0-\zeta^0)\Delta_+(x-y-\eta-\zeta)\cr
&\qquad\qquad\quad +\vartheta(y^0+\eta^0+\zeta^0-x^0)\Delta_+(x)(y+\eta+\zeta-x)\big\}\cr
&=-\delta(x-y-\eta-\zeta)
\end{align}
and thus
\begin{align}
&(\Box_x+m^2)\langle 0|T\big(\sigma(x)X\big)|0\rangle\cr
&=-g\int dy~\langle 0|T\Big(
e^{\frac{i}{2}\theta^{\mu\nu}\partial_\mu^\xi\partial_\nu^\eta}e^{\frac{i}{2}\theta^{\rho\sigma}\partial_\rho^\zeta\partial_\sigma^\chi}(-1)\delta(x-y-\eta-\zeta) \cr
&\qquad\qquad \phi^\dag(y+\xi)\phi(y+\eta+\chi)X\Big)|0\rangle\Big|_{\xi,\eta,\zeta,\chi=0} \qquad + \text{c.t.}\cr
&=g\langle 0|T\Big(
e^{\frac{i}{2}\theta^{\mu\nu}\partial_\mu^\xi\partial_\nu^\eta}e^{\frac{i}{2}\theta^{\rho\sigma}\partial_\rho^\zeta\partial_\sigma^\chi} \cr
&\qquad\qquad \phi^\dag(x-\eta-\zeta+\xi)\phi(x-\eta-\zeta+\eta+\chi)X\Big)|0\rangle\Big|_{\xi,\eta,\zeta,\chi=0} \qquad + \text{c.t.}\cr
&=g\langle 0|T\Big(
e^{\frac{i}{2}\theta^{\mu\nu}\partial_\mu^\xi\partial_\nu^\eta}e^{\frac{i}{2}\theta^{\rho\sigma}\partial_\rho^\zeta\partial_\sigma^\chi} \cr
&\qquad \qquad\phi^\dag(x-\eta-\zeta+\xi)\phi(x-\zeta+\chi)X\Big)|0\rangle\Big|_{\xi,\eta,\zeta,\chi=0} \qquad + \text{c.t.}\cr
&=g\langle 0|T\Big(
e^{\frac{i}{2}\theta^{\rho\sigma}\partial_\rho^\zeta\partial_\sigma^\chi} \phi^\dag(x-\zeta)\phi(x+\chi)X\Big)|0\rangle\Big|_{\xi,\eta,\zeta,\chi=0} \qquad + \text{c.t.}\cr
&=g\langle 0|T\big(
(\phi\star\phi^\dag)(x)X\big)|0\rangle \qquad + \text{c.t.}
\end{align}
which we recognize to be up to contact terms the classical equation of motion for the field $\sigma$, agreeing with the result in the chapter on only spatial noncommutativity.
\newline

\subsection{Additional terms in the case $\theta^{0i}\neq 0$}

Whether or not the star product acts on the stepfunctions makes a difference for $\theta^{0i}\neq0$. The classical equation of motion will  be violated by the appearance of additional terms. We want to derive these and show that they can be written as a series expansion in $\theta^{0i}$ starting at first order. \\

 In (\ref{last_line}) the star product is not applied to the stepfunctions, however we learn from the above calculation that we obtain the classical equation of motion if we allow the star product to act on the step functions: exactly as above we compute
\begin{align}
&-g\int dy~(\Box_x+m^2)\langle 0|T\Big(
e^{\frac{i}{2}\theta^{\mu\nu}\partial_\mu^\xi\partial_\nu^\eta}e^{\frac{i}{2}\theta^{\rho\sigma}\partial_\rho^\zeta\partial_\sigma^\chi} \cr
&\qquad \big\{\vartheta(x^0-y^0-\eta^0-\zeta^0)\Delta_+(x-y-\eta-\zeta)+\vartheta(y^0+\eta^0+\zeta^0-x^0)\Delta_+(y+\eta+\zeta-x)\big\}\cr
&\qquad \phi^\dag(y+\xi)\phi(y+\eta+\chi)X\Big)|0\rangle\Big|_{\xi,\eta,\zeta,\chi=0} \quad + \text{c.t.}\cr
=~&...\cr
=~&g\langle 0|T\big(
(\phi\star\phi^\dag)(x)X\big)|0\rangle \quad + \text{c.t.} \quad .
\end{align}
 The difference between the two scenarios, to be calculated in the following, will be nonvanishing and appears as additional terms in the equation of motion for the field $\sigma$.
 \begin{align}
&(\Box_x+m^2)\langle 0|T\big(\sigma(x)X\big)|0\rangle ~~ - ~~ g\langle 0|T\big(
(\phi\star\phi^\dag)(x)X\big)|0\rangle \cr
=&-g\int dy~(\Box_x+m^2)\langle 0|T\Big(
e^{\frac{i}{2}\theta^{\mu\nu}\partial_\mu^\xi\partial_\nu^\eta}e^{\frac{i}{2}\theta^{\rho\sigma}\partial_\rho^\zeta\partial_\sigma^\chi} \cr
&\qquad \big\{\vartheta(x^0-y^0)\Delta_+(x-y-\eta-\zeta)+\vartheta(y^0-x^0)\Delta_+(y+\eta+\zeta-x)\big\}\cr
&\qquad \phi^\dag(y+\xi)\phi(y+\eta+\chi)X\Big)|0\rangle\Big|_{\xi,\eta,\zeta,\chi=0} \cr
&+g\int dy~(\Box_x+m^2)\langle 0|T\Big(
\underbrace{e^{\frac{i}{2}\theta^{\mu\nu}\partial_\mu^\xi\partial_\nu^\eta}e^{\frac{i}{2}\theta^{\rho\sigma}\partial_\rho^\zeta\partial_\sigma^\chi}}_{(\star)} \cr
&\qquad \big\{\vartheta(x^0-y^0-\eta^0-\zeta^0)\Delta_+(x-y-\eta-\zeta)+\vartheta(y^0+\eta^0+\zeta^0-x^0)\Delta_+(y+\eta+\zeta-x)\big\}\cr
&\qquad \phi^\dag(y+\xi)\phi(y+\eta+\chi)X\Big)|0\rangle\Big|_{\xi,\eta,\zeta,\chi=0}\quad + \text{c.t.} \label{two_terms}
\end{align}
The time derivatives $\partial_0^\eta, \partial_0^\zeta$ in the expression $(\star)$  either act on the step functions or on the part behind. We decompose them into
\begin{align}
\partial_0^\eta&=\partial_0^{\eta(\vartheta)}+\partial_0^{\eta(b)}\cr
\partial_0^\zeta&=\partial_0^{\zeta(\vartheta)}+\partial_0^{\zeta(b)}\nonumber
\end{align}
where the first derivative in the sum only acts on the step function and the second derivative only on the part behind. So $(\star)$ can be written as
\begin{align}
e^{\frac{i}{2}\theta^{\mu\nu}\partial_\mu^\xi\partial_\nu^\eta}e^{\frac{i}{2}\theta^{\rho\sigma}\partial_\rho^\zeta\partial_\sigma^\chi}&=e^{\frac{i}{2}(\theta^{i0}\partial_i^\xi\partial_0^\eta+\theta^{\mu i}\partial_\mu^\xi\partial_i^\eta+\theta^{0j}\partial_0^\zeta\partial_j^\chi+\theta^{j\sigma}\partial_j^\zeta\partial_\sigma^\chi)}\cr
&=e^{\frac{i}{2}(\theta^{i0}\partial_i^\xi\partial_0^{\eta(\vartheta)}+\theta^{0j}\partial_0^{\zeta(\vartheta)}\partial_j^\chi)}e^{\frac{i}{2}(\theta^{i0}\partial_i^\xi\partial_0^{\eta(b)}+\theta^{\mu i}\partial_\mu^\xi\partial_i^\eta+\theta^{0j}\partial_0^{\zeta(b)}\partial_j^\chi+\theta^{j\sigma}\partial_j^\zeta\partial_\sigma^\chi)}
\nonumber
\end{align}
and expanding the first factor
\beq
e^{\frac{i}{2}(\theta^{i0}\partial_i^\xi\partial_0^{\eta(\vartheta)}+\theta^{0j}\partial_0^{\zeta(\vartheta)}\partial_j^\chi)}=\sum_{l=0}^\infty\frac{1}{l!}\big(\frac{i}{2}\big)^l\Big(\theta^{i0}\partial_i^\xi\partial_0^{\eta(\vartheta)}+\theta^{0j}\partial_0^{\zeta(\vartheta)}\partial_j^\chi\Big)^l
\eeq
we find
\begin{align}
e^{\frac{i}{2}\theta^{\mu\nu}\partial_\mu^\xi\partial_\nu^\eta}e^{\frac{i}{2}\theta^{\rho\sigma}\partial_\rho^\zeta\partial_\sigma^\chi}=&
\sum_{l=0}^\infty\frac{1}{l!}\big(\frac{i}{2}\big)^l\Big(\theta^{i0}\partial_i^\xi\partial_0^{\eta(\vartheta)}+\theta^{0j}\partial_0^{\zeta(\vartheta)}\partial_j^\chi\Big)^l\cr
&\quad e^{\frac{i}{2}(\theta^{i0}\partial_i^\xi\partial_0^{\eta(b)}+\theta^{\mu i}\partial_\mu^\xi\partial_i^\eta+\theta^{0j}\partial_0^{\zeta(b)}\partial_j^\chi+\theta^{j\sigma}\partial_j^\zeta\partial_\sigma^\chi)} \quad .
\end{align}
Replacing $(\star)$ in (\ref{two_terms}) by this expression, the contribution coming from $l=0$ has no time derivative acting on a step function and is seen to exactly cancel with the first term. Thus
\begin{align}
&(\Box_x+m^2)\langle 0|T\big(\sigma(x)X\big)|0\rangle ~~ - ~~ g\langle 0|T\big(
(\phi\star\phi^\dag)(x)X\big)|0\rangle \cr
=~&g\int dy~(\Box_x+m^2)\langle 0|T\Big(\sum_{l=1}^\infty\frac{1}{l!}\big(\frac{i}{2}\big)^l\Big(\theta^{i0}\partial_i^\xi\partial_0^{\eta(\vartheta)}+\theta^{0j}\partial_0^{\zeta(\vartheta)}\partial_j^\chi\Big)^l \cr
&\qquad
e^{\frac{i}{2}(\theta^{i0}\partial_i^\xi\partial_0^{\eta(b)}+\theta^{\mu i}\partial_\mu^\xi\partial_i^\eta+\theta^{0j}\partial_0^{\zeta(b)}\partial_j^\chi+\theta^{j\sigma}\partial_j^\zeta\partial_\sigma^\chi)} \cr
&\qquad \big\{\vartheta(x^0-y^0-\eta^0-\zeta^0)\Delta_+(x-y-\eta-\zeta)+\vartheta(y^0+\eta^0+\zeta^0-x^0)\Delta_+(y+\eta+\zeta-x)\big\}\cr
&\qquad \phi^\dag(y+\xi)\phi(y+\eta+\chi)X\Big)|0\rangle\Big|_{\xi,\eta,\zeta,\chi=0}\quad + \text{c.t.} \cr
=~&g\int dy~(\Box_x+m^2)\langle 0|T\Big(\sum_{l=1}^\infty\frac{1}{l!}\big(\frac{i}{2}\big)^l\Big(\theta^{i0}\partial_i^\xi\partial_0^{\eta(\vartheta)}+\theta^{0j}\partial_0^{\zeta(\vartheta)}\partial_j^\chi\Big)^l \cr
&\qquad \big\{\vartheta(x^0-y^0-\eta^0-\zeta^0)e^{\frac{i}{2}\theta^{\mu\nu}\partial_\mu^\xi\partial_\nu^\eta}e^{\frac{i}{2}\theta^{\rho\sigma}\partial_\rho^\zeta\partial_\sigma^\chi}\Delta_+(x-y-\eta-\zeta)\cr
&\qquad +\vartheta(y^0+\eta^0+\zeta^0-x^0)e^{\frac{i}{2}\theta^{\mu\nu}\partial_\mu^\xi\partial_\nu^\eta}e^{\frac{i}{2}\theta^{\rho\sigma}\partial_\rho^\zeta\partial_\sigma^\chi}\Delta_+(y+\eta+\zeta-x)\big\}\cr
&\qquad \phi^\dag(y+\xi)\phi(y+\eta+\chi)X\Big)|0\rangle\Big|_{\xi,\eta,\zeta,\chi=0}\quad + \text{c.t.}\quad .
\end{align}
Note that it does not make a difference whether we apply $\partial_0^{\eta(\vartheta)}$ or $\partial_0^{\zeta(\vartheta)}$ to the step functions, we also get the same result by taking $\partial_0^{y(\vartheta)}$ which we do in the following:
\begin{align}
&(\Box_x+m^2)\langle 0|T\big(\sigma(x)X\big)|0\rangle ~~ - ~~ g\langle 0|T\big(
(\phi\star\phi^\dag)(x)X\big)|0\rangle \cr
=~&g\int dy~(\Box_x+m^2)\langle 0|T\Big(\sum_{l=1}^\infty\frac{1}{l!}\big(\frac{i}{2}\big)^l\Big(\theta^{i0}\partial_i^\xi+\theta^{0j}\partial_j^\chi\Big)^l\big(\partial_0^{y(\vartheta)}\big)^l \cr
&\qquad \big\{\vartheta(x^0-y^0)e^{\frac{i}{2}\theta^{\mu\nu}\partial_\mu^\xi\partial_\nu^\eta}e^{\frac{i}{2}\theta^{\rho\sigma}\partial_\rho^\zeta\partial_\sigma^\chi}\Delta_+(x-y-\eta-\zeta)\cr
&\qquad +\vartheta(y^0-x^0)e^{\frac{i}{2}\theta^{\mu\nu}\partial_\mu^\xi\partial_\nu^\eta}e^{\frac{i}{2}\theta^{\rho\sigma}\partial_\rho^\zeta\partial_\sigma^\chi}\Delta_+(y+\eta+\zeta-x)\big\}\cr
&\qquad \phi^\dag(y+\xi)\phi(y+\eta+\chi)X\Big)|0\rangle\Big|_{\xi,\eta,\zeta,\chi=0}\quad + \text{c.t.} \quad .
\end{align}
In each term of this sum exists at least one derivative
$\partial_0^{y(\vartheta)}$ which acting on the step functions yields a one-dimensional $\delta$-distribution:
\begin{align}
&(\Box_x+m^2)\langle 0|T\big(\sigma(x)X\big)|0\rangle ~~ - ~~ g\langle 0|T\big(
(\phi\star\phi^\dag)(x)X\big)|0\rangle \cr
=~&g\int dy~(\Box_x+m^2)\langle 0|T\Big(\sum_{l=1}^\infty\frac{1}{l!}\big(\frac{i}{2}\big)^l\Big(\theta^{i0}\partial_i^\xi-\theta^{io}\partial_i^\chi\Big)^l \cr
&\qquad \big\{-\big(\partial_{y_0}^{l-1} \delta^{(1)}(x^0-y^0)\big)e^{\frac{i}{2}\theta^{\mu\nu}\partial_\mu^\xi\partial_\nu^\eta}e^{\frac{i}{2}\theta^{\rho\sigma}\partial_\rho^\zeta\partial_\sigma^\chi}\Delta_+(x-y-\eta-\zeta)\cr
&\qquad +\big(\partial_{y_0}^{l-1} \delta^{(1)}(y^0-x^0)\big)e^{\frac{i}{2}\theta^{\mu\nu}\partial_\mu^\xi\partial_\nu^\eta}e^{\frac{i}{2}\theta^{\rho\sigma}\partial_\rho^\zeta\partial_\sigma^\chi}\Delta_+(y+\eta+\zeta-x)\big\}\cr
&\qquad \phi^\dag(y+\xi)\phi(y+\eta+\chi)X\Big)|0\rangle\Big|_{\xi,\eta,\zeta,\chi=0}\quad + \text{c.t.} \quad .
\end{align}
With the help of $\delta^{(1)}(x^0-y^0)=\delta^{(1)}(y^0-x^0)$ and $\Delta_+(x)-\Delta_+(-x)=\Delta(x)$, the commutator function, we can write this more compact by
\begin{align}
&(\Box_x+m^2)\langle 0|T\big(\sigma(x)X\big)|0\rangle ~~ - ~~ g\langle 0|T\big(
(\phi^\dag\star\phi^\dag)(x)X\big)|0\rangle \cr
=&-g\int dy~(\Box_x+m^2)\langle 0|T\Big(\sum_{l=1}^\infty\frac{1}{l!}\big(\frac{i}{2}\big)^l(\theta^{i0})^l\Big(\partial_i^\xi-\partial_i^\chi\Big)^l \cr
&\qquad\big\{\big(\partial_{y_0}^{l-1} \delta^{(1)}(x^0-y^0)\big)e^{\frac{i}{2}\theta^{\mu\nu}\partial_\mu^\xi\partial_\nu^\eta}e^{\frac{i}{2}\theta^{\rho\sigma}\partial_\rho^\zeta\partial_\sigma^\chi}\Delta(x-y-\eta-\zeta)\big\}\cr
&\qquad \phi^\dag(y+\xi)\phi(y+\eta+\chi)X\Big)|0\rangle\Big|_{\xi,\eta,\zeta,\chi=0}\quad + \text{c.t.} \quad .
\end{align}
The time derivative $\partial_{x_0}$ in $\Box_x+m^2$ can act on the $\delta^{(1)}$-distributions or on the commutator. With the help of $(\Box_x+m^2)\Delta(x)=0$ we find that only the terms contribute where at least one time derivative is applied to the $\delta^{(1)}$-distributions:
\begin{align}
&(\Box_x+m^2)\langle 0|T\big(\sigma(x)X\big)|0\rangle ~~ - ~~ g\langle 0|T\big(
(\phi\star\phi^\dag)(x)X\big)|0\rangle \cr
=&-g\int dy~\langle 0|T\Big(\sum_{l=1}^\infty\frac{1}{l!}\big(\frac{i}{2}\big)^l(\theta^{i0})^l\Big(\partial_i^\xi-\partial_i^\chi\Big)^l \cr
&\qquad\big\{\big(\partial_{x_0}^2\partial_{y_0}^{l-1} \delta^{(1)}(x^0-y^0)\big)e^{\frac{i}{2}\theta^{\mu\nu}\partial_\mu^\xi\partial_\nu^\eta}e^{\frac{i}{2}\theta^{\rho\sigma}\partial_\rho^\zeta\partial_\sigma^\chi}\Delta(x-y-\eta-\zeta)\cr
&\qquad +\big(\partial_{x_0}\partial_{y_0}^{l-1} \delta^{(1)}(x^0-y^0)\big)e^{\frac{i}{2}\theta^{\mu\nu}\partial_\mu^\xi\partial_\nu^\eta}e^{\frac{i}{2}\theta^{\rho\sigma}\partial_\rho^\zeta\partial_\sigma^\chi}\partial_{x_0}\Delta(x-y-\eta-\zeta)\big\}\cr
&\qquad \phi^\dag(y+\xi)\phi(y+\eta+\chi)X\Big)|0\rangle\Big|_{\xi,\eta,\zeta,\chi=0}\quad + \text{c.t.} \cr
=&-g\int dy~\langle 0|T\Big(\sum_{l=1}^\infty\frac{1}{l!}\big(\frac{i}{2}\big)^l(\theta^{i0})^l\Big(\partial_i^\xi-\partial_i^\chi\Big)^l \cr
&\qquad\big\{\big(\partial_{y_0}^{l+1} \delta^{(1)}(x^0-y^0)\big)e^{\frac{i}{2}\theta^{\mu\nu}\partial_\mu^\xi\partial_\nu^\eta}e^{\frac{i}{2}\theta^{\rho\sigma}\partial_\rho^\zeta\partial_\sigma^\chi}\Delta(x-y-\eta-\zeta)\cr
&\qquad +\big(\partial_{y_0}^{l} \delta^{(1)}(x^0-y^0)\big)e^{\frac{i}{2}\theta^{\mu\nu}\partial_\mu^\xi\partial_\nu^\eta}e^{\frac{i}{2}\theta^{\rho\sigma}\partial_\rho^\zeta\partial_\sigma^\chi}\partial_{y_0}\Delta(x-y-\eta-\zeta)\big\}\cr
&\qquad \phi^\dag(y+\xi)\phi(y+\eta+\chi)X\Big)|0\rangle\Big|_{\xi,\eta,\zeta,\chi=0}\quad + \text{c.t.}
\end{align}
where in the second equality we replaced the partial derivatives $\partial_{x_0}$ by $\partial_{y_0}$, such that we can now integrate by parts with respect to $dy_0$. We do not expect contributions from the boundary and use
\beq
\partial_{y_0}^l\big(\Delta\phi^\dag\phi\big)=\sum_{m=1}^l \binom{l}{m} \big(\partial_{y_0}^m\Delta\big)\big(\partial_{y_0}^{l-m}\phi^\dag\phi\big)
\eeq
 to arrive at
\begin{align}
&(\Box_x+m^2)\langle 0|T\big(\sigma(x)X\big)|0\rangle ~~ - ~~ g\langle 0|T\big(
(\phi\star\phi^\dag)(x)X\big)|0\rangle \cr
=&-g\int dy~\delta^{(1)}(x^0-y^0)\langle 0|T\Big(\sum_{l=1}^\infty\frac{1}{l!}\big(\frac{i}{2}\big)^l(\theta^{i0})^l\Big(\partial_i^\xi-\partial_i^\chi\Big)^l e^{\frac{i}{2}\theta^{\mu\nu}\partial_\mu^\xi\partial_\nu^\eta}e^{\frac{i}{2}\theta^{\rho\sigma}\partial_\rho^\zeta\partial_\sigma^\chi}\cr
&\qquad\Big\{(-1)^{l+1}\sum_{m=1}^{l+1} \binom{l+1}{m}\big(\partial_{y_0}^m\Delta(x-y-\eta-\zeta)\big)\partial_{y_0}^{l-m+1}\big(\phi^\dag(y+\xi)\phi(y+\eta+\chi)\big)\cr
&\qquad +(-1)^{l}\sum_{m=1}^{l} \binom{l}{m}\big(\partial_{y_0}^{m+1}\Delta(x-y-\eta-\zeta)\big)\partial_{y_0}^{l-m}\big(\phi^\dag(y+\xi)\phi(y+\eta+\chi)\big) \Big\}\cr
&\qquad \qquad X\Big)|0\rangle\Big|_{\xi,\eta,\zeta,\chi=0}\quad + \text{c.t.}\quad .
\end{align}
We group the terms slightly different in order to recognize them as a series expansion in $\theta^{0i}$, starting at first order:
\begin{align}
&(\Box_x+m^2)\langle 0|T\big(\sigma(x)X\big)|0\rangle ~~ - ~~ g\langle 0|T\big(
(\phi\star\phi^\dag)(x)X\big)|0\rangle \cr
=&\sum_{l=1}^\infty\frac{1}{l!}\big(\frac{i}{2}\big)^l(\theta^{0i})^lg\int dy~\delta^{(1)}(x^0-y^0)\langle 0|T\Big(\big(\partial_i^\xi-\partial_i^\chi\big)^l e^{\frac{i}{2}\theta^{\mu\nu}\partial_\mu^\xi\partial_\nu^\eta}e^{\frac{i}{2}\theta^{\rho\sigma}\partial_\rho^\zeta\partial_\sigma^\chi}\cr
&\qquad\Big\{\sum_{m=1}^{l+1} \binom{l+1}{m}\big(\partial_{y_0}^m\Delta(x-y-\eta-\zeta)\big)\partial_{y_0}^{l-m+1}\big(\phi^\dag(y+\xi)\phi(y+\eta+\chi)\big)\cr
&\qquad -\sum_{m=1}^{l} \binom{l}{m}\big(\partial_{y_0}^{m+1}\Delta(x-y-\eta-\zeta)\big)\partial_{y_0}^{l-m}\big(\phi^\dag(y+\xi)\phi(y+\eta+\chi)\big) \Big\} X\Big)|0\rangle\Big|_{\xi,\eta,\zeta,\chi=0}\cr
&+ \text{c.t.} \quad .
\end{align}

\subsection{Modified phase factors}

We have just shown that the classical equation of motion is violated in the case $\theta^{0i}\neq 0$ and derived the additional terms as a series expansion in $\theta^{0i}$. The aim of this subsection is to write the whole quantized equation of motion in a form that resembles the classical one. We will achieve that by introducing modified phase factors. \\

Let us start again from expression (\ref{last_line}):
\begin{align}
&(\Box_x+m^2)\langle 0|T\big(\sigma(x)X\big)|0\rangle\cr
&=-g\int dy~(\Box_x+m^2)\langle 0|T\Big(
e^{\frac{i}{2}\theta^{\mu\nu}\partial_\mu^\xi\partial_\nu^\eta}e^{\frac{i}{2}\theta^{\rho\sigma}\partial_\rho^\zeta\partial_\sigma^\chi} \cr
&\qquad \big\{\vartheta(x^0-y^0)\Delta_+(x-y-\eta-\zeta)+\vartheta(y^0-x^0)\Delta_+(y+\eta+\zeta-x)\big\}\cr
&\qquad \phi^\dag(y+\xi)\phi(y+\eta+\chi)X\Big)|0\rangle\Big|_{\xi,\eta,\zeta,\chi=0} \qquad + \text{c.t.} \quad .
\end{align}
We will first apply $(\Box+m^2)$, where the part $(\Box+m^2)$ fully acting on $\Delta_+$ gives $0$, such that
\begin{align}
&(\Box_x+m^2)\langle 0|T\big(\sigma(x)X\big)|0\rangle\cr
&=-g\int dy~\langle 0|T\Big(
e^{\frac{i}{2}\theta^{\mu\nu}\partial_\mu^\xi\partial_\nu^\eta}e^{\frac{i}{2}\theta^{\rho\sigma}\partial_\rho^\zeta\partial_\sigma^\chi} \cr
&\qquad \big\{(\partial_{x_0}^2\vartheta(x^0-y^0))\Delta_+(x-y-\eta-\zeta)+
(\partial_{x_0}\vartheta(x^0-y^0))\partial_{x_0}\Delta_+(x-y-\eta-\zeta)   \cr
&\qquad +(\partial_{x_0}^2\vartheta(y^0-x^0))\Delta_+(y+\eta+\zeta-x)
+(\partial_{x_0}\vartheta(y^0-x^0))\partial_{x_0}\Delta_+(y+\eta+\zeta-x)
\big\}\cr
&\qquad \phi^\dag(y+\xi)\phi(y+\eta+\chi)X\Big)|0\rangle\Big|_{\xi,\eta,\zeta,\chi=0} \qquad + \text{c.t.} \quad .
\end{align}
The fields $\phi^\dag, \phi$ as well as $\Delta_+$ will now be expressed by their Fourier transforms
\begin{align}
\phi^\dag(x)&=\frac{1}{(2\pi)^4}\int d^4p~ \tilde{\phi}^\dag(p) e^{-ipx}  \cr
\phi(x)&=\frac{1}{(2\pi)^4}\int d^4q~ \tilde{\phi}^\dag(q) e^{-iqx}  \cr
i\Delta_+(x)&=\int \frac{d^3k}{2E_k(2\pi)^3}~ e^{-ikx}
\end{align}
 which results in
\begin{align}
&(\Box_x+m^2)\langle 0|T\big(\sigma(x)X\big)|0\rangle\cr
&=ig\frac{1}{(2\pi)^8}\int dy ~d^4p ~d^4q ~\frac{d^3k}{2E_k(2\pi)^3}~\langle 0|T\Big(
e^{\frac{i}{2}\theta^{\mu\nu}\partial_\mu^\xi\partial_\nu^\eta}e^{\frac{i}{2}\theta^{\rho\sigma}\partial_\rho^\zeta\partial_\sigma^\chi} \cr
&\qquad \big\{(\partial_{x_0}^2\vartheta(x^0-y^0))e^{-ik(x-y-\eta-\zeta)}+
(\partial_{x_0}\vartheta(x^0-y^0))(-iE_k)e^{-ik(x-y-\eta-\zeta)}   \cr
&\qquad +(\partial_{x_0}^2\vartheta(y^0-x^0))e^{-ik(y-x+\eta+\zeta)}
+(\partial_{x_0}\vartheta(y^0-x^0))iE_k e^{-ik(y-x+\eta+\zeta)}
\big\}\cr
&\qquad \tilde{\phi}^\dag(p)\tilde{\phi}(q)e^{-ip(y+\xi)}e^{-iq(y+\eta+\chi)}X\Big)|0\rangle\Big|_{\xi,\eta,\zeta,\chi=0} \qquad + \text{c.t.}\quad .
\end{align}
It is now possible to perform the derivatives with respect to $\xi,\eta,\zeta,\chi$ and we obtain
\begin{align}
&(\Box_x+m^2)\langle  0|T\big(\sigma(x)X\big)|0\rangle\cr
&=ig\frac{1}{(2\pi)^8}\int dy ~d^4p ~d^4q ~\frac{d^3k}{2E_k(2\pi)^3}~ \cr
&\qquad\langle 0|T\Big(
\big\{\big[(\partial_{x_0}^2\vartheta(x^0-y^0))+
(\partial_{x_0}\vartheta(x^0-y^0))(-iE_k)\big]e^{-ik(x-y)}e^{-i(k\wedge p+q\wedge k)}   \cr
&\qquad \qquad +\big[(\partial_{x_0}^2\vartheta(y^0-x^0))
+(\partial_{x_0}\vartheta(y^0-x^0))iE_k \big]e^{-ik(y-x)}e^{-i(p\wedge k+k\wedge q)}
\big\}\cr
&\qquad\qquad \tilde{\phi}^\dag(p)\tilde{\phi}(q)e^{-ipy}e^{-iqy}e^{-ip\wedge q}X\Big)|0\rangle \qquad + \text{c.t.} \quad .
\end{align}
Carrying out the integration over $d^3y$ we obtain three-dimensional $\delta$-distributions and a factor $(2\pi)^3$, and from a time derivative acting on a step-function we get a one-dimensional $\delta$-distribution:
\begin{align}
&(\Box_x+m^2)\langle  0|T\big(\sigma(x)X\big)|0\rangle\cr
&=ig\frac{1}{(2\pi)^8}\int dy^0 ~d^4p ~d^4q ~\frac{d^3k}{2E_k}~ \cr
&\qquad\langle 0|T\Big(
\big\{\big[(\partial_{x_0}\delta(x^0-y^0))+
\delta(x^0-y^0)(-iE_k)\big]e^{-ikx}e^{iE_ky^0}\delta^{(3)}(p^i+q^i-k^i)e^{-i(k\wedge p+q\wedge k)}   \cr
&\qquad \qquad +\big[-(\partial_{x_0}\delta(x^0-y^0))
-\delta(x^0-y^0)iE_k \big]e^{ikx}e^{-iE_ky^0}\delta^{(3)}(p^i+q^i+k^i)e^{-i(p\wedge k+k\wedge q)}
\big\}\cr
&\qquad\qquad \tilde{\phi}^\dag(p)\tilde{\phi}(q)e^{-i(p^0+q^0)y^0}e^{-ip\wedge q}X\Big)|0\rangle \qquad + \text{c.t.}
\end{align}
and we perform the integration over $d^3k$:
\begin{align}
&(\Box_x+m^2)\langle  0|T\big(\sigma(x)X\big)|0\rangle\cr
&=ig\frac{1}{(2\pi)^8}\int dy^0 ~d^4p ~d^4q ~\frac{1}{2E_{p+q}}~ \cr
&\qquad\langle 0|T\Big(
\big\{\big[(\partial_{x_0}\delta(x^0-y^0))-iE_{p+q}
\delta(x^0-y^0))\big]e^{-i(p+q)^+x}e^{iE_{p+q}y^0}e^{-i((p+q)^+\wedge p+q\wedge (p+q)^+)}   \cr
&\qquad \qquad +\big[-(\partial_{x_0}\delta(x^0-y^0))
-iE_{p+q}\delta(x^0-y^0) \big]e^{-i(p+q)^-x}e^{-iE_{p+q}y^0}e^{-i((p+q)^-\wedge p+q\wedge (p+q)^-)}
\big\}\cr
&\qquad\qquad \tilde{\phi}^\dag(p)\tilde{\phi}(q)e^{-i(p^0+q^0)y^0}e^{-ip\wedge q}X\Big)|0\rangle \qquad + \text{c.t.}
\end{align}
where, as already introduced before, $p^\pm\equiv (\pm E_p,{\bf p})$.
The one-dimensional $\delta$-distributions allow us to carry out the $dy^0$ integration:
\begin{align}
&(\Box_x+m^2)\langle  0|T\big(\sigma(x)X\big)|0\rangle\cr
&=ig\frac{1}{(2\pi)^8}\int d^4p ~d^4q ~\frac{1}{2E_{p+q}}~ \cr
&\qquad\langle 0|T\Big(
\big\{\big[-i(p^0+q^0)-iE_{p+q}\big]e^{-i(p+q)^ix_i}e^{-i((p+q)^+\wedge p+q\wedge (p+q)^+)}   \cr
&\qquad \qquad +\big[i(p^0+q^0)-iE_{p+q}\big]e^{-i(p+q)^ix_i}e^{-i((p+q)^-\wedge p+q\wedge (p+q)^-)}
\big\}\cr
&\qquad\qquad \tilde{\phi}^\dag(p)\tilde{\phi}(q)e^{-i(p^0+q^0)x^0}e^{-ip\wedge q}X\Big)|0\rangle \qquad + \text{c.t.}\quad .
\end{align}
The two terms in the curly brackets differ mainly by their phase factors, so let us investigate these closer:
\begin{align}
e^{-i((p+q)^\pm\wedge p+q\wedge (p+q)^\pm)}= e^{-\frac{i}{2}(\pm E_{p+q}\theta^{oi}p_i+(p_i+q_i)\theta^{i0}p_0+q_i\theta^{ij}p_j+q^0\theta^{0i}(p_i+q_i)\pm q_i\theta^{i0}E_{p+q}+q_i\theta^{ij}p_j)}
\end{align}
which we use to write
\begin{align}
&(\Box_x+m^2)\langle  0|T\big(\sigma(x)X\big)|0\rangle\cr
&=ig\frac{1}{(2\pi)^8}\int d^4p ~d^4q ~\frac{1}{2E_{p+q}}~ \cr
&\qquad\langle 0|T\Big(
\big\{\big[-i(p^0+q^0)-iE_{p+q}\big]e^{-i(E_{p+q}\theta^{oi}p_i+ q_i\theta^{i0}E_{p+q})}   \cr
&\qquad \qquad +\big[i(p^0+q^0)-iE_{p+q}\big]e^{-i(-E_{p+q}\theta^{oi}p_i- q_i\theta^{i0}E_{p+q})}
\big\}\cr
&\qquad\qquad \tilde{\phi}^\dag(p)\tilde{\phi}(q)e^{-i(p+q)x}e^{-\frac{i}{2}(2q_i\theta^{ij}p_j+ (p_i+q_i)\theta^{i0}p_0+ q^0\theta^{0i}(p_i+q_i) }e^{-ip\wedge q}X\Big)|0\rangle \quad + \text{c.t.}\cr
&=2g\frac{1}{(2\pi)^8}\int d^4p ~d^4q ~\frac{1}{2E_{p+q}}~ \cr
&\qquad\langle 0|T\Big(
\big\{-(p^0+q^0)\sin{(E_{p+q}\theta^{oi}p_i+ q_i\theta^{i0}E_{p+q})}\cr
&\qquad \qquad \qquad +E_{p+q}\cos{(E_{p+q}\theta^{oi}p_i+ q_i\theta^{i0}E_{p+q})}\big\}\cr
&\qquad\qquad \tilde{\phi}^\dag(p)\tilde{\phi}(q)e^{-i(p+q)x}e^{-\frac{i}{2}(2q_i\theta^{ij}p_j+ (p_i+q_i)\theta^{i0}p_0+ q^0\theta^{0i}(p_i+q_i) }e^{-ip\wedge q}X\Big)|0\rangle \quad + \text{c.t.} \quad .
\cr
\end{align}
Further simplifications of the overall phase factor are possible, if we note
\beq
e^{-ip \wedge q}=e^{-\frac{i}{2}(p_0\theta^{0i}q_i+p_i\theta^{i0}q_0+p_i\theta^{ij}q_j)}
\eeq
leading us to
\begin{align}
&(\Box_x+m^2)\langle  0|T\big(\sigma(x)X\big)|0\rangle\cr
&=g\frac{1}{(2\pi)^8}\int d^4p ~d^4q ~\langle 0|T\Big(
\big\{-\frac{p^0+q^0}{E_{p+q}}\sin{(E_{p+q}\theta^{oi}p_i+ q_i\theta^{i0}E_{p+q})}\cr
&\hspace{5cm} +\cos{(E_{p+q}\theta^{oi}p_i+ q_i\theta^{i0}E_{p+q})}\big\}\cr
&\qquad\qquad \tilde{\phi}^\dag(p)\tilde{\phi}(q)e^{-i(p+q)x}e^{-\frac{i}{2}(q_i\theta^{ij}p_j+ p_i\theta^{i0}p_0+ q^0\theta^{0i}q_i}X\Big)|0\rangle \quad + \text{c.t.}
\end{align}
which we leave as our final result. We encounter two terms on the right hand side instead of one, as it would be the case if the classical equation of motion was recovered.  They come with modified phase factors, the alterations in there depend on $\theta^{0i}$.
We may check that we obtain the classical equation of motion if $\theta^{0i}$ vanishes. In this case we recognize the first term to vanish, the second simplifies to
\begin{align}
&(\Box_x+m^2)\langle  0|T\big(\sigma(x)X\big)|0\rangle\cr
&=g\frac{1}{(2\pi)^8}\int d^4p ~d^4q ~\langle 0|T\Big( \tilde{\phi}^\dag(p)\tilde{\phi}(q)e^{-i(p+q)x}e^{-iq\wedge p}X\Big)|0\rangle \quad + \text{c.t.}
\end{align}
which is indeed seen to be the desired equation of motion (\ref{quant_eq_mo}), with the fields being expressed by their Fourier transforms.

\subsection{Implication for currents and Ward identities}

The reason for the additional terms arising in the quantized equation of motion is seen to come from the definition of the time-ordering, where the star product is carried out before. From these considerations, it is clear that similar terms will appear in addition to a classical current if considered on the quantized level. However, these calculations are more lengthy, as we have to consider a composed operator at place $x$ instead of a single field. Therefore, we have chosen not to present them here. \\
In  section \ref{BRS} it has been shown how the Slavnov-Taylor identities are easily found from the  BRS current conservation law on the level of Green's functions. It seems obvious that the appearance of additional terms in the BRS current, which will emerge in TOPT,  violates the Slavnov Taylor identities. This is the reason for the violation of Ward identities in TOPT.

\newpage

\section{The violation of remaining Lorentz invariance}

In this section we want to investigate the question of remaining Lorentz invariance if TOPT is applied. We show the violation in a simple scattering process and argue that the general reason lies in the special form of the phase factors in TOPT. Note that the violation only occurs for $\theta^{0i}\neq 0$, as only then this characteristic appearance of the phase factors matters. \\

Though a noncommutative background described by $\theta^{\mu\nu}$ explicitly breaks Lorentz-invariance, we may find a subgroup that leaves it invariant, e.g. in the case of
\beq
\theta^{\mu\nu}= \begin{pmatrix} 0&\theta_e&0&0 \\ -\theta_e&0&0&0 \\ 0&0&0&\theta_m \\ 0&0&-\theta_m&0 \end{pmatrix}
\label{SpecForm}
\eeq

this is $SO(1,1)\times SO(2)\subset \mathcal{L_+^\uparrow}$.
Such a subgroup is therefore expected to be a remaining symmetry group of the theory, such that physical amplitudes should be left invariant under group transformations. However, we will show that this is not the case in TOPT.\\

Let us first demonstrate this in a specific example. We choose a two by two scattering process in $\varphi^3$ theory on tree-level for incoming on-shell momenta $p_1, p_2$ and outgoing momenta again $p_1,p_2$.
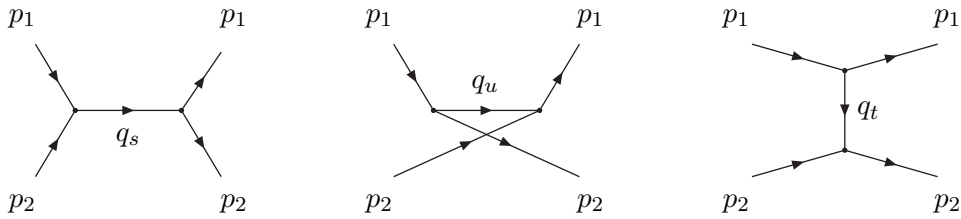
\begin{figure}[h]
\begin{center}
\begin{picture}(300,100)(0,0)
\Vertex(-5,50){1}
\Vertex(35,50){1}
\ArrowLine(-20,75)(-5,50)
\ArrowLine(-20,25)(-5,50)
\ArrowLine(35,50)(50,72)
\ArrowLine(35,50)(50,25)
\ArrowLine(-5,50)(35,50)
\Text(-25,85)[]{$p_1$}
\Text(-25,15)[]{$p_2$}
\Text(55,85)[]{$p_1$}
\Text(55,15)[]{$p_2$}
\Text(15,40)[]{$q_s$}
\Vertex(130,50){1}
\Vertex(170,50){1}
\ArrowLine(115,75)(130,50)
\ArrowLine(115,25)(170,50)
\ArrowLine(170,50)(185,75)
\ArrowLine(130,50)(185,25)
\ArrowLine(130,50)(170,50)
\Text(110,85)[]{$p_1$}
\Text(110,15)[]{$p_2$}
\Text(190,85)[]{$p_1$}
\Text(190,15)[]{$p_2$}
\Text(150,60)[]{$q_u$}
\Vertex(285,65){1}
\Vertex(285,35){1}
\ArrowLine(250,75)(285,65)
\ArrowLine(250,25)(285,35)
\ArrowLine(285,65)(320,75)
\ArrowLine(285,35)(320,25)
\ArrowLine(285,65)(285,35)
\Text(245,85)[]{$p_1$}
\Text(245,15)[]{$p_2$}
\Text(325,85)[]{$p_1$}
\Text(325,15)[]{$p_2$}
\Text(295,50)[]{$q_t$}
\end{picture}
\parbox{12cm}{\caption{\label{scatt}  A scattering process in $\varphi^3$ theory: s-, u- and t-channel}}
\end{center}
\end{figure}
According to TOPT, the amplitude is diagrammatically given by the graphs in Fig. \ref{scatt} and reads
\begin{align}
i\mathcal{M}&=i\mathcal{M}_s+i\mathcal{M}_u+i\mathcal{M}_t \cr
i\mathcal{M}_s&=g^2\sum_{\lambda=\pm 1}\frac{1}{2E_{q_s}}\frac{\lambda}{q^0_s-\lambda(E_{q_s}-i\epsilon)}V(p_1,p_2,-q_s^\lambda)^2\Big|_{q_s=p_1+p_2}  \cr
i\mathcal{M}_u&=g^2\sum_{\lambda=\pm 1}\frac{1}{2E_{q_u}}\frac{\lambda}{q^0_u-\lambda(E_{q_u}-i\epsilon)}V(p_1,p_2,-q_u^\lambda)^2\Big|_{q_u=p_1-p_2} \cr
i\mathcal{M}_t&=g^2\sum_{\lambda=\pm 1}\frac{1}{2E_{q_t}}\frac{\lambda}{q^0_t-\lambda(E_{q_t}-i\epsilon)}V(p_1,p_2,-q_t^\lambda)^2\Big|_{q_t=0}
\label{iM}
\end{align}
where
\begin{align}
E_q&=\sqrt{m^2+\underline{q}^2} \cr
q^\lambda&=(\lambda E_q,\underline{q}) \cr
V(p_1,p_2,p_3)&= \frac{1}{6}\sum_{\pi\epsilon S_3} e^{-i(p_{\pi(1)}, p_{\pi(2)}, p_{\pi(3)})}
\end{align}
and $g$ is the coupling constant. \newline

Now, we will choose specific $p_1, p_2$ and $\theta^{\mu\nu}$ of type (\ref{SpecForm}) in \emph{frame 1}, calculate $i\mathcal{M}=i\mathcal{M}_s+i\mathcal{M}_u+i\mathcal{M}_t$ there and compare to $i\mathcal{M}'$ which we compute in \emph{frame 2} being related to \emph{frame 1} by the transformation
\beq
G=\begin{pmatrix} \cosh\beta&\sinh\beta&0&0 \\ \sinh\beta&\cosh\beta&0&0\\ 0&0&1&0\\ 0&0&0&1 \end{pmatrix}
\quad \in\quad SO(1,1)\times SO(2) \quad .
\eeq

The following configuration is chosen in \emph{frame 1:}
\begin{align}
p_1&=(E_p,0,0,p)\quad, \qquad E_p=\sqrt{m^2+p^2} \cr
p_2&=(E_p,0,0,-p) \cr
\theta^{\mu\nu}&=\theta_e\begin{pmatrix} 0&1&0&0 \\ -1&0&0&0 \\ 0&0&0&0 \\ 0&0&0&0 \end{pmatrix} \cr
q_s&=p_1+p_2=(2E_p,0,0,0) \cr
q_s^\lambda&=(\lambda m,0,0,0) \cr
q_u&=p_1-p_2=(0,0,0,2p) \cr
q_u^\lambda&=(\lambda E_{2p},0,0,2p)\quad ,\qquad E_{2p}=\sqrt{m^2+(2p)^2} \cr
q_t&=p_1-p_1=(0,0,0,0) \cr
q_t^\lambda&=(\lambda m,0,0,0)
\end{align}
implying the configuration in \emph{frame 2:}
\begin{align}
p_1'&=(E_p\cosh\beta,E_p\sinh\beta,0,p)  \cr
p_2'&=(E_p\cosh\beta,E_p\sinh\beta,0,-p) \cr
\theta^{'\mu\nu}&=\theta^{\mu\nu}=\theta_e\begin{pmatrix} 0&1&0&0 \\ -1&0&0&0 \\ 0&0&0&0 \\ 0&0&0&0 \end{pmatrix} \cr
q'_s&=p_1'+p_2'=(2E_p\cosh\beta,2E_p\sinh\beta,0,0) \cr
(q'_s)^\lambda &= (\lambda E_{q'_s}, 2E_p\sinh\beta,0,0)\quad, \quad E_{q'_s}=\sqrt{m^2+4E_p^2\sinh^2\beta} \cr
q'_u&=p_1'-p_2'=(0,0,0,2p) \cr
(q'_u)^\lambda&=(\lambda E_{2p},0,0,2p)\cr
q'_t&=p_1'-p_1'=(0,0,0,0) \cr
(q'_t)^\lambda&=(\lambda m,0,0,0)
\qquad .
\end{align}
To compute $i\mathcal{M}$ in \emph{frame 1} we first notice that $\theta^{\mu\nu}$ does not appear in the noncommutative phase factor:
\begin{align}
p_1\wedge p_2=p_1\wedge q_j^\lambda&=p_2\wedge q_j^\lambda=0 \cr
\Longrightarrow\quad  V(p_1,p_2,-q_j^\lambda)&=1 \qquad \text{for}\quad j=s,u,t \nonumber
\end{align}
allowing us to easily carry out the summation over $\lambda$ in (\ref{iM})
\begin{align}
i\mathcal{M}_s&=g^2\sum_{\lambda=\pm 1}\frac{1}{2E_{q_s}}\frac{\lambda}{q_s^0-\lambda(E_{q_s}-i\epsilon)}\Big|_{q_s=p_1+p_2} \cr
&=g^2\frac{1}{2E_{q_s}}\big(\frac{1}{q_s^0-(E_{q_s}-i\epsilon)}+\frac{-1}{q_s^0+(E_{q_s}-i\epsilon)}\big)\Big|_{q_s=p_1+p_2} \cr
&=g^2\frac{1}{2E_{q_s}}\frac{2E_{q_s}}{(q_s^0)^2-E_{q_s}^2+i\epsilon}\big)\Big|_{q_s=p_1+p_2} \cr
&=g^2\frac{1}{(p_1+p_2)^2-m^2}
\end{align}
and analogously
\begin{align}
i\mathcal{M}_u&=g^2\frac{1}{(p_1-p_2)^2-m^2} \\
 i\mathcal{M}_t&=-g^2\frac{1}{m^2} \quad .
\end{align}
In \emph{frame 2} we start again with the computation of the phase factor:
\begin{align}
p_1'\wedge p_2'&=\frac{1}{2}\theta_e E_p\cosh\beta\cdot E_p\sinh\beta-\frac{1}{2}\theta_e E_p\sinh\beta \cdot E_p\cosh\beta=0 \cr
p_1'\wedge(q'_s)^\lambda&=\frac{1}{2}\theta_e E_p\cosh\beta\cdot 2E_p\sinh\beta -\frac{1}{2}\theta_e E_p\sinh\beta\cdot\lambda E_{q'_s}\cr
&=\theta_e E_p^2\sinh\beta\cosh\beta-\frac{1}{2}\theta_e\lambda E_p E_{q'_s}\sinh\beta\cr
p_2'\wedge(q'_s)^\lambda&=\frac{1}{2}\theta_e E_p \cosh\beta\cdot 2E_p\sinh\beta -\frac{1}{2}\theta_e E_p\sinh\beta\cdot\lambda E_{q'_s}\cr
&=\theta_e E_p^2\sinh\beta\cosh\beta-\frac{1}{2}\theta_e\lambda E_p E_{q'_s}\sinh\beta\cr
&=p_1'\wedge(q_s')^\lambda \cr
p_1\wedge(q'_u)^\lambda&=-\frac{1}{2}\lambda\theta_e E_pE_{2p}\sinh\beta \cr
p_2\wedge(q'_u)^\lambda&=-\frac{1}{2}\lambda\theta_e E_pE_{2p}\sinh\beta \cr
&=p_1\wedge(q'_u)^\lambda \cr
p_1\wedge(q'_u)^\lambda&=-\frac{1}{2}\lambda\theta_e E_pm\sinh\beta \cr
p_2\wedge(q'_t)^\lambda&=-\frac{1}{2}\lambda\theta_e E_pm\sinh\beta \cr
&=p_1\wedge(q'_t)^\lambda
\nonumber
\end{align}
and thus
\begin{align}
V(p_1',p_2',-q_s'^\lambda)&=\frac{1}{6}\Big(e^{-i2p_1'\wedge(q'_s)^\lambda}+e^{-i0}+e^{-i2p_1'\wedge(q'_s)^\lambda}\cr
&\qquad +e^{-i0}+e^{i2p_1'\wedge(q'_s)^\lambda}+e^{i2p_1'\wedge(q'_s)^\lambda}\Big)\cr
&=\frac{1}{3}+\frac{2}{3}\cos(2p_1'\wedge(q'_s)^\lambda)\cr
&=\frac{1}{3}+\frac{2}{3}\cos(2\theta_e E_p^2\sinh\beta\cosh\beta-\lambda\theta_e E_p
 E_{q'_s}\sinh\beta) \cr
V(p_1',p_2',-q_u'^\lambda)&=
\frac{1}{3}+\frac{2}{3}\cos(2p_1'\wedge(q'_u)^\lambda) \cr
&=\frac{1}{3}+\frac{2}{3}\cos(\lambda\theta_e E_p E_{2p}\sinh\beta)
\end{align}
which we will now expand to $2^{nd}$ order in $\theta_e$:
\begin{align}
V(p_1',p_2',-q'^\lambda_s)&=1+\frac{2}{3}(-\frac{1}{2})\theta_e^2\big(2E_p^2\sinh\beta\cosh\beta-\lambda E_pE_{q'_s}\sinh\beta\big)^2 +o(\theta_e^2)\cr
&=1-\frac{1}{3}\theta_e^2\big(4E_p^4\sinh^2\beta\cosh^2\beta-4\lambda E_p^3E_{q'_s}\sinh^2\beta\cosh\beta \cr
& \hspace{6cm} +E_p^2E_{q'_s}^2\sinh^2\beta\big)+o(\theta_e^2)
\end{align}
\begin{align}
V^2(p_1',p_2',-q'^\lambda_s)&=1-\frac{2}{3}\theta_e^2\big(4E_p^4\sinh^2\beta\cosh^2\beta-4\lambda E_p^3E_{q'_s}\sinh^2\beta\cosh\beta \cr
& \hspace{6cm} +E_p^2E^2_{q'_s}\sinh^2\beta\big)+o(\theta_e^2) \cr
&=1-\frac{2}{3}\theta_e^2\big(4E_p^4\sinh^2\beta\cosh^2\beta+E_p^2E^2_{q'_s}\sinh^2\beta\big) \cr
& \hspace{4cm}+\frac{8}{3}\lambda\theta_e^2E_p^3E_{q'_s}\sinh^2\beta\cosh\beta+o(\theta_e^2) \cr
&=V^a_s+\lambda V^b_s +o(\theta_e^2)
\end{align}
where we introduced $V^a_s, V^b_s$ to shorten our expressions
\begin{align}
V^a_s&=1-\frac{2}{3}\theta_e^2\big(4E_p^4\sinh^2\beta\cosh^2\beta+E_p^2E^2_{q'_s}\sinh^2\beta\big)\cr
V^b_s&=\frac{8}{3}\theta_e^2E_p^3E_{q'_s}\sinh^2\beta\cosh\beta \quad ,
\end{align}
note that both $V^a_s$ and $V^b_s$ do not depend on $\lambda$.\\
The expressions for the u-channel are slightly simpler:
\begin{align}
V(p_1',p_2',-q'^\lambda_u)&=1+\frac{2}{3}(-\frac{1}{2})\theta_e^2\big(\lambda E_p E_{2p}\sinh\beta \big)^2 +o(\theta_e^2)\cr
&=1-\frac{1}{3}\theta_e^2 E_p^2 E_{2p}^2\sinh^2\beta +o(\theta_e^2)\cr
V^2(p_1',p_2',-q'^\lambda_u)&=1-\frac{2}{3}\theta_e^2 E_p^2 E_{2p}^2\sinh^2\beta +o(\theta_e^2)\cr
&=V^a_u+\lambda V^b_u +o(\theta_e^2)
\end{align}
with
\begin{align}
V^a_u&=1-\frac{2}{3}\theta_e^2 E_p^2 E_{2p}^2\sinh^2\beta \cr
V^b_u&=0
\end{align}
from where we obtain the expression for the t-channel by replacing $E_{2p}$ by $m$:
\begin{align}
V^2(p_1',p_2',-q'^\lambda_t)&=V^a_t+\lambda V^b_t +o(\theta_e^2) \cr
V^a_t&=1-\frac{2}{3}\theta_e^2 E_p^2 m^2\sinh^2\beta \cr
V^b_t&=0 \qquad .
\end{align}
We are now able to simplify $i\mathcal{M}'_j$ ($j=s,u,t$) as follows:
\begin{align}
i\mathcal{M'}_j=~&g^2\sum_{\lambda=\pm 1}\frac{1}{2E_{q'_j}}\frac{\lambda}{q'^0_j-\lambda(E_{q'_j}-i\epsilon)}V(p'_1,p'_2,-(q'_j)^\lambda)^2 \cr
=~&g^2\sum_{\lambda=\pm 1}\frac{1}{2E_{q'_j}}\frac{\lambda}{q'^0_j-\lambda(E_{q'_j}-i\epsilon)}V^a_j \cr
&+\sum_{\lambda=\pm 1}\frac{1}{2E_{q'_j}}\frac{1}{q'^0_j-\lambda(E_{q'_j}-i\epsilon)}V^b_j +o(\theta_e^2)\cr
=~&g^2\frac{1}{2E_{q'_j}}\Big(\frac{1}{q'^0_j-(E_{q'_j}-i\epsilon)}+\frac{-1}{q'^0_j+(E_{q'_j}-i\epsilon)}\Big)V^a_j \cr
&+\frac{1}{2E_{q'_j}}\Big(\frac{1}{q'^0_j-(E_{q'_j}-i\epsilon)}+\frac{1}{q'^0_j+(E_{q'_j}-i\epsilon)}\Big)V^b_j +o(\theta_e^2)\cr
=~&g^2\frac{1}{(q'^0_j)^2-E_{q'_j}^2+i\epsilon}V^a_j
+\frac{1}{2E_{q'_j}}\frac{2q'^0_j}{(q'^0_j)^2-E_{q'_j}^2+i\epsilon}V^b_j +o(\theta_e^2)\cr
\end{align}
so for the s-channel:
\begin{align}
i\mathcal{M'}_s=~&g^2\frac{1}{q'^2_s-m^2}V^a_s
+\frac{1}{2E_{q'_s}}\frac{2E_p\cosh\beta}{q'^2_s-m^2}V^b_s +o(\theta_e^2)\cr
=~&g^2\frac{1}{(p_1+p_2)^2-m^2}\Big(V^a_s+\frac{2E_p\cosh\beta}{2E_{q'_s}}V^b_s\Big)+o(\theta_e^2)\cr
=~&g^2\frac{1}{(p_1+p_2)^2-m^2}\Big(1-\frac{2}{3}\theta_e^2\big(4E_p^4\sinh^2\beta\cosh^2\beta+E_p^2E_{q'_s}^2\sinh^2\beta\big) \cr
& \hspace{6cm} +\theta_e^2\frac{8}{3}E_p^4\sinh^2\beta\cosh^2\beta\Big)+o(\theta_e^2)\cr
=~&g^2\frac{1}{(p_1+p_2)^2-m^2}\Big(1-\frac{2}{3}\theta_e^2E_p^2E_{q'_s}^2\sinh^2\beta\Big)+o(\theta_e^2)\cr
=~&g^2\frac{1}{(p_1+p_2)^2-m^2} \cr
&-\frac{2}{3}g^2\theta_e^2\frac{1}{(p_1+p_2)^2-m^2}E_p^2\big(m^2+4E_p^2\sinh^2\beta\big)\sinh^2\beta +o(\theta_e^2)\cr
\end{align}
and for the u-channel:
\begin{align}
i\mathcal{M'}_u=~&g^2\frac{1}{q'^2_u-m^2}V^a_u +o(\theta_e^2)\cr
=~&g^2\frac{1}{(p_1-p_2)^2-m^2}-\frac{2}{3}g^2\theta_e^2\frac{1}{(p_1-p_2)^2-m^2} E_p^2 E_{2p}^2\sinh^2\beta +o(\theta_e^2)
\end{align}
from where we easily obtain the t-channel:
\begin{align}
i\mathcal{M'}_t&=-g^2\frac{1}{m^2}+\frac{2}{3}g^2\theta_e^2\frac{1}{m^2}E_p^2 m^2\sinh^2\beta +o(\theta_e^2)\quad .
\end{align}
We read off the differences in the amplitudes calculated in the two frames:
\begin{align}
i\mathcal{M}_s-i\mathcal{M}'_s&=\frac{2}{3}g^2\theta_e^2\frac{1}{(p_1+p_2)^2-m^2}E_p^2\big(m^2+4E_p^2\sinh^2\beta\big)\sinh^2\beta+o(\theta_e^2) \cr
i\mathcal{M}_u-i\mathcal{M}'_u&=\frac{2}{3}g^2\theta_e^2\frac{1}{(p_1-p_2)^2-m^2} E_p^2 E_{2p}^2\sinh^2\beta+o(\theta_e^2)\cr
i\mathcal{M}_t-i\mathcal{M}'_t&=-\frac{2}{3}g^2\theta_e^2 E_p^2 \sinh^2\beta+o(\theta_e^2)
\end{align}
which are nonvanishing for $p,\theta_e,\beta \neq 0$. \newline
The complete amplitude differs as follows:
\begin{align}
i\mathcal{M}-i\mathcal{M}'&=\frac{2}{3}g^2\theta_e^2\frac{1}{(p_1+p_2)^2-m^2}E_p^2\big(m^2+4E_p^2\sinh^2\beta\big)\sinh^2\beta \cr
& \quad +\frac{2}{3}g^2\theta_e^2\frac{1}{(p_1-p_2)^2-m^2} E_p^2 E_{2p}^2\sinh^2\beta\cr
& \quad -\frac{2}{3}g^2\theta_e^2 E_p^2\sinh^2\beta+o(\theta_e^2)\cr
&=\frac{2}{3}g^2\theta_e^2 E_p^2\sinh^2\beta\Big(\frac{m^2+4E_p^2\sinh^2\beta}{3m^2+4p^2}-\frac{m^2+4p^2}{4p^2+m^2}-1\Big)+o(\theta_e^2)\cr
&=\frac{2}{3}g^2\theta_e^2 E_p^2\sinh^2\beta\Big(\frac{m^2+4E_p^2\sinh^2\beta}{3m^2+4p^2}-2\Big)+o(\theta_e^2)\\
&< 0 \qquad \quad \text{for}\quad p,\theta_e,\beta \neq 0 \quad \text{and} ~ \beta ~\text{sufficiently small.} \nonumber
\end{align}

The last inequality is easily verified if we notice that for small enough $\beta$ the term $\frac{m^2+4E_p^2\sinh^2\beta}{3m^2+4p^2}$ is less or equal $\frac{1}{3}$.\\

The origin of the above demonstrated violation of remaining symmetry in TOPT is the following: an amplitude calculated according to this prescription exhibits noncommutative phase factors depending on $p_i$ and $q_j^{\lambda_j}$, if $p_i$ are the external, $q_j$ the internal momenta and $\lambda_j=\pm 1$. More precisely, these phase factors are functions of the complex numbers $p_i\wedge p_j, p_i\wedge q_j^{\lambda_j}$ and $ q_i^{\lambda_i}\wedge q_j^{\lambda_j}$. Under a transformation that leaves $\theta^{\mu\nu}$ unchanged and takes $p_i \rightarrow p_i',~ q_j \rightarrow q_j'$ we have
\begin{eqnarray}
p_i\wedge p_j \rightarrow& p_i'\wedge p_j'&=p_i\wedge p_j \cr
p_i\wedge q_j^{\lambda_j} \rightarrow &p_i'\wedge (q_j')^{\lambda_j}& \neq p_i'\wedge (q_j^{\lambda_j})'\hspace{6mm}=p_i\wedge q_j^{\lambda_j} \cr
q_i^{\lambda_i}\wedge q_j^{\lambda_j} \rightarrow &(q_i')^{\lambda_i}\wedge (q_j')^{\lambda_j} &\neq (q_i^{\lambda_i})'\wedge (q_j^{\lambda_j})'=q_i^{\lambda_i}\wedge q_j^{\lambda_j}
\end{eqnarray}
where the inequalities in the last two lines arise because the internal momenta $q_i$ are in general \emph{not} on-shell and therefore $(q_i')^{\lambda_i}\neq (q_i^{\lambda_i})'$. This means that the noncommutative phase factor is not left invariant by the transformation and can lead, as demonstrated above, to different amplitudes.

\newpage

\newpage
\thispagestyle{plain}
\mbox{}       

\chapter{The formulation of perturbation theory using retarded functions}

The formulation of quantized field theories with the help of retarded functions instead of time-ordered products was introduced in \cite{LSZ2} and further elaborated in \cite{Symanzik}, a nice presentation can also be found in \cite{Rzewuski}. In the following, we first present this approach for commutative theories, which we will do in such a way that the alterations for the noncommutative case can be easily seen. We discuss unitarity as well as equations of motion and currents, where we will face difficulties for $\theta^{0i}\neq 0$. These are analyzed and shown to come from a certain type of diagrams. The latter point suggests a modified theory which is unitary and preserves the classical equations of motion and currents on the quantized level.

\section{Introduction to retarded functions in commutative theories}

\subsection{Retarded functions and the generating functional}

Consider a field theory with a single hermitian field $\phi$ of mass $m$. The retarded products are then given by retarded multiple commutators of $\phi$:
\beq
R(x;x_1...x_n)= (-i)^n\sum_{\text{perm}}\vartheta(x^0-x^0_1)...\vartheta(x^0_{n-1}-x^0_n)\big[...[\phi(x),\phi(x_1)]...\phi(x_n)\big]
\eeq
where the summation is taken over all permutations of the $n$ coordinates $x_i$, $\vartheta$ denotes the step function.
The retarded functions are now defined as the vacuum expectation values of the retarded products:
\beq
r(x;x_1...x_n)=\langle 0|R(x;x_1...x_n)|0\rangle
\eeq
and can be generated from the functional
\begin{align}
\mathcal{R}[j',j]=&\exp\Big\{2\int dx\sin\Big(\frac{1}{2}\frac{\delta}{\delta j(x)}\frac{\delta}{\delta\frac{\delta}{\delta j'(x)}}\Big)I\big[\frac{\delta}{\delta j'}\big]\Big\}\times \cr
&  \quad \exp\Big\{\int dydz\Big(\frac{1}{4}j'(y)\Delta^{(1)}(y-z)j'(z)-j'(y)\Delta^{ret}(y-z)j(z)\Big)\Big\}
\label{ret_func}
\end{align}
by means of functional differentiation:
\beq
r(x;x_1...x_n)=\frac{\delta}{\delta j'(x)}\frac{\delta^n}{\delta j(x_1)...\delta j(x_n)}\mathcal{R}[j',j]\Big|_{j'=j=0}    \qquad . \label{ret_fcts}
\eeq

We need to comment on some expressions appearing in equation (\ref{ret_func}).

With $I$ we denoted the interaction functional, e.g in $\phi^3$-theory this would be \\
$I[q]= \frac{g}{3!}\int dx~ q^3(x)$.

$\Delta^{ret}$ is a Green's function to the Klein-Gordon equation
\beq
\Delta^{ret}(x)=\lim_{\epsilon\rightarrow +0} \frac{-1}{(2\pi)^4}\int d^4k\frac{e^{-ikx}}{(k+i\epsilon)^2-m^2}
\eeq
with the support property $\Delta^{ret}(x)=0$ for $x_0<0$.
$\Delta^{(1)}$ is given by
\beq
\Delta^{(1)}(x)=\frac{1}{(2\pi)^3}\int d^4k~ \delta(k^2+m^2)e^{-ikx}
\eeq
 being a solution to the homogeneous Klein-Gordon equation: $(\Box+m^2)\Delta^{(1)}(x)=0$.
\newline

The S-matrix can be constructed with the help of retarded functions as done by H. Lehmann, K. Symanzik and W. Zimmermann in \cite{LSZ2}. The reduction formula in the case of a scattering process with $n$ incoming particles of momenta $p_1,...,p_n$ and two outgoing particles of momenta $q_1,q_2$ reads
\begin{align}
S(p_1,...,p_n;q_1,q_2)=&\frac{-i}{(2\pi)^{\frac{3}{2}(n+2)}}\int dx_1...dx_n dy_1dy_2\exp\big\{i(\sum_{i=1}^np_ix_i-\sum_{i=1}^2q_iy_i)\big\}\times \cr
&\quad \times K_{x_1}...K_{x_n}K_{y_1}K_{y_2}r(y_2;y_1x_1...x_n)  \label{LSZ2}
\end{align}
where $K_x=\Box_x+m^2$.

 \subsection{Diagrammatic rules \label{dia1}}

Computing retarded functions by using equation (\ref{ret_fcts}) on recognizes that one can cast the outcome in the form of diagrams. Its lines will obviously carry $\Delta^{ret}$ or $\Delta^{(1)}$, and for\\
 $r(x;x_1,...,x_n)$ there will be endpoints $x, x_1,...,x_n$.\newline

To see which diagrams are allowed according to (\ref{ret_fcts}), we expand the first exponential in (\ref{ret_func}) in the example of $I[q]=gq^m$:
\begin{align}
\mathcal{R}[j',j]=1+&\sum_{n=1}^\infty \frac{1}{n!}\prod_{i=1}^n\int dy_i~2\sin\Big(\frac{1}{2}\frac{\delta}{\delta j(y_i)}\frac{\delta}{\delta\frac{\delta}{\delta j'(y_i)}}\Big)\int dz_i~g^m\frac{\delta^m}{\delta j'(z_i)^m} \times \cr
& \times\exp\Big\{\int dydz\Big(\frac{1}{4}j'(y)\Delta^{(1)}(y-z)j'(z)-j'(y)\Delta^{ret}(y-z)j(z)\Big)\Big\} \quad .
\end{align}
Recalling that
\beq
r(x;x_1...x_n)=\frac{\delta}{\delta j'(x)}\frac{\delta^n}{\delta j(x_1)...\delta j(x_n)}\mathcal{R}[j',j]\Big|_{j'=j=0}    \qquad .
\eeq
we see that $x$ is connected by $\Delta^{ret}(x-a)$ or $\Delta^{(1)}(z-a)=\Delta^{(1)}(a-z)$, the points $x_i$ are connected by $\Delta^{ret}(a_i-x_i)$; $a, a_i$ being some inner or outer points.

The $\frac{\delta}{\delta\frac{\delta}{\delta j'(y_i)}}$ in the $\sin$ can only act on $\frac{\delta^m}{\delta j'(z_i)^m}$, such that by expanding $\sin$ we can make the replacement
\begin{align}
&\int dy_i~2\sin\Big(\frac{1}{2}\frac{\delta}{\delta j(y_i)}\frac{\delta}{\delta\frac{\delta}{\delta j'(y_i)}}\Big)\int dz_i~g^m\frac{\delta^m}{\delta j'(z_i)^m} \cr
\equiv&~
2\sum_{j\leq [\frac{m-1}{2}]}g^m\int dz_i\frac{1}{(2j+1)!}\big(\frac{1}{2}\big)^{2j+1}\frac{\delta^{2j+1}}{\delta j(z_i)^{2j+1}}\frac{\delta^{m-2j-1}}{\delta j'(z_i)^{m-2j-1}}
\end{align}
such that at the vertex $z_i$ we have an odd power of $\frac{\delta}{\delta j(z_i)}$. As an incoming $\Delta^{ret}(a-z_i)$ at vertex $z_i$ can only be created by $\frac{\delta}{\delta j(z_i)}$ and vice versa, we find that the number of incoming $\Delta^{ret}$-functions at each vertex must be odd.\newline

One checks that there are no further restrictions to diagrams as the ones mentioned above, so we have found the diagrammatic rules for the retarded function $r(x;x_1...x_n)$:
\begin{enumerate}
\item $x,x_1,...,x_n$ are the endpoints of the diagram, inner points are called vertices.
\item $\Delta^{ret}(x-y)$ is symbolized by
\begin{picture}(50,10)(0,0)
\Vertex(5,5){1}
\Vertex(45,5){1}
\ArrowLine(5,5)(45,5)
\Text(5,-2)[]{$x$}
\Text(45,-2)[]{$y$}
 \end{picture},$\quad$
  $\Delta^{(1)}(x-y)=\Delta^{(1)}(y-x)$ by
\begin{picture}(50,10)(0,0)
\Vertex(5,5){1}
\Vertex(45,5){1}
\Line(5,5)(45,5)
\Text(5,-2)[]{$x$}
\Text(45,-2)[]{$y$}
 \end{picture} \quad .
\item $x$ is connected by one line, $\Delta^{ret}(x-a)$ or $\Delta^{(1)}(x-a)$. The points $x_i$ are also connected by one line each, $\Delta^{ret}(a_i-x_i)$.
\item The number of lines at each vertex is $m$ for $\phi^m$-theory, the contributing factor g, one integrates over the vertices.
\item The number of incoming functions $\Delta^{ret}(a_i-z_i)$ at each vertex $z_i$ is odd.
\item When put into the reduction formula (\ref{LSZ2}) only diagrams contribute where $x$ is connected by $\Delta^{ret}(x-a)$.
\end{enumerate}
The last rule follows from $K_x\Delta^{(1)}(x-a)=0$.

\subsection{The tree level}

What is the relation between the standard perturbation theory for the S-matrix, in terms of diagrams with ordinary propagators, and the above one? In the following we will show that they agree on tree-level. \newline

For this purpose we first proof the following
\begin{prop}
 On tree level all diagrams contributing to $r(x;x_1...x_n)$ only contain the function $\Delta^{ret}$ and  no function $\Delta^{(1)}$.
\end{prop}
\emph{Proof:} Denote the number of vertices of a contributing diagram by $V$ and its number of internal lines, i.e. those not connected to $x,x_1,...,x_n$, by $I$. To satisfy rule 5, we need at least V functions $\Delta^{ret}$. At least V-1 of them must be internal ones, as all external except the one at $x$ are directed to the outer points.
But on tree-level we have the relation $V=I+1$ which follows by induction over $V$: for $V=1$ we have $I=0$ and for $V\rightarrow V+1$ we find $I\rightarrow I+1$. So rule 5 can only be fulfilled if every internal line carries the function $\Delta^{ret}$, and if $x$ is also connected by $\Delta^{ret}$.
\newline

To see that this gives the standard result when put into the reduction formula (\ref{LSZ2}), we notice that $\Delta^{ret}$ and $\Delta_C$ coincide if no singularities appear. We do not expect singularities at the tree-level and find the desired result.

\subsection{Unitarity in terms of the generating functional}

This analysis of unitarity follows closely the presentation in \cite{Rzewuski}, and will be of fundamental importance when we come to noncommutative theories.

Rzewuski derives the generalized unitarity condition
\beq
\mathcal{R}[0,j]=1
\eeq
which implies unitarity for the S-matrix \cite{Rzewuski}.
To verify that this condition is fulfilled for the functional (\ref{ret_func}) in the case of $\phi^m$-theory we perform again a Taylor expansion of the first exponential:
\begin{align}
\mathcal{R}[0,j]=1+&\sum_{n=1}^\infty \frac{1}{n!}\prod_{i=1}^n\int dy_i2\sin\Big(\frac{1}{2}\frac{\delta}{\delta j(y_i)}\frac{\delta}{\delta\frac{\delta}{\delta j'(y_i)}}\Big)\int dz_ig^m\frac{\delta^m}{\delta j'(z_i)^m} \times \cr
& \times\exp\Big\{\int dydz\Big(\frac{1}{4}j'(y)\Delta^{(1)}(y-z)j'(z)-j'(y)\Delta^{ret}(y-z)j(z)\Big)\Big\}\Bigg|_{j'=0}
\end{align}
Each factor in the n-th term ($n \geq1$) of the sum contains at least one functional derivative $\frac{\delta}{\delta (z_i)}$ such that we obtain $\prod_{i=1}^n \int dx_i j'(x_i)\Delta^{ret}(x_i-z_i)$ in front of the exponential, which does not vanish for $j'=0$ only if every factor is differentiated with some $\frac{\delta}{\delta j(z_j)}$. This means that at each vertex $z_i$ we have an ending $\Delta^{ret}(a-z_i)$, and the point $a$  must be again out of the $\{z_i\}_{i=1}^n$, which implies that we have a closed cycle of $\Delta^{ret}$-functions, i.e. an expression of the form $\Delta^{ret}(z_{i_1}-z_{i_2})\Delta^{ret}(z_{i_2}-z_{i_3})...\Delta^{ret}(z_{i_k}-z_{i_1})$ with $i_1,i_2,...,i_k\in \{1,...,n\}$.

The last statement can be seen as follows: choose $z_{i_1}$, which appears in a function $\Delta^{ret}(a-z_{i_1}), a$ among the $z_i$'s, say $a=z_{i_2}$. Either $z_{i_2}= z_{i_1}$ and we have found a closed cycle, or $z_{i_2}\neq z_{i_1}$ in which case we proceed  by finding $z_{i_3}$ such that $\Delta^{ret}(z_{i_3}-z_{i_2})$ appears. In the case $z_{i_3}= z_{i_1}$ or $z_{i_3}= z_{i_2}$ we are finished, otherwise we go on in the same way. The limited number of points $\{z_i\}_{i=1}^n$ implies that the procedure will stop and yield a closed cycle of $\Delta^{ret}$-functions.

It follows from the support properties of the $\Delta^{ret}$-function that a closed cycle of them vanishes, such that every term in the above sum except the first one is zero, giving us the fulfilled unitarity condition $\mathcal{R}[0,j]=1$.

\subsection{Composite operators: equations of motion and currents \label{comp}}

To derive equations of motion and current conservation laws on the level of Green's functions, we define retarded functions $r^\mathcal{O}(x;x_1...x_n)$ for a composite operator $\mathcal{O}$ at place $x$ and single fields at $x_1,...,x_n$ in the following way. We differentiate the generating functional by $\frac{\delta}{\delta j'(x)}$ once for every single field appearing in  $\mathcal{O}$ and by $\frac{\delta^n}{\delta j(x_1)...\delta j(x_n)}$. For $\mathcal{O}$ in the form $\mathcal{O}=D_1\phi D_2\phi...D_k\phi$ with $D_i$ differential operators this means
\beq
r^{D_1\phi D_2\phi...D_k\phi}(x;x_1...x_n) \equiv D_1\frac{\delta}{\delta j'(x)}D_2\frac{\delta}{\delta j'(x)}...D_k\frac{\delta}{\delta j'(x)}\frac{\delta^n}{\delta j(x_1)...\delta j(x_n)}\mathcal{R}[j',j]\Big|_{j'=j=0}
\eeq
e.g.
\beq
r^{\phi (\Box+m^2)\phi}(x;x_1...x_n) \equiv \frac{\delta}{\delta j'(x)}(\Box_x+m^2)\frac{\delta}{\delta j'(x)}\frac{\delta^n}{\delta j(x_1)...\delta j(x_n)}\mathcal{R}[j',j]\Big|_{j'=j=0} \quad .
\eeq

Diagrammatic rules for $r^\mathcal{O}(x;x_1...x_n)$ with $\mathcal{O}=D_1\phi D_2\phi...D_k\phi$  can be easily read off, the only change to the previous rules lies in how the point $x$ is treated, we therefore replace rule 3 by
\begin{itemize}
\item[3'.] $x$ is connected by $k$ lines; the $i^{\text{th}}$ line carries $D_i\Delta^{ret}(x-a_i)$ or $D_i\Delta^{(1)}(x-a_i)$. The points $x_i$ are connected by one line each, $\Delta^{ret}(b_i-x_i)$.
\end{itemize}

As an example of equations of motion and current conservation laws we now want to prove the bilinear equation of motion in $\phi^3$-theory ($I[q]=\frac{g}{3!}q^3$) on the level of retarded functions, i.e. show that
\beq
r^{\phi (\Box+m^2)\phi}(x;x_1...x_n)=r^{g\phi^3}(x;x_1...x_n)+ \text{c.t.}   \label{eqmo}
\eeq
with c.t. meaning contact terms.
We will evaluate both sides of the above equation diagrammatically:
\begin{align}
r^{\phi (\Box+m^2)\phi}(x;x_1...x_n)=&
\begin{picture}(100,50)(0,40)
\Vertex(50,85){1}
\Vertex(30,0){1}
\Vertex(70,0){1}
\GOval(50,42.5)(17,30)(0){0.75}
\DashArrowLine(50,85)(30,55){2}
\DashArrowLine(50,85)(70,55){2}
\ArrowLine(30,30)(30,0)
\ArrowLine(70,30)(70,0)
\Text(50,95)[]{$x$}
\Text(30,-10)[]{$x_1$}
\Text(50,-10)[]{...}
\Text(70,-10)[]{$x_n$}
\Text(67,75)[l]{$(\Box+m^2)$}
\end{picture}
 \cr \cr \cr \cr
=&\int dy
\begin{picture}(100,70)(10,40)
\Vertex(50,85){1}
\Vertex(30,0){1}
\Vertex(70,0){1}
\Vertex(60,70){1}
\GOval(50,42.5)(17,30)(0){0.75}
\DashArrowLine(50,85)(30,55){2}
\ArrowLine(50,85)(60,70)
\ArrowLine(30,30)(30,0)
\ArrowLine(70,30)(70,0)
\DashLine(60,70)(50,60){2}
\DashLine(60,70)(70,55){2}
\Text(50,95)[]{$x$}
\Text(70,70)[]{$y$}
\Text(30,-10)[]{$x_1$}
\Text(50,-10)[]{...}
\Text(70,-10)[]{$x_n$}
\Text(60,85)[l]{$(\Box+m^2)$}
\end{picture} +
\quad \sum_{k=1}^n
\begin{picture}(100,70)(0,40)
\Vertex(65,85){1}
\Vertex(30,0){1}
\Vertex(70,0){1}
\Vertex(110,0){1}
\GOval(50,42.5)(17,30)(0){0.75}
\DashArrowLine(65,85)(30,55){2}
\ArrowLine(65,85)(110,0)
\ArrowLine(30,30)(30,0)
\ArrowLine(70,30)(70,0)
\Text(65,95)[]{$x$}
\Text(30,-10)[]{$x_1$}
\Text(50,-10)[]{...}
\Text(50,0)[]{$\check{x_k}$}
\Text(70,-10)[]{$x_n$}
\Text(110,-10)[]{$x_k$}
\Text(93,50)[l]{$(\Box+m^2)$}
\end{picture} \nonumber
\end{align}
\\ \\ \\

where the dashed arrow line
\begin{picture}(50,10)(0,0)
\Vertex(5,5){1}
\Vertex(45,5){1}
\DashArrowLine(5,5)(45,5){2}
 \end{picture}
can be
\begin{picture}(50,10)(0,0)
\Vertex(5,5){1}
\Vertex(45,5){1}
\ArrowLine(5,5)(45,5)
 \end{picture}or
\begin{picture}(50,10)(0,0)
\Vertex(5,5){1}
\Vertex(45,5){1}
\Line(5,5)(45,5)
\end{picture}
and the dashed line
\begin{picture}(50,10)(0,0)
\Vertex(5,5){1}
\Vertex(45,5){1}
\DashLine(5,5)(45,5){2}
 \end{picture}
stands for
\begin{picture}(50,10)(0,0)
\Vertex(5,5){1}
\Vertex(45,5){1}
\ArrowLine(5,5)(45,5)
 \end{picture},
\begin{picture}(50,10)(0,0)
\Vertex(5,5){1}
\Vertex(45,5){1}
\ArrowLine(45,5)(5,5)
 \end{picture}
 or
\begin{picture}(50,10)(0,0)
\Vertex(5,5){1}
\Vertex(45,5){1}
\Line(5,5)(45,5)
\end{picture}.
We have used $(\Box+m^2)\Delta^{(1)}(x)=0$ to skip diagrams that have a line $\Delta^{(1)}$ between $x$ and $y$ resp. $x$ and $x_k$. \\Applying   $(\Box+m^2)\Delta^{ret}(x)=\delta(x)$ we recognize the last diagram as contact terms, such that
\begin{align}
r^{\phi (\Box+m^2)\phi}(x;x_1...x_n)=&
\begin{picture}(100,70)(0,40)
\Vertex(50,85){1}
\Vertex(30,0){1}
\Vertex(70,0){1}
\GOval(50,42.5)(17,30)(0){0.75}
\DashArrowLine(50,85)(30,55){2}
\DashLine(50,85)(70,55){2}
\DashLine(50,85)(50,60){2}
\ArrowLine(30,30)(30,0)
\ArrowLine(70,30)(70,0)
\Text(50,95)[]{$x$}
\Text(30,-10)[]{$x_1$}
\Text(50,-10)[]{...}
\Text(70,-10)[]{$x_n$}
\end{picture}
\quad + \quad \text{c.t.} \label{eqmol}
\end{align}
\\
\\ \\ \\
The right hand side of equation (\ref{eqmo}) yields in terms of diagrams
\begin{align}
r^{g\phi^3}(x;x_1...x_n)=&
\begin{picture}(100,70)(0,40)
\Vertex(50,85){1}
\Vertex(30,0){1}
\Vertex(70,0){1}
\GOval(50,42.5)(17,30)(0){0.75}
\DashArrowLine(50,85)(30,55){2}
\DashArrowLine(50,85)(70,55){2}
\DashArrowLine(50,85)(50,60){2}
\ArrowLine(30,30)(30,0)
\ArrowLine(70,30)(70,0)
\Text(50,95)[]{$x$}
\Text(30,-10)[]{$x_1$}
\Text(50,-10)[]{...}
\Text(70,-10)[]{$x_n$}
\end{picture}
\end{align}
\\ \\
To verify that both sides are equal up to contact terms, we need to show that diagrams belonging to (\ref{eqmol}) with a dashed line being a $\Delta^{ret}$-function that points to $x$ are zero. But this results from the following Lemma, when we consider the point $x$ not as an outer point but a vertex of the diagram and remember that a closed cycle of $\Delta^{ret}$-functions vanishes.
\begin{lemma}
 A diagram having at each vertex at least one incoming $\Delta^{ret}$-function attached and the outer points connected by outgoing $\Delta^{ret}$-functions contains a closed cycle of $\Delta^{ret}$-functions.
\end{lemma}
\emph{Proof:} Let $\{z_i\}_{i=1}^n$ be the set of vertices, at each $z_i$ we have a function $\Delta^{ret}(a_i-z_i)$, and $a_i$ must, as the outer points are connected by outgoing $\Delta^{ret}$-functions, be itself out of $\{z_i\}_{i=1}^n$. We can now use the same argumentation as in the proof of unitarity to obtain a closed cycle of $\Delta^{ret}$-functions.

\section{Retarded functions in noncommutative theories}

We implement noncommutativity by bringing the star product into the interaction functional, e.g. in noncommutative $\phi^3$-theory $I[q]=\frac{g}{3!} \int dx (q\star q\star q)(x)$. This results in star multiplication at each vertex, in the Fourier representation meaning that we associate to every vertex a noncommutative phase factor $V(\pm p_1,...,\pm p_n)$ if $p_1,...,p_n$ are the momenta flowing
${\scriptsize \bigl\{
       \begin{aligned}&\text{in}\\
                 &\text{out}\end{aligned}\bigr\}}$
of the vertex. This phase factor is given by the n-point-function at first order, e.g.  in $\phi^3$-theory it reads
\beq
V(p_1,p_2,p_3)= \frac{1}{6}\sum_{\pi\epsilon S_3} e^{-i(p_{\pi(1)}\wedge p_{\pi(2)}+ p_{\pi(1)}\wedge p_{\pi(3)} +  p_{\pi(2)}\wedge p_{\pi(3)})} \qquad .
\eeq
This reminds us of the modified Feynman rules presented in chapter \ref{spatial}, in fact, they are similar in the sense that also there one takes the star product after a time-ordering is performed.\\

We formulate the diagrammatic rules in momentum space, which are essentially the ones as in subsection \ref{dia1}, the only change being the additional phase factor:
\begin{enumerate}
\item $x,x_1,...,x_n$ are the endpoints of the diagram, inner points are called vertices
\item $\tilde{\Delta}^{ret}(p)$ is symbolized by
\begin{picture}(50,10)(0,0)
\ArrowLine(5,5)(45,5)
\Text(25,-2)[]{$p$}
\end{picture},$\quad$
  $\tilde{\Delta}^{(1)}(p)$ by
\begin{picture}(50,10)(0,0)
\Line(5,5)(45,5)
\LongArrow(15,0)(35,0)
\Text(25,-7)[]{$p$}
 \end{picture}\quad .
\item $x$ is connected by one line,
\begin{picture}(50,10)(0,0)
\Vertex(5,5){1}
\ArrowLine(5,5)(45,5)
\Text(5,-2)[]{$x$}
\Text(25,-2)[]{$p$}
\end{picture}
 or
\begin{picture}(50,15)(0,5)
\Vertex(5,10){1}
\Line(5,10)(45,10)
\LongArrow(15,5)(35,5)
\Text(25,-2)[]{$p$}
\Text(5,3)[]{$x$}
 \end{picture}
 and carries a factor $e^{-ipx}$.\\
  The points $x_i$ are also connected by one line each,
\begin{picture}(50,10)(0,0)
\Vertex(5,5){1}
\ArrowLine(45,5)(5,5)
\Text(5,-2)[]{$x_i$}
\Text(25,-2)[]{$p_i$}
\end{picture}
 and contribute a factor $e^{ip_ix_i}$.
  \item A vertex with momenta $p_1,...,p_n$ flowing
${\scriptsize \bigl\{
       \begin{aligned}&\text{in}\\
                 &\text{out}\end{aligned}\bigr\}}$
contributes the coupling constant $g$, the noncommutative phase factor $V(\pm p_1,...,\pm p_n)$ and the momentum conservation $\delta(\pm p_1 ...\pm p_n)$.
\item The number of incoming $\Delta^{ret}$-functions at each vertex is odd.
\item When put into the reduction formula (\ref{LSZ2}) only diagrams contribute where $x$ is connected by a $\Delta^{ret}$-function.
\end{enumerate}

\subsection{Unitarity of the fishgraph}

Applying the diagrammatic rules for retarded functions, we will discuss a two by two scattering process in $\phi^4$-theory in second order perturbation theory and show the validity of the optical theorem
\begin{align}
2~\text{Im}
\begin{picture}(100,30)(0,25)
\ArrowArcn(50,0)(40,45,20)
\ArrowArcn(50,0)(40,160,135)
\CArc(50,0)(40,45,135)
\ArrowArc(50,60)(40,315,340)
\ArrowArc(50,60)(40,200,225)
\DashLine(50,55)(50,5){2}
\CArc(50,60)(40,225,315)
\Text(10,55)[]{$$}
\Text(10,5)[]{$$}
\Text(93,55)[]{$$}
\Text(93,5)[]{$$}
\end{picture}
=\int d\mu ~~\Bigg|
\begin{picture}(50,30)(10,45)
\ArrowLine(15,70)(35,50)
\ArrowLine(15,30)(35,50)
\ArrowLine(35,50)(55,70)
\ArrowLine(35,50)(55,30)
\end{picture}\Bigg|^2 \qquad .
\cr
\end{align}
The S-matrix element $S(p_1,p_2;q_1,q_2)$ can be computed from retarded functions by (\ref{LSZ2}), the only diagram contributing to $r(y_2;y_1x_1x_2)$ is
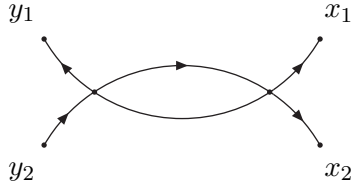
\begin{center}
\begin{picture}(300,90)(0,0)
\Vertex(98,70){1}
\Vertex(98,30){1}
\Vertex(202,70){1}
\Vertex(202,30){1}
\Vertex(117,50){1}
\Vertex(183,50){1}
\ArrowArcn(150,0)(60,55,30)
\ArrowArcn(150,0)(60,125,55)
\ArrowArcn(150,0)(60,150,125)
\CArc(150,100)(60,235,305)
\ArrowArcn(150,100)(60,235,210)
\ArrowArc(150,100)(60,305,330)
\Text(90,80)[]{$y_1$}
\Text(90,20)[]{$y_2$}
\Text(210,80)[]{$x_1$}
\Text(210,20)[]{$x_2$}
\end{picture}
\end{center}
such that
\begin{align}
r(y_2;y_1x_1x_2)=&
\begin{picture}(300,30)(0,50)
\Vertex(98,70){1}
\Vertex(98,30){1}
\Vertex(202,70){1}
\Vertex(202,30){1}
\Vertex(117,50){1}
\Vertex(183,50){1}
\ArrowArcn(150,0)(60,55,30)
\ArrowArcn(150,0)(60,125,55)
\ArrowArcn(150,0)(60,150,125)
\CArc(150,100)(60,235,305)
\ArrowArcn(150,100)(60,235,210)
\ArrowArc(150,100)(60,305,330)
\Text(90,80)[]{$y_1$}
\Text(90,20)[]{$y_2$}
\Text(210,80)[]{$x_1$}
\Text(210,20)[]{$x_2$}
\Text(117,60)[]{$z_1$}
\Text(183,60)[]{$z_2$}
\Text(98,58)[]{$p_1'$}
\Text(112,33)[]{$p_2'$}
\Text(150,70)[]{$k_1$}
\LongArrow(140,35)(160,35)
\Text(150,25)[]{$k_2$}
\Text(202,58)[]{$q_1'$}
\Text(188,33)[]{$q_2'$}
\end{picture} \cr\cr\cr\cr
=&
\int dz_1dz_2dp_1'dp_2'dq_1'dq_2'dk_1dk_2\frac{1}{(2\pi)^{24}}\tilde{\Delta}^{ret}(p_1')\tilde{\Delta}^{ret}(p_2')\tilde{\Delta}^{ret}(k_1)\tilde{\Delta}^{(1)}(k_2) \cr
&\quad \tilde{\Delta}^{ret}(q_1')\tilde{\Delta}^{ret}(q_2') V(-p_1',-k_1,-k_2,p_2')V(-q_1',-q_2',k_2,k_1)\cr
&\quad  e^{-ip_1'(z_1-y_1)}e^{-ip_2'(y_2-z_1)}e^{-iq_1'(z_2-x_1)}e^{-iq_2'(z_2-x_2)}e^{-ik_1(z_1-z_2)}e^{-ik_2(z_1-z_2)}\cr
=&
\int dp_1'dp_2'dq_1'dq_2'dk_1dk_2\frac{1}{(2\pi)^{16}}\tilde{\Delta}^{ret}(p_1')\tilde{\Delta}^{ret}(p_2')\tilde{\Delta}^{ret}(k_1)\tilde{\Delta}^{(1)}(k_2) \cr
&\quad \tilde{\Delta}^{ret}(q_1')\tilde{\Delta}^{ret}(q_2') V(-p_1',-k_1,-k_2,p_2')V(-q_1',-q_2',k_2,k_1)\cr
&\quad  e^{ip_1'y_1}e^{-ip_2'y_2}e^{iq_1'x_1}e^{iq_2'x_2}\delta(p_1'-p_2'+k_1+k_2)\delta(q_1'+q_2'-k_1-k_2)
\end{align}
and so for the S-matrix element according to formula (\ref{LSZ2}), if we make use of\\
 $K_x\tilde{\Delta}^{ret}(p)e^{-ipx}=e^{-ipx}$~:
\begin{align}
S(p_1,p_2;q_1,q_2)=&\frac{-1}{(2\pi)^8}\int dk_1dk_2\tilde{\Delta}^{ret}(k_1)\tilde{\Delta}^{(1)}(k_2) \cr
&V(p_1,-k_1,-k_2,p_2)V(-q_1,-q_2,k_2,k_1)\cr
&\delta(-p_1-p_2+k_1+k_2)\delta(q_1+q_2-k_1-k_2) \quad .
\end{align}
Applying
\begin{align}
\tilde{\Delta}^{ret}(k)&=\frac{-1}{(k+i\epsilon)^2-m^2}=\sum_{\lambda=\pm 1}\frac{-\lambda}{2E_k(k^0-\lambda E_k+i\epsilon}\cr
\Rightarrow&\quad \text{Im}\tilde{\Delta}^{ret}(k)=-\sum_{\lambda=\pm1}\frac{2\pi\lambda}{2E_k}\delta(k^0-\lambda E_k)\cr
\tilde{\Delta}^{(1)}(k)&=2\pi\delta(k^2+m^2)=\sum_{\lambda=\pm 1}\frac{2\pi}{2E_k}\delta(k^0-\lambda E_k) \cr
\Rightarrow &\quad\text{Im}\tilde{\Delta}^{(1)}(k)=0\nonumber
\end{align}
and the reality of the phase factors V we obtain
\begin{align}
\text{Im}S(p_1,p_2;q_1,q_2)=~&\frac{1}{(2\pi)^6}\int dk_1dk_2\sum_{\lambda_1,\lambda_2=\pm 1}\frac{\lambda_1}{2E_{k_1}2E_{k_2}}\delta(k_1^0-\lambda_1E_{k_1})\delta(k_2^0-\lambda_2E_{k_2})\cr
&\quad V(p_1,-k_1,-k_2,p_2)V(-q_1,-q_2,k_2,k_1)\cr
&\quad\delta(-p_1-p_2+k_1+k_2)\delta(q_1+q_2-k_1-k_2)\cr
=~&\frac{1}{(2\pi)^6}\int dk^3_1dk^3_2\sum_{\lambda_1,\lambda_2=\pm 1}\frac{\lambda_1}{2E_{k_1}2E_{k_2}}
V(p_1,-k_1^{\lambda_1},-k_2^{\lambda_2},p_2)\cr
&\quad V(-q_1,-q_2,k_2^{\lambda_2},k_1^{\lambda_1})
\delta(-p_1-p_2+k_1^{\lambda_1}+k_2^{\lambda_2})\delta(q_1+q_2-k_1^{\lambda_1}-k_2^{\lambda_2}) \cr
\end{align}
where the functions $\delta(k_1^0-\lambda_1E_{k_1}),\delta(k_2^0-\lambda_2E_{k_2})$ allowed us to carry out the integration over $k^0_1$ and $k^0_2$.
It is now a kinematic argument that for on-shell momenta $p_1,p_2,q_1,q_2$ the conditions
$\delta(-p_1-p_2+k_1^{\lambda_1}+k_2^{\lambda_2})$ and $\delta(q_1+q_2-k_1^{\lambda_1}-k_2^{\lambda_2})$ can only be fulfilled if $\lambda_1=\lambda_2=+1$, which leads to
\begin{align}
\text{Im}S(p_1,p_2;q_1,q_2)=~&\frac{1}{(2\pi)^6}\int\frac{dk^3_1}{2E_{k_1}}\frac{dk^3_2}{2E_{k_2}}
V(p_1,-k_1^+,-k_2^+,p_2) V(-q_1,-q_2,k_2^+,k_1^+)\cr
&\quad \delta(-p_1-p_2+k_1^++k_2^+)\delta(q_1+q_2-k_1^+-k_2^+)\quad .
\end{align}

The other side of the unitarity relation reads
\begin{align}
&\int\frac{dk^3_1}{(2\pi)^32E_{k_1}}\frac{dk^3_2}{(2\pi)^32E_{k_2}}
\begin{picture}(100,50)(0,50)
\Vertex(50,50){1}
\ArrowLine(30,70)(50,50)
\ArrowLine(30,30)(50,50)
\ArrowLine(50,50)(70,70)
\ArrowLine(50,50)(70,30)
\Text(25,75)[]{$p_1$}
\Text(25,25)[]{$p_2$}
\Text(75,75)[]{$k_1$}
\Text(75,25)[]{$k_2$}
\end{picture}
\begin{picture}(100,50)(0,50)
\Vertex(50,50){1}
\ArrowLine(30,70)(50,50)
\ArrowLine(30,30)(50,50)
\ArrowLine(50,50)(70,70)
\ArrowLine(50,50)(70,30)
\Text(25,75)[]{$k_1$}
\Text(25,25)[]{$k_2$}
\Text(75,75)[]{$q_1$}
\Text(75,25)[]{$q_2$}
\end{picture}
\cr\cr\cr
=~&\frac{1}{(2\pi)^6}\int\frac{dk^3_1}{2E_{k_1}}\frac{dk^3_2}{2E_{k_2}}V(p_1,-k_1^+,-k_2^+,p_2) V(-q_1,-q_2,k_2^+,k_1^+)\cr
&\quad \delta(-p_1-p_2+k_1^++k_2^+)\delta(q_1+q_2-k_1^+-k_2^+) \quad,
\end{align}
coinciding with the left hand side.

\subsection{The question of unitarity in general}

In a previous subsection we proved the generalized unitarity condition
\beq
\mathcal{R}[0,j]=1
\eeq
in the case of commutative $\phi^m$-theory, the proof easily generalizes to other commutative theories. The subtle point was to see that the only diagrams appearing in $\mathcal{R}[0,j]$ at higher orders in the coupling constant than zero necessarily involve a closed cycle of $\Delta^{ret}$-functions, i.e. an expression of the type
\beq
\Delta^{ret}(z_{1}-z_{2})\Delta^{ret}(z_{2}-z_{3})\cdots\Delta^{ret}(z_{k}-z_{1}) \quad \label{closed_cycle}.
\eeq
From the support property of the $\Delta^{ret}$-functions
\beq
\Delta^{ret}(x)\neq0 \quad \text{only for} \quad x^0>0
\eeq
we find as a condition that (\ref{closed_cycle}) does not vanish
\beq
z_1^0>z_2^0>...>z_k^0>z_1^0
\eeq
which cannot be fulfilled, meaning that (\ref{closed_cycle}) is zero. \newline

In noncommutative theories the above arguments that lead us to higher-order diagrams all involving a closed cycle of $\Delta^{ret}$-functions go through, one only has to consider the star multiplication at each vertex, so that a closed cycle of $\Delta^{ret}$-functions now has the form
\beq
\begin{picture}(0,0)(0,0)
\LongArrow(5,-10)(5,-3)
\Line(5,-10)(246,-10)
\Line(246,-5)(246,-10)
\end{picture}
\Delta^{ret}(z_{1}-z_{2})\stackrel{z_2}{\star}\Delta^{ret}(z_{2}-z_{3})\stackrel{z_3}{\star}...\stackrel{z_k}{\star}\Delta^{ret}(z_{k}-z_{1})\stackrel{z_1}{\star} \quad . \label{star_cycle}
\eeq
If time is not involved in the star product, the earlier argumentation using the support properties of the $\Delta^{ret}$-functions, formulated in terms of the time coordinate, to conclude that this expression vanishes, still holds. This means that for $\theta^{0i}=0$ we have unitarity.\newline

 But in the case $\theta^{0i}\neq 0$ this can no longer be maintained, as we then also smear over the time coordinate. In fact, it was argued in \cite{BahnsDiss}, that e.g.
$\Delta^{ret}(x)\star\Delta^{ret}(-x)\neq 0$. The diagrams involving expressions (\ref{star_cycle}) thus are the ones which violate unitarity if time does not commute with space. \\

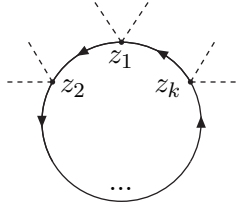
\begin{figure}[h]
\begin{center}
\begin{picture}(150,90)(0,50)
\Vertex(75,105){1}
\DashLine(75,105)(65,120){2}
\DashLine(75,105)(85,120){2}
\Vertex(49,90){1}
\DashLine(49,90)(40,105){2}
\DashLine(49,90)(32,90){2}
\Vertex(101,90){1}
\DashLine(101,90)(110,105){2}
\DashLine(101,90)(118,90){2}
\ArrowArc(75,75)(30,90,150)
\ArrowArc(75,75)(30,30,90)
\ArrowArc(75,75)(30,150,210)
\ArrowArc(75,75)(30,-30,30)
\CArc(75,75)(30,30,-30)
\Text(75,99)[]{$z_1$}
\Text(57,88)[]{$z_2$}
\Text(93,88)[]{$z_k$}
\Text(75,50)[]{$...$}
\end{picture}
\parbox{8cm}{\caption{\label{diag_cycle}A closed cycle of $\Delta^{ret}$-functions}}
\end{center}
\end{figure}

Let us closer examine one-loop diagrams exhibiting (\ref{star_cycle}), for which the graph is drawn in Fig. \ref{diag_cycle}.
As we are on the one-loop level, the graphs attached to the $z_i$'s different from the closed cycle are tree graphs, we refer to them as the tree-part of the diagram.

The following proposition holds:
\begin{prop}
A one-loop diagram contributing to $r(x;x_1,...x_n)$ and containing a closed cycle of $\Delta^{ret}$-functions
contains at least one function $\Delta^{(1)}$ in the tree-part.
\end{prop}
 We first show the following
\begin{lemma}
\begin{itemize}
\item[(i)] In a tree diagram with outer points $x_1,...,x_n,z$ where $x_1,...,x_n$ are connected by ingoing $\Delta^{ret}$-functions and every vertex has an odd number of incoming $\Delta^{ret}$-functions the point $z$ is connected by an outgoing $\Delta^{ret}$-function.
\item[(ii)] In a tree diagram with outer points $x,x_1,...,x_n,z$ where x is connected by an outgoing $\Delta^{ret}$-function, $x_1,...,x_n$ by ingoing $\Delta^{ret}$-functions, $z$ \emph{not} by an ingoing $\Delta^{ret}$ and where each vertex has an odd number of incoming $\Delta^{ret}$-functions we find a $\Delta^{(1)}$-function appearing.
\end{itemize}
\end{lemma}
\emph{Proof:}
Let $V$ be the number of vertices of the diagram, $I$ the number of internal lines, on tree-level we have the relation $V=I+1$ (see the proof of the generalized unitarity condition in commutative theories).\\
(i) To have at least one incoming $\Delta^{ret}$ at each vertex, we need at least $V$ lines with $\Delta^{ret}$-functions flowing into a vertex, from $V=I+1$ we conclude that at least one must be an external line, this can only be the line connecting the point $z$, meaning that $z$ has to be connected by an outgoing $\Delta^{ret}$-function.\\
(ii) $z$ is not connected by an ingoing $\Delta^{ret}$-function, that means it can be connected by
$\Delta^{(1)}$, in which case we are done, or by an outgoing $\Delta^{ret}$. Suppose no $\Delta^{(1)}$ would appear in the second case, then we would have I+2=V+1 lines with $\Delta^{ret}$-functions flowing into vertices.
But to have an odd number of incoming $\Delta^{ret}$-functions at each vertex, the diagram may only contain $V, V+2,V+4,...$ lines with $\Delta^{ret}$ flowing into a vertex, yielding the contradiction.\\

We now come back to prove the proposition. If in such a diagram the point $x$ is connected by  $\Delta^{(1)}$ we are finished. Otherwise we have $x$ connected via an outgoing $\Delta^{ret}$ and $x_1,...,x_n$ connected by ingoing$ \Delta^{ret}$-functions.
To apply the lemma we notice that every line going out of the circle in (\ref{diag_cycle}) from the vertices $z_1,...,z_k$ is connected to a tree diagram, with outer points among the $x,x_1,...,x_n$. Only one of these lines can thus be connected to a tree diagram having $x$ as outer point, and using the lemma we find that the other lines  must be outgoing $\Delta^{ret}$-functions (treat the vertex $z_i$ as the outer point $z$ in part (i)).
The missing line, say at point $z_1$, connected to a tree diagram having $x$ as outer point cannot be an ingoing $\Delta^{ret}$, as we then would have two incoming $\Delta^{ret}$-functions at $z_1$. Applying part (ii) of the lemma (consider $z_1$ as the outer point $z$) we see that a $\Delta^{(1)}$-function must appear in this tree diagram.\\

We now leave the question of the validity of the generalized unitarity condition and focus on the unitarity on the level of S-matrix elements, which is the only physically relevant one. We encounter
\begin{prop}
If we consider an S-matrix element on one-loop level for a scattering process of momenta $(p_1,...,p_m) \rightarrow(q_1,...,q_n)$ 
 the optical theorem is valid up to a set of momentum configurations
which has measure zero within the set of all possible configurations\\
$\{(p_1,...,p_m,q_1,...,q_n)\in\mathbb{R}^{m+n}~|~p_1+...+p_m=q_1+...+q_n\}$.
\end{prop}
\emph{Proof:} It follows from the above consideration of the generalized unitarity condition that the violation of the optical theorem can only occur for diagrams that contain a closed cycle of $\Delta
^{ret}$-functions, and so have a $\Delta^{(1)}$ in the tree part. Consider this diagram in the momentum space representation: let $\Delta^{(1)}$ be connected to momentum $k$. Because this appears in the tree part of the diagram, $k$ is given by a sum of external momenta:
\begin{align}
\pm k=p_{i_1}+...+ p_{i_r}- q_{j_1}-...-q_{j_s}\quad; \quad &i_1,...,i_r\epsilon\{1,...,m\}\cr &j_1,...,j_s\in\{1,...,n\}
\end{align}
We thus notice the function
\beq
\tilde{\Delta}^{(1)}(k)=2\pi\delta(k^2+m^2)
\eeq
to put an extra condition on the above sum of external momenta, in fact being only fulfilled on a hypersurface of one dimension less than the space of all possible external momenta. The set of momentum configurations where the optical theorem may be violated has thus measure zero in the set of all possible momentum configurations.

\subsection{Equations of motion and currents}

We now want to discuss the question whether the classical equations of motion and currents are still valid on the quantized level in noncommutative  theories. From the considerations in subsection \ref{comp} of the bilinear equation of motion in the commutative case we learn that the only diagrams which might spoil the classical equations contain a closed cycle of $\Delta^{ret}$-functions. As this still vanishes for $\theta^{0i}=0$ we conclude that we recover the classical equations in this case. \\
However, if $\theta^{0i}$ does not vanish, these diagrams will be different from zero. Notice that the violation of the classical equations of motion and currents due to these terms may earliest occur on one-loop level, on the tree-level we still recover the classical results.

\subsection{A modified theory \label{modified}}

Let us first summarize our results so far. For space-time noncommutativity unitarity has turned out to be violated and the classical equations of motion and currents do not hold on the quantized level. In both cases these unpleasant outcomes are exactly due to diagrams which contain a closed cycle of $\Delta^{ret}$-functions. Their vanishing for $\theta^{0i}=0$ is the reason that in this case the approach via retarded functions yields a unitary theory and respects the classical equations. \\
It is thus obvious that we encounter unitarity as well as validity of classical equations of motion and currents on the quantized level if we modify the theory by the requirement that we do not allow diagrams which exhibit a closed cycle of $\Delta^{ret}$-functions. This altered theory can probably not be derived from a functional like (\ref{ret_func}), instead it is  defined by the diagrammatical rules of subsection \ref{dia1} together with the rule of subsection \ref{comp} if we impose the additional requirement
\begin{itemize}
\item[7.] A diagram may not contain a closed cycle of $\Delta^{ret}$-functions.
\end{itemize}
As such diagrams vanish for $\theta^{0i}=0$ the equivalence of the modified theory with the ordinary one derived from (\ref{ret_func}) in the case of only spatial noncommutativity is evident.

\newpage

\newpage
\thispagestyle{plain}
\mbox{}       

\chapter*{Outlook}

\markboth{Outlook}{}

Time-space noncommutativity poses remarkable difficulties.
To formulate a perturbative approach which satisfies the fundamental requirement of unitarity and respects the classical equations of motion and currents on the quantized level is still a task to work on. Our suggestion of such a theory is formulated in subsection \ref{modified}, future work in this approach could focus on the divergence properties of the one and higher loop levels. \\
TOPT has been shown to suffer from fundamental problems. These are the violation of Ward identities which could be traced back to additional terms in the quantized current, and the no longer covariant behaviour under remaining Lorentz symmetry. The latter may suggest a formulation of time-ordering that leads to modifications in such a way that remaining Lorentz symmetry is preserved, such work has been done very recently \cite{HeslopSibold}. The new formulation seems, apart from its unitarity, to respect classical equations of motion and currents and should be closer studied.  \\
The validity of equations of motion and currents within the Yang-Feldman approach was not investigated and should also be a subject of further studies. In this respect one will have to define time-ordered Green's functions, first steps in this direction were carried out in \cite{BahnsDiss}.\\

\addcontentsline{toc}{chapter}{Outlook}
 
\newpage
\thispagestyle{plain}
\mbox{}       
\newpage

\begin{appendix}

\chapter{Notations and useful relations}

We write noncommutative phase factors most often in a form that uses the wedge product
\beq
p\wedge q=\frac{1}{2}\theta^{\mu\nu}p_\mu q_\nu
\eeq
and the brief notation
\beq
(p_1,p_2,...,p_n)=\sum_{i<j\leq n}p_i\wedge p_j \quad .
\eeq
\\

In TOPT we do not longer encounter causal propagators, but parts of it. In the scalar case these are the quantities $P_\lambda(k)$ with $\lambda=\pm 1$
\beq
P_\lambda(k)=\frac{i\lambda}{2E_k(k^0-\lambda(E_k-i\epsilon))}
\eeq
which represent the Fourier transforms of the positive and negative energy contribution of the  propagator
\beq
\Delta_C(x-y)=\sum_{\lambda=\pm1}\int \frac{d^4k}{(2\pi)^4}P_\lambda(k)e^{-ik(x-y)}
\quad .
\eeq
For spin-$\frac{1}{2}$ particles the quantities $S_\lambda(k), \lambda=\pm 1$
\beq
S_\lambda(k)=\frac{i\lambda(\not\!k^\lambda+m)}{2E_k(k^0-\lambda(E_k-i\epsilon))} \quad
\eeq
 are important, which come from the decomposition
\beq
S_C(x-y)=\sum_{\lambda=\pm1}\int \frac{d^4k}{(2\pi)^4}S_\lambda(k)e^{-ik(x-y)} \quad .
\eeq
In the definition of $S_\lambda(k)$ we have just used the often met notation
\beq
k^\lambda=(\lambda E_k,k_1,k_2,k_3) \quad .
\eeq
\\

Discussing the approach via retarded functions we encounter two quantities that can be assigned to lines, at first the retarded Green's function
 \beq
\Delta^{ret}(x)=\lim_{\epsilon\rightarrow +0} \frac{-1}{(2\pi)^4}\int d^4k\frac{e^{-ikx}}{(k+i\epsilon)^2-m^2}
\eeq
which vanishes for $x^0\leq 0$ and fullfills $(\Box+m^2)\Delta^{ret}(x)=\delta(x)$; and at second  the function
\beq
\Delta^{(1)}(x)=\frac{1}{(2\pi)^3}\int d^4k~ \delta(k^2+m^2)e^{-ikx}
\eeq
which is a solution to the homogeneous Klein-Gordon equation: $(\Box+m^2)\Delta^{(1)}(x)=0$.

\end{appendix}

\newpage

\thispagestyle{plain}

\section*{Acknowledgments}

\hspace{1.5 cm}

I wish to express my gratitude to Prof. Dr. Klaus Sibold for introducing  me to noncommutative theories and for his decisive hints leading me to the study of quantized equations of motion and current conservation laws. With constant support, he has guarded me during my first steps in scientific work. \\

Warm thanks is also given to the other members in our group, Yi Liao, Christoph Dehne and Paul Heslop for the often vital and inspiring discussions, from which I benefited a lot. As only one example, the result on violation  of remaining Lorentz invariance goes back to a lively debate on the transformation properties of $\theta^{\mu\nu}$.\\

All members at the institute have created a motivating and pleasant working atmosphere, which I am thankful for.\\

Financial support of the Carl-von-Bach Stiftung is gratefully acknowledged, it has enabled me to study unoffended of economic worries.\\

Last but not least, I am grateful to my family and my friends. The time of recreation that I could spend with them and the stimulations which I received were essential during the writing of this thesis.

\newpage
\thispagestyle{plain}
\mbox{}       

\backmatter

\end{document}